\documentclass[12pt]{article}
\usepackage{amsmath,amssymb,array,cite,latexsym, arydshln}

\newcommand{\be}{\begin{equation}}
\newcommand{\ee}{\end{equation}}
\def\bea{\begin{eqnarray}}
\def\eea{\end{eqnarray}}

\begin{document}


\thispagestyle{empty}

\begin{center}
\Large{\bf Kronecker product in terms of Hubbard operators and the Clebsch-Gordan decomposition of $SU(2) \times SU(2)$ }

\end{center}

\vskip2ex
\begin{center}
Marco Enr\'{\i}quez and Oscar Rosas-Ortiz\footnote{Corresponding Author. Email: orosas@fis.cinvestav.mx. 
}\\[2ex]
Departamento de F\'{\i}sica, Cinvestav,\\
AP 14-740, 07000 M\'exico~DF, Mexico
\end{center}

\begin{center}
\begin{minipage}{14cm}
{\footnotesize {\bf Abstract.} We review the properties of the Kronecker (direct, or tensor) product of square matrices $A \otimes B \otimes C \cdots$ in terms of Hubbard operators. In its simplest form, a Hubbard operator $X_n^{i,j}$ can be expressed as the $n$-square matrix which has entry 1 in position $(i,j)$ and zero in all other entries. The algebra and group properties of the observables that define a multipartite quantum system are notably straightforward in such a framework. In particular, we use the Kronecker product in Hubbard notation to get the Clebsch-Gordan decomposition of the product group $SU(2) \times SU(2)$. Finally, the $n$-dimensional irreducible representations so obtained are used to derive closed forms of the Clebsch-Gordan coefficients that rule the addition of angular momenta. Our results can be further developed in many different directions.}
\end{minipage}
\end{center}

\vskip2ex
\noindent
{\footnotesize
}

\vskip1cm
\tableofcontents
\vskip1cm


\section{Introduction}
\setcounter{page}{1}
\setcounter{footnote}{0}

The Kronecker product, represented by the symbol $\otimes$, has attracted the attention of researchers in diverse areas of mathematics and theoretical physics over the last decades \cite{Bre78,Gra81,Hen83,Bla83,Hor91,Van00,Lan04,Ste11,Ste12}. Introduced by Zehfuss in 1858 (see the historical review given in \cite{Hen83}), this is a matrix operation also known as direct or tensor product, defined for matrices $A=[a_{i,j}]$ and $B$ of any order to be $A \otimes B =[a_{i,j} B]$. That is, for instance 
\[
A \otimes B= \left(
\begin{array}{ccc}
a_{11} & a_{12} & a_{13}\\
a_{21} & a_{22} & a_{23}
\end{array} \right) \otimes B  = \left(
\begin{array}{c|c|c}
a_{11}B & a_{12}B & a_{13}B\\
\hline
a_{21}B & a_{22}B & a_{23}B
\end{array} \right).
\]
Matrix calculus includes the derivatives of a matrix with respect to a scalar, a scalar with respect to a matrix, and a matrix with respect to a matrix; all these operations are defined in terms of the product $\otimes$ \cite{Gra81,Hor91,Ste11} (other interesting applications can be found in \cite{Van00}). In physics, this product arises quite naturally if the group properties of the dynamical variables of a given system are considered \cite{Wey31,Wig59,Rac65}, and it is fundamental in the study of multipartite systems \cite{Per95,Nie00}. Indeed, the Clebsch-Gordan decomposition of the Kronecker product of two irreducible representations is one of the most useful problems in group theory, since the reduction of such a product into the sum of irreducible representations confirms the unicity of the representation for the simple reducible groups \cite{Bac77}. It is then quite natural to find immediate applications of both, the product $\otimes$ and the Clebsch-Gordan decomposition, in the addition of angular momenta \cite{Bie81} as well as in the identification of symmetries in quantum physics \cite{Fon70}.

Despite the simplicity of its definition, calculating the tensor product $A\otimes B\otimes C \otimes \cdots$ becomes cumbersome for large matrix sizes and/or for a large number of factors. This fact is particularly notable in the design of fast Fourier transform algorithms where the factorisation of the discrete Fourier matrix $\widetilde F_n$ is relevant. Namely, for $n=2m$ with $m \in \mathbb N$, the $n$-square (Fourier) matrix\footnote{A Fourier matrix $\widetilde F_n = F_n/\sqrt n$ is unitary. Here we also refer to (\ref{fourier}) as Fourier matrix although it is a rescaled version of $\widetilde F_n$.}:
\be
F_n=[f_{i,j}], \quad f_{i,j}= w^{(j-1)(i-1)}, \quad i,j=1,2,\ldots, n, \quad w = e^{2\pi i /n},
\label{fourier}
\ee
can be expressed as the product
\[
F_n = B_n (\mathbb I_2 \otimes F_m) \Pi_n^T,
\]
where
\[
B_n = \left(
\begin{array}{rr}
\mathbb I_m & \Omega_m\\
\mathbb I_m & -\Omega_m
\end{array}
\right), \qquad \mathbb I_k = \mbox{diag} \underbrace{(1,1,\ldots,1)}_{k-\mbox{times}}, \qquad \Omega_k= \mbox{diag}(1,w, w^2, \ldots, w^{k-1}),
\]
and $\Pi_k$ is the $k\times k$-permutation matrix obtained by grouping the odd columns of the identity $\mathbb I_k$ first, and the even columns second. Hereafter $A^T$ is the matrix transpose of $A$. The procedure can be repeated if $m$ is even. Indeed, if  $n=2^t$, $t\in \mathbb N$, then $F_n$ can be factorized into the product of $t=\log_2 n$ matrix factors (Cooley-Tukey factorisation):
\be
F_n = (\mathbb I_1 \otimes B_n) (\mathbb I_2 \otimes B_{n/2})  (\mathbb I_4 \otimes B_{n/4}) \cdots  (\mathbb I_{n/2} \otimes B_2) P_n^T,
\label{tukey}
\ee
with $P_n^T$ the bit-reversing permutation matrix \cite{Van92}. Note that each factor $\mathbb I_k \otimes B_{n/k}$ in (\ref{tukey})  has only two nonzero entries per row. Thus, only $2n$ of the $n^2$ entries associated to $\mathbb I_k \otimes B_{n/k}$ are different from zero. This fact suggests that there must be a better and simpler form of calculating the Kronecker product in the factorisation algorithms. 

On the other hand, the entries of the first row and column of the Fourier matrix ({\ref{fourier}) are all equal to unity while the other entries are either $\pm 1$ or $\pm i$. Thus, $F_n$ is a dephased, complex Hadamard matrix \cite{Tad06}. The relevance of a dephased matrix is that only its lower right $(n-1)$-square sub-matrix is necessary in the calculations where it is involved.    In this subject, it can be shown that the product $D_r H D_c$ brings any Hadamard matrix $H$  into the dephased form for a pair of uniquely determined diagonal unitary matrices $D_r$ and $D_c$ \cite{Tad06}. The construction can be simplified since two Hadamard matrices, $H_1$ and $H_2$, are equivalent if there exist diagonal unitary matrices $D_1$ and $D_2$, and permutation matrices $P_1$ and $P_2$, such that \cite{Tad06,Haa96}:
\be
H_1 = D_1 P_1 H_2 P_2 D_2.
\label{inteq}
\ee
According to the former property, $H_1$ is dephased if $D_1=D_r$ and $D_2 =D_c$, no matter the form of $P_1H_2P_2$. In this case dephasing is equivalent to the permutation of rows and columns defined by $P_1$ and $P_2$. However, using conventional approaches, it is not easy to verify whether there exist such permutations \cite{Tad06}. The problem becomes even more complicated for large size matrices since the permutations grow as $N!$ Then, it is apparent the necessity of a proper framework in which the determination of the above described permutations becomes a tractable problem.

In general, the matrix algebra includes algorithms that are fairly complicated and cumbersome for either matrices of large sizes or a large number of matrices to operate with. It would therefore be desirable to construct a mathematical framework in which the problems like those aforementioned are feasible; no matter the number or the size of the matrices involved. To get a suitable approach it is useful to consider the operators $X^{p,q}$ introduced by Hubbard \cite{Hub63,Hub64,Hub65} (three of six papers). These obey the multiplication rule
\be
X^{i,j} X^{k,m} = \delta_{jk} X^{i,m},
\label{halg1}
\ee
and have the properties
\be
(X^{i,j})^{\dagger} = X^{j,i}, \quad \sum_{k} X^{k,k} = \mathbb I, \quad [X^{i,j}, X^{k,m}]_{\pm} = \delta_{jk} X^{i,m} \pm \delta_{mi} X^{k,j}.
\label{halg2}
\ee
Hereafter the sub-label in $[A,B]_{\pm}$ stands for either the commutator ($-$) or the anticommutator ($+$) of $A$ and $B$.

The Hubbard operators provide a way to study groups of particles that interact strongly one with each other in such a way that a weak interaction between the groups is also allowed\footnote{To distinguish between the particles of different groups, Hubbard used $X^{(i)}_{pq}$ \cite{Hub64} as well as $X_i^{pq}$ \cite{Hub65} to denote the operator $\vert i,p \rangle \langle i,q \vert$, with $i$ labelling a given group while $\vert i, p \rangle$ and $\vert i, q \rangle$ represent two different states of a particle in that group. In contrast, we use a sub-label ``$n$'' to denote the matrix order of the linear representation of the operator $X^{i,j}$, as this is done in equation (\ref{halg3}). An exception is done in Section~\ref{new} where the Hubbard model of strong interacting electrons is revisited.}\cite{Hub63,Hub64,Hub65,Ovc97,Ovc04,Dag94,Kor93,Kor94,Col01,Izy92,Sav96,Kos05,Kis06,Bul07}. For example, they are useful in the description of atoms in which Coulomb repulsion prevents double-occupancy of a given orbital \cite{Col01}. In such cases, the strong interactions determine the energy of the groups of particles and can be included in the Hamiltonian as linear combinations of the Hubbard operators \cite{Hub63,Hub64,Hub65}. This is the situation for strongly correlated electrons \cite{Hub63,Hub64,Hub65,Izy92}, whether they are in a cavity \cite{Sav96} or in a two-atom molecule \cite{Kos05}, as well as for double quantum dots \cite{Kis06,Bul07} among other systems \cite{Ovc97,Ovc04,Dag94,Kor93,Kor94}.

Given an $n$-dimensional vector space ${\cal H}_n$ with orthonormal basis $\{ \vert \psi_k \rangle \}_{k=1}^n$, the Hubbard operators are written in terms of the outer products of the basis elements:
\be
X_n^{i,j} := \vert \psi_i \rangle \langle \psi_j \vert, \quad i,j =1,2, \ldots, n.
\label{halg3}
\ee
That is, the operator $X_n^{i,j}=\vert \psi_i \rangle \langle \psi_j \vert$ is a representation of $X^{i,j}$ in the space ${\cal H}_n$. This causes a transition from the state $\vert \psi_j \rangle$ to the state $\vert \psi_i \rangle$ of the system that is described by the vectors in ${\cal H}_n$. In general, any linear operator $O : {\cal H}_n \rightarrow {\cal H}_n$ can be represented in terms of the Hubbard operators
\be
O = \sum_{i,j}o_{i,j} X_n^{i,j}, \quad o_{i,j} = \langle \psi_i \vert O \vert \psi_j \rangle.
\label{halg4}
\ee
This property plays a central role in what follows since the operators $X_n^{i,j}$ are the cornerstone of our approach. Using the $X$-operator representation (\ref{halg4}) one can address the algebra of square matrices in compact form, no matter the size or the number of the factors. Problems like the determination of the permutation matrices fulfilling (\ref{inteq}) become simpler in this notation. Indeed, if the basis vector $\vert \psi_k \rangle$ is the $n$-tuple that has a unity in the $k$-th position and all other entries equal to zero, the Hubbard operator (\ref{halg3}) is in its simplest form:
\be
X_n^{i,j} = {\footnotesize
\left(
\begin{array}{ccccccc}
&&& 0 &&&\\
& 0_{(i-1) \times (j-1)} & & \vdots & & 0_{(i-1) \times (n-j)} &\\
&& & 0 &&&\\
0 & \cdots & 0 & 1 & 0 & \cdots & 0  \\
&& & 0 &&&\\
& 0_{(n-i) \times (j-1)} &  & \vdots & & 0_{(n -i) \times (n-j)} & \\
&&& 0 &&&
\end{array}
\right),
}
\label{hub00}
\ee
where $0_{s\times t}$ is the null matrix of order $s\times t$. That is, in the representation defined by the basis vectors $\vert \psi_k \rangle$ the Hubbard operator $X_n^{i,j}$ corresponds to the $n$-square matrix for which all the entries are zero except the one at the $i$-th row and the $j$-th column, where it takes the value 1. Such an array of zeros and a single unit is appropriate to operate the Kronecker products $\mathbb I_k \otimes B_{n/k}$ of Eq.~(\ref{tukey}) in plain notation; these products are but the linear combination of only $2n$ Hubbard operators.

The present work attempts to stimulate further progress in the applications of the Hubbard operators by introducing a useful manner to calculate the Kronecker product. The paper is structured in two main parts. First, in Section~\ref{tithub} some basic definitions and the Hubbard notation are introduced. A simple model of interacting atoms is discussed in Section~\ref{new} where the corresponding Hamiltonian is expressed as a combination of linear and quadratic forms of the Hubbard operators. The simplest case corresponds to a net of atoms which have a single energy level of interest each one. Then, in Section~\ref{hubmodels} the Pauli principle and the notion of a unique orbital per site are used to recover the Hubbard model of strongly correlated electrons as well as its strong-coupling limit known as the t-J model. In Section~\ref{titkro} we review some of the most important properties of the direct product by using the Hubbard operators as the building-blocks of the Kronecker algebra of square matrices. Properties just as the composition of permutations are nicely worked in Hubbard notation. Some other properties as the Kronecker powers of operators $A^{\otimes k}$ are explicitly developed for their possible application in quantum control of multipartite systems. The Kronecker algebra of the $X$-operators is applied to solve two particular problems of physical interest in Section~\ref{phys}. We first discuss the problem of diagonalizing a given $n$-level Hamiltonian (Section~\ref{diagon}), and pay particular attention to a system of interacting spin-1/2 particles that is described by the Heisenberg model (a limit case of the models discussed in Section~\ref{new}). In Section~\ref{jay} we use the $X$-operator representation to solve the Jaynes-Cummings problem associated to a single-atom in a single-mode of quantised electromagnetic fields in a cavity. Some comments concerning the generalisations to include an arbitrary number of cavities are also given. In Section~\ref{titbas} the useful notation of the direct sum of vector spaces and the linear representation of groups are also revisited while the Clebsch-Gordan problem is stated in general form. The second part of the paper  is devoted to the applications involving angular momenta. We first review the construction of irreducible representations of the $SU(2)$ Lie group in Hubbard notation (Section~\ref{titsu2}), then general expressions are derived for the representation of $SU(2) \times SU(2)$ in Hubbard notation (Section~\ref{titsu2-2}). In Section~\ref{titcle2} the Clebsch-Gordan coefficients of the $SU(2) \times SU(2)$ Lie group are derived and written in a definite form using the Hypergeometric function ${}_3F_2$. We close the paper with some concluding remarks. An appendix is added to analyze some basic properties of the ceiling and floor functions that are required along the paper.


\section{The Hubbard framework}
\label{tithub}

Matrix calculus was developed for square matrices \cite{Tur27}, and finds a lot of applications in quantum theory where the observables are represented by Hermitian operators. The latter are expressed as square matrices according to the representation defined by the mensurable physical quantities and the related eigenvectors, see e.g. \cite{Mie07}. Any observable $O$ of a quantum multipartite system ${\cal S} ={\cal S}_1 + {\cal S}_2 + \cdots$ is defined on the entire Hilbert space ${\cal H} = {\cal H}_1 \otimes {\cal H}_2 \otimes \cdots$, and can be expressed as matrix Kronecker products of the observables $O_k$ belonging to the subsystems ${\cal S}_k$ \cite{Per95,Nie00}. We shall focus on the properties of square matrices in the understanding that they give a suitable linear representation of the group of observables defining a quantum system. Yet, most of the results we are going to derive can be immediately extended to the case of $n\times m$ matrices. However, some specific properties of square matrices require caution to be promoted to (or they simply can not be applied in) the rectangular case.


\subsection{Notation and basic properties}
\label{titnot-sub}

Let the ket $\vert x \rangle$ be an element of the vector space $\mathbb K^n$, with $\mathbb K$ a field which could be either $\mathbb R$ or $\mathbb C$. This will be represented as a single-column matrix containing $n$ numbers $x_i \in \mathbb K$. The latter will be indexed from 1 unless otherwise stated. Thus,
\be
\vert x \rangle \in \mathbb K^n \quad \Rightarrow \quad \vert x \rangle =(x_1, x_2, \ldots, x_n)^T, \quad x_k \in \mathbb K.
\label{ket}
\ee
The size $n$ of any $n$-tuple $\vert x \rangle$ will be implied whenever $\vert x \rangle$ be written as a linear combination of the orthonormal basis $\{ \vert e_k^n \rangle\}_{k=1}^n$ of $\mathbb K^n$, otherwise this will be explicitly stated if necessary. Here $\vert e_k^n \rangle$ means the $n$-tuple which has a unity in the $k$-th position and all other entries equal to zero. The Hermitian transpose $\vert x \rangle^{\dagger}$ of any ket $\vert x \rangle \in \mathbb K^n$ will be represented by the bra $\langle x \vert$, this last is also called the dual of $\vert x \rangle$, defined as
\be
\vert x \rangle^{\dagger} := \langle x \vert =(x_1^{\dagger}, x_2^{\dagger}, \ldots, x_n^{\dagger}),
\label{bra}
\ee
with $x_k^{\dagger} = x_k$ if $\mathbb K = \mathbb R$, and $x_k^{\dagger} = \overline x_k$ for $\mathbb K = \mathbb C$. The symbol $\overline z$ stands for the complex conjugate of $z \in \mathbb C$. Note that the basis vectors are real, i.e. $\vert e_k^n \rangle^{\dagger} = \vert e_k^n \rangle^T = \langle e_k^n \vert$ is the $n$-dimensional row vector having 1 in the $k$-th position and 0 in all other entries. Therefore, the inner product between arbitrary basis elements is non-negative 
\be
\langle e_k^n \vert e_j^n \rangle = \delta_{kj}, \quad k,j \in \{1, 2, \ldots, n\}.
\label{inner}
\ee
So that any ket $\vert x \rangle \in \mathbb K^n$ can be also expressed as the linear combination
\be
\vert x \rangle = \sum_{k=1}^n x_k \vert e_k^n \rangle, \quad x_{\ell} = \langle e_{\ell}^n \vert x \rangle \in \mathbb K.
\label{ket1}
\ee
In similar form,
\be
\langle x \vert = \sum_{k=1}^n x_k^{\dagger} \langle e_k^n \vert, \quad x_{\ell}^{\dagger} = \langle x \vert e_{\ell}^n \rangle \in \mathbb K.
\label{bra1}
\ee
Hence, the inner product between $\vert x \rangle$ and $\vert y \rangle$, both arbitrary vectors in $\mathbb K^n$, is given by
\be
\langle x \vert y \rangle = \sum_{k, \ell =1}^n x_k^{\dagger} y_{\ell} \langle e_k^n \vert e_{\ell}^n \rangle = \sum_{k=1}^n x_k^{\dagger} y_k.
\label{inner1}
\ee
Since $(\langle x \vert y \rangle)^{\dagger} = \langle y \vert x \rangle$, the space $\mathbb K^n$ is Euclidean (Hermitian) with linear (sesquilinear) metric if $\mathbb K = \mathbb R$ ($\mathbb K = \mathbb C$) \cite{Bac77}. In general we shall write $\mathbb K^n = \mbox{Sp} \{ \vert e_i^n \rangle \}_{i=1}^n$ to denote that $\mathbb K^n$ is spanned by the orthonormal set $\{ \vert e_1^n \rangle, \vert e_2^n \rangle, \ldots, \vert e_n^n \rangle \}$, with $n \in \mathbb N$. In turn, $\mbox{Sp} \{ \vert e_k^n \rangle \}$ will denote the one-dimensional space spanned by the single basis ket $\vert e_k^n \rangle$. From the inner product (\ref{inner1}) one can identify every $\langle x \vert$ with a given mapping of $\mathbb K^n$ into $\mathbb K$. The set of all these mappings is spanned by the duals of the basis vectors $\vert e_k^n \rangle$ and is included in the set of all the functionals $\mathbb K^n \rightarrow \mathbb K$. We write ${\cal K}^n = \mbox{Sp} \{ \langle e_i^n \vert \}_{i=1}^n$ for such a dual vector space.

Now, using (\ref{ket1}) and (\ref{bra}), the outer product between $\vert x \rangle$ and $\vert y \rangle$ yields the dyad
\be
\vert x \rangle \langle y \vert =\sum_{i,j=1}^n x_i y_j^{\dagger} \vert e_i^n \rangle \langle e_j^n \vert \equiv \sum_{i,j=1}^n x_i y_j^{\dagger} X_n^{i,j},
\label{out}
\ee
where the ``dyadic'' operators
\be
X_n^{i,j} = \vert e_i^n \rangle \langle e_j^n \vert, \quad i,j=1,2,\ldots, n
\label{hub1}
\ee
are represented by the matrices (\ref{hub00}). The action of $\vert x \rangle \langle y \vert$ on $\mathbb K^n$ produces the transition from the ket $\vert y \rangle$ to $\vert x \rangle$. In this sense, the matrix array
\be
\vert x \rangle \langle y \vert = \left(
\begin{array}{c}
x_1\\ x_2\\ \vdots\\ x_n
\end{array} 
\right) (y_1^{\dagger}, y_2^{\dagger}, \ldots, y_n^{\dagger}) = \left(
\begin{array}{cccc}
x_1 y_1^{\dagger} & x_1 y_2^{\dagger} & \cdots & x_1 y_n^{\dagger}\\
x_2 y_1^{\dagger} & x_2 y_2^{\dagger} & \cdots & x_2 y_n^{\dagger}\\
\vdots & \vdots & \ddots & \vdots\\
x_n y_1^{\dagger} & x_n y_2^{\dagger} & \cdots & x_n y_n^{\dagger}
\end{array}
\right)
\label{out2}
\ee
is the linear representation of the transition operator $\vert x \rangle \langle y \vert$ in the vector space  $\mathbb K^n$. In turn, the action of $X_n^{i,j}$ on the basis vector $\vert e_k^n \rangle$ gives 
\be
X_n^{i,j} \vert e_k^n \rangle = \delta_{jk} \vert e_i^n \rangle,
\label{hub3}
\ee
so that its action on any ket $\vert x \rangle$ in $\mathbb K^n$ reads
\be
X_n^{i,j} \vert x \rangle = x_j \vert e_i^n \rangle,
\label{hub3b}
\ee
and its matrix elements are easily calculated
\be
\langle e_i^n \vert X_n^{k, \ell} \vert e_j^n \rangle= \delta_{\ell j} \langle e_i^n \vert e_k^n \rangle = \delta_{\ell j} \delta_{i k}.
\label{hub3a}
\ee
It is then clear that $X_n^{i,j}$ projects $\mathbb K^n$ into the one-dimensional space $\mbox{Sp} \{ \vert e_i^n \rangle \}$. The algebra of these operators is defined by the inner product, which can be set to coincide with the conventional matrix product, but it is simpler to use the algebraic  rule
\be
X_n^{i,j}X_n^{k,\ell}=\vert e_i^n \rangle \langle e_j^n \vert e_k^n \rangle \langle e_\ell^n \vert=\delta_{jk} X_n^{i,\ell}.
\label{hub2}
\ee
From (\ref{hub2}), the following result is immediate
\be
[X_n^{i,j}, X_n^{k,m}]_{\pm} = X_n^{i,j} X_n^{k,m} \pm X_n^{k,m} X_n^{i,j} = \delta_{jk} X_n^{i,m} \pm \delta_{mi} X_n^{k,j}.
\label{hub2a}
\ee
On the other hand, the operators $X_n^{i,j}$ are noninvertible real matrices (i.e., $\mbox{det} X_n^{i,j} =0$, and $\overline{X_n^{i,j}} = X_n^{i,j}$) such that their transpose and adjoint (conjugate transpose) versions coincide
\be
\left( X_n^{i,j} \right)^{\dagger} = \left( X_n^{i,j} \right)^T = X_n^{j,i}.
\label{hub4}
\ee
This last expression is easily verified by using (\ref{hub1}) as follows
\[
\left( X_n^{i,j} \right)^T = (\vert e_i^n \rangle \langle e_j^n \vert)^T = (\langle e_j^n \vert)^T  (\vert e_i^n \rangle)^T= \vert e_j^n \rangle \langle e_i^n \vert = X_n^{j,i}.
\]
Note that the symmetric operators $X_n^{i,i}$ are Hermitian and satisfy the completeness relation
\be
\mathbb I_n = \sum_{i =1}^n X_n^{i,i}.
\label{hub5}
\ee
Then, the matrices $X_n^{i,j}$ correspond to the linear representation of the Hubbard operators in the space $\mathbb K^n$, as all the properties (\ref{halg2}) are verified.

The introduction of Hubbard operators in the algebra of operators representing quantum dynamical variables is very advantageous since substantial simplifications are achieved with the properties (\ref{hub2})--(\ref{hub5}). Concrete realisations will be presented in the next sections, with special emphasis in the square matrix representation. Before that, some words concerning the case of rectangular matrices are necessary. 


\subsection{Rectangular matrices}
\label{titrec-sub}

To generalise the results of the previous section one would consider rectangular matrices. In contrast with the square matrices, a rectangular matrix transforms a vector in the space ${\cal H}_n$ into a vector in the space ${\cal H}_m$ where, in general, ${\cal H}_n \neq {\cal H}_m$. It is also well known that the multiplication of two rectangular matrices is defined only if the amount of columns of the first factor is equal to the number of rows of the second factor. Concerning the equivalent of the Hubbard operators in the rectangular case, let $E_{n\times m}^{i,j}$ be the $n\times m$-elementary matrix having entry 1 in position $(i,j)$ and all other entries equal to zero \cite{Gra81}. This can be expressed as
\be
E_{n\times m}^{i,j} = \vert e_i^n \rangle \langle e_j^m \vert.
\label{elem1}
\ee
Note that $\left( E_{n\times m}^{i,j} \right)^{\dagger} = \left( E_{n\times m}^{i,j} \right)^T = E_{m\times n}^{j,i}$, so $\left( E_{n\times m}^{i,j} \right)^{\dagger}$ and $E_{n\times m}^{i,j}$ are defined to act on different vector spaces for $n\neq m$, no matter the values of $i$ and $j$. In the same context, the product between $n\times m$-elementary matrices is constrained to the multiplication rule
\be
E_{n\times m}^{i,j} E_{m\times p}^{k, \ell} = \delta_{jk} E_{n\times p}^{i,\ell}.
\label{elem2}
\ee
Thus, expressions like $E_{m\times p}^{k, \ell} E_{n\times m}^{i,j}$ are meaningless if $p \neq n$ since the number of columns of $E_{m\times p}^{k, \ell}$ differs from the number of rows of $E_{n\times m}^{i,j}$. In spite of these apparent complications, the theorems of matrix calculus deduced for square matrices may be modified for the rectangular case. This is particularly right for the ``square matrices in the broader sense'' defined in \cite{Wig59}, Ch. 2 (see also general expressions in \cite{Gra81}). As we have indicated, our interest is addressed to $n$-square matrices since they represent the most general linear operators in the vector space $\mathbb K^n$. The outline above is to stress that care must be taken with regard to the generalisations of our results to the rectangular case.


\subsection{Square and permutation matrices}
\label{titsqu-sub}

Using (\ref{hub1}) it is easy to see that any $n$-square matrix $A=[a_{i,j}]$ can be expressed in terms of the Hubbard operators
\be
A = \sum_{i,j=1}^n a_{i,j} X_n^{i,j}, \quad a_{i,j} \in \mathbb K,
\label{hub6}
\ee
so that the conventional matrix product $AB$ is expressed as follows
\be
AB = \left( \sum_{i,j=1}^n a_{i,j} X_n^{i,j} \right)  \left( \sum_{k,\ell=1}^n b_{k,\ell} X_n^{k,\ell} \right) =  \sum_{i,\ell =1}^n \left( \sum_k^n  a_{i,k} b_{k,\ell} \right) X_n^{i,\ell} = C,
\label{hub7}
\ee
where $C$ is the $n$-square matrix 
\be
C= \sum_{i,\ell =1}^n c_{i,\ell} X_n^{i,\ell}, \quad c_{i,\ell}= \sum_{k=1}^n a_{i,k} b_{k,\ell}.
\label{hub7b}
\ee
The complex conjugate $\overline A$, the transpose $A^T$, and the adjoint $A^{\dagger}$ of a matrix $A$ read as
\be
\overline A = \sum_{i,j=1}^n \overline a_{i,j} X_n^{i,j}, \qquad A^T = \sum_{i,j=1}^n a_{i,j} X_n^{j,i}, \qquad A^{\dagger}= \sum_{i,j=1}^n a_{i,j}^{\dagger} X_n^{j,i}.
\label{hub8}
\ee
On the other hand, the action of $A$ on the basis vectors $\vert e^n_j \rangle$ is derived from (\ref{hub6}) and (\ref{hub3}), this gives
\be
A \vert e^n _j \rangle = \sum_{k=1}^n a_{k,j} \vert e^n_k \rangle.
\label{abas}
\ee
Then, for an arbitrary vector $\vert x \rangle$ in $\mathbb K^n$ we have
\be
A \vert x \rangle = \sum_{k,j,\ell=1}^n a_{k,j} x_{\ell} X_n^{k,j} \vert e_{\ell}^n \rangle =\sum_{k,\ell=1}^n a_{k,\ell} x_{\ell} \vert e^n_k \rangle.
\label{abas2}
\ee
The expression for the trace of a matrix is easily recovered:
\be
\langle e_i^n \vert A \vert e_j^n \rangle= \sum_{k,\ell =1}^n a_{k,\ell} \delta_{i,k} \delta_{\ell j} = a_{i,j} \quad \Rightarrow \quad \mbox{Tr} A = \sum_{i=1}^n \langle e_i^n \vert A \vert e_i^n \rangle = \sum_{i=1}^n a_{i,i}.
\label{hub7c}
\ee
To give an example, consider a square matrix $H =[h_{i,j}]$ of size $n$ consisting of unimodular entries, $\vert h_{i,j} \vert =1$, and fulfilling $HH^{\dagger} = n\mathbb I_n$. This is called a Hadamard matrix\footnote{More precisely, the square matrices with $\pm 1$ entries and having pairwise orthogonal rows are named after Hadamard  \cite{Had93}. These are included in the set of self-reciprocal matrices introduced by Sylvester \cite{Syl67}. Then, the definition above corresponds to a generalisation of what is commonly known as a Hadamard matrix \cite{Tad06,Ban12}.}. In the simplest case, with $n=2$ and real entries, we have 
\be
H= \frac{1}{\sqrt 2} \sum_{i,j=1}^2 (-1)^{(i-1)(j-1)} X_2^{i,j} = \frac{1}{\sqrt 2} \left(
\begin{array}{rr}
1 & 1\\
1 & -1
\end{array}
\right),
\label{had1}
\ee
where the factor $1/\sqrt 2$ has been introduced to make $H$ unitary. Using (\ref{abas2}) the action of $H$ on $\vert x \rangle \in \mathbb K^2$ reads as
\be
H \vert x \rangle = \frac{1}{\sqrt 2} \sum_{i,k=1}^2 (-1)^{(i-1)(k-1)} x_k \vert e_i^2 \rangle = \frac{1}{\sqrt 2} \left[ (x_1 + x_2) \vert e_1^2 \rangle + (x_1 -x_2) \vert e_2^2 \rangle \right].
\label{had1a}
\ee
From (\ref{hub7}), the multiplication of $H$ with itself gives
\be
H^2 = HH = \sum_{i, \ell =1}^2 c_ {i, \ell} X_2^{i, \ell} = \mathbb I_2 , \qquad c_{i, \ell} = \frac{1 + (-1)^{i+ \ell}}{2} =\delta_{i,\ell}.
\label{had2}
\ee
As another example of interest consider a permutation defined by the bijection $\pi$ of the set of natural numbers $S=\{1, \ldots,n\}$ onto itself. In the Cauchy's two-line notation this map reads as
\[
\pi= \left(
\begin{array}{cccc}
1 & 2 & \cdots & n\\
\pi(1) & \pi(2) & \cdots & \pi(n)
\end{array}
\right).
\]
In particular, the identical permutation $\pi_e$ is such that $\pi_e (k) =k$ for all $k$ in $S$. The set of all $n!$ permutations of $S$ forms the symmetric (or permutation) group $S_n$ of order $n$ with the identity $\pi_e$ as the unit element and the composition of maps as the product. Such a group plays an important role in quantum physics (see e.g. Ch. 13 of Ref. \cite{Wig59}  and Ref. \cite{Bac77}). For example, the Schr\"odinger equation is invariant under the permutation of electrons since the physical equivalence of all these particles. A linear representation of $S_n$ is obtained by assigning a matrix $P_{\pi}$ per each permutation $\pi$. This is a square matrix of order $n$ that has only one entry 1 per row and column, and is zero elsewhere. In Hubbard notation the matrix $P_{\pi}$ reads in simple form
\be
P_{\pi} = \sum_{j=1}^n X_n^{j, \pi(j)}.
\label{per1}
\ee
In this representation the properties of permutation matrices can be studied in compact form. To give a pair of examples consider first the product of $P_{\sigma}$ and $P_{\pi}$, two permutation matrices of order $n$. From (\ref{hub2}) we have
\be
P_{\sigma} P_{\pi} = \sum_{k. \ell=1}^n X_n^{k, \sigma(k)} X_n^{\ell, \pi(\ell)} = \sum_{k. \ell =1}^n \delta_{\sigma(k), \ell} X_n^{k, \pi(\ell)} =\sum_{k=1}^n X_n^{k, \pi( \sigma(k))} = P_{\pi \circ \sigma}.
\label{per2}
\ee
Thus, the composition $\pi \circ \sigma$ of permutations $\pi$ and $\sigma$ is obtained from the product of the corresponding matrices. It is clear that the product of permutation matrices is non-commutative as $P_{\pi} P_{\sigma} = P_{\sigma \circ \pi}$ and $\sigma \circ \pi \neq \pi \circ \sigma$ in general. As a second example let us verify the orthogonality of permutation matrices $P_{\pi}^{-1} = P_{\pi^{-1}} =P_{\pi}^T$. From (\ref{hub8}) and (\ref{hub2}) one arrives at
\be
P_{\pi} P_{\pi}^T= \sum_{k,\ell =1}^n \delta_{\pi(k), \pi(\ell)} X_n^{k,\ell}= \sum_{k=1}^n X_n^{k,k} = \mathbb I_n.
\label{per3}
\ee
Similarly, $P_{\pi}^T P_{\pi} = \mathbb I_n$. From these results it follows the rule $(P_{\sigma} P_{\pi})^{-1} = P_{\pi}^{-1} P_{\sigma}^{-1}$.

To close this section we emphasise that the action of $P_{\pi}$ on any ket $\vert x \rangle \in \mathbb K^n$ is immediately calculated in Hubbard representation
\be
P_{\pi} \vert x \rangle = \sum_{j,k=1}^n x_k X_n^{j, \pi(j)} \vert e_k^n \rangle = \sum_{j,k=1}^n x_k \delta_{\pi(j), k} \vert e_j^n \rangle = \sum_{j=1}^n x_{\pi(j)} \vert e_j^n \rangle.
\label{per1a}
\ee
In the next sections some of the properties of the Kronecker product of permutation matrices are going to be discussed.


\subsection{Basic physical models}
\label{new}

As a first physical example consider an atom having $n$ energy levels. The spectral decomposition of the Hamiltonian can be written as
\be
h = \sum_{k=1}^n E_k X^{k,k}_n,
\ee
with $h \vert \psi_{\ell} \rangle = E_{\ell} \vert \psi_{\ell} \rangle$, $\ell =1, \ldots, n$. For a set of $N$ widely separated (isolated) atoms of this same sort one can write
\be
H_0= \sum_{i=1}^N h_i = \sum_{i=1}^N \sum_{k=1}^n E_{i;k} X^{k,k}_i.
\ee
Henceforth in this section we omit the letter $n$ labelling the order of the Hubbard operators and use the latin sub-label ``$i$'' to denote the atom (site) to which they belong. Thus, the $n \times n$-matrix operator $X_i^{k, \ell} =\vert i, \psi_k \rangle \langle i, \psi_{\ell} \vert$ transforms the local state $\vert i, \psi_{\ell} \rangle$ into the local state $\vert i, \psi_k \rangle$, both of the same site $i$. If the interaction between these atoms is allowed, in a first approach we can write 
\be
H=H_0+H_1, \qquad H_1 = \sum_{i,j =1}^N \sum_{k, \ell, r,s=1}^n \lambda_{i,j; k,\ell, r,s} X_i^{k,\ell} X_j^{r,s}.
\label{model}
\ee
Here $H_1$ is a quadratic form in $X$-operators that represents the atom-atom interaction and refers to the energy involved with the movement of electrons between the sites $i$ and $j$ when all possible sites are considered. Terms corresponding to three- and four-atom interactions can be added by using cubic and quartic forms in $X$-operators if necessary. The coefficients $\lambda_{i,j; k,\ell, r,s}$ in (\ref{model}) must be determined by direct calculation according to the parameters that define the system under consideration. Thus, the Hamiltonian of a set of interacting atoms can be written in terms of linear and quadratic forms of the operators that produce transitions between the local energy states of the system. 

To get some insight of the usefulness of the Hamiltonian (\ref{model}) first notice that the $X$-operators reported in the previous sections correspond to the single site case of the present model. The product rule (\ref{halg1}) still holds if this is evaluated at the same site
\be
X_i^{k, \ell} X_i^{r,s} = \delta_{\ell r} X_i^{k,s}.
\label{refe1}
\ee
On the other hand, from (\ref{hub6}) we know that any operator $A$ acting on the states belonging to the $i$-th atom can be written in terms of the Hubbard operators. Thus, using (\ref{hub5}) with $\mathbb I_i$ the identity operator in site $i$, we get
\be
A_i = \mathbb I_i A_i \mathbb I_i =\sum_{k,\ell =1}^n (a_i)_{k,\ell} X_i^{k,\ell}.
\label{ai}
\ee
Here $(a_i)_{k,\ell} = \langle i, \psi_k \vert A_i \vert  i, \psi_{\ell} \rangle$ gives the probability that the system be in the (final) state $\vert i, \psi_k \rangle$ after the action of $A_i$ on the (initial) state $\vert  i, \psi_{\ell} \rangle$. In particular, given a set of orthonormal orbitals $\left\{ \vert \phi_{i,\mu} \rangle \right\}$ centered on the related atom sites $\vec r_i$, the creation and annihilation operators $c^{\dagger}_{i\mu}$ and $c_{i\mu}$ acquire the $X$-representation
\be
c^{\dagger}_{i\mu}= \sum_{k,\ell =1}^n \langle c^{\dagger}_{i\mu} \rangle_{k,\ell} X_i^{k,\ell}, \quad \langle c^{\dagger}_{i\mu} \rangle_{k,\ell} = \langle i, \psi_k \vert c^{\dagger}_{i\mu} \vert i, \psi_{\ell} \rangle, \quad c_{i\mu} = h.c. (c^{\dagger}_{i\mu}).
\label{amas}
\ee
These relationships can be reversed to express $X_i^{k,\ell}$ as a linear combination of products of the ladder operators $c^{\dagger}_{i\mu}$ and $c_{i\mu}$. Indeed, one requires $n_{ik}$ and $n_{i\ell}$ annihilation and creation operators respectively. If the difference $n_{ik} -n_{i\ell}$ is even (odd) the operator $X_i^{k,\ell}$ will have boson (fermion) character \cite{Hub65}. In other words, for $i \neq j$ one has $[X_i^{k, \ell}, X_j^{r,s}]_{\pm}=0$, with ``$+$'' if both operators are fermion-like and ``$-$'' otherwise. In this context the commutation (\ref{halg2}) is generalised as follows
\be
[X_i^{k, \ell}, X_j^{r,s}]_{\pm}= \delta_{ij} (\delta_{\ell r} X_i^{k,s} \pm \delta_{sk} X_i^{r,\ell}).
\label{refe2}
\ee


\subsubsection{Hubbard and t-J models}
\label{hubmodels}

In the simplest case, for atoms having only one energy level of interest, there will be at most two electrons per site, so the unique orbital is defined by the spin $\sigma$ ($\pm1/2$) of each electron that can occupy the energy level. We will have four different states: No electrons (vacuum) $\vert i, 0 \rangle$, a single electron with either spin up $\vert i, + \rangle$ or spin down $\vert i, - \rangle$, and two electrons that obey the Pauli principle $\vert i, 2 \rangle=\vert  i, + -\rangle$. Let $E_{i;0}, E_{i;1}$, and $E_{i;2}$ be the corresponding energies ($\vert i, \pm \rangle$ sharing the same energy $E_{i;1}$). The operators $c^{\dagger}_{i\sigma}$ annihilate any state including at least one electron with spin $\sigma$ and creates an state with an additional electron of spin $\sigma$ otherwise. The operators $c_{i\sigma}$ annihilate the state $\vert i, 0 \rangle$ as well as the states that have a single electron of spin $-\sigma$. In all other cases the $c_{i\sigma}$ eliminate the electron of spin $\sigma$. For instance, in the site $i$ one has $c^{\dagger}_{-} \vert 2 \rangle =0$ and $c_{-} \vert 2 \rangle = \vert + \rangle$. The Hubbard operators in the site $i$ will be represented by square matrices (\ref{hub00}) of order 4, as the vector representation is of dimension 4. We have 8 fermion-like Hubbard operators $X^{0,\sigma}, X^{\sigma,0}, X^{\sigma,2}, X^{2,\sigma}$ and 8 boson-like operators that include 4 diagonal matrices $X^{0,0}, X^{+,+}, X^{-,-}, X^{2,2}$ and 4 nondiagonal matrices $X^{+,-}, X^{-,+}, X^{2,0}, X^{0,2}$. According to (\ref{hub6}) and (\ref{amas}) the ladder operators have fermion character and are non-diagonal
\be
c^{\dagger}_{i\sigma} =X_i^{\sigma,0}+2\sigma X_i^{2,-\sigma}, \quad c_{i\sigma} = X_i^{0,\sigma}+2\sigma X_i^{-\sigma, 2}.
\label{ces}
\ee
In a similar manner one realises that the Hamiltonian (\ref{model}) includes a diagonal boson-like and a non-diagonal fermion-like parts respectively given by
\be
\begin{array}{c}
H_0= \displaystyle\sum_{i=1}^N \left[E_{i;0} X_i^{0,0}+ E_{i;1} \left( X_i^{+,+}+X_i^{-,-} \right) + E_{i;2} X_i^{2,2}
\right],\\[3ex]
H_1= \displaystyle\sum_{i,j=1}^N t_{i,j} \left[ \left(X_i^{+,0} + X_i^{2,-} \right) \left( X_j^{+,0} + X_j^{-,2} \right) + \left(X_i^{-,0} - X_i^{2,+} \right) \left( X_j^{0,+} - X_j^{+,2} \right)
\right].
\end{array}
\label{hubx}
\ee
Using $E_{i;0} =0$, $E_{i; 1}= \varepsilon -\mu$, and $E_{i;2} = 2 E_{i;1}+U$, with $N \rightarrow +\infty$, $\varepsilon$ the single electron energy in a crystal field, $\mu$ the chemical potential parameter that controls the electron density, and $U$ the repulsive Coulomb interaction between electrons on the same site, the operator $H=H_0+H_1$ so constructed can be identified with the Hubbard Hamiltonian \cite{Hub63,Hub64,Hub65} (see also \cite{Ovc97}) in the $X$-operator representation. In that case $t_{ij}$ corresponds to the hopping parameter between adjacent sites $i$ and $j$.  This can be also shown that, in the strong-coupling limit $U/t>>1$, the operators $X_i^{0,\sigma}$ and $X_i^{-\sigma,2}$ are associated to a fermion-like quasiparticle in the lower and upper Hubbard bands respectively \cite{Ovc04}, so the ladder operators (\ref{ces}) represent the decoupling of the free electron band onto the lower and upper Hubbard sub-bands.

The Hubbard Hamiltonian (\ref{hubx}) corresponds to the simplest model of strongly correlated electrons \cite{Ovc97,Ovc04,Dag94,Kor93,Kor94}. This means that ``the influence of the interactions of electrons on the same atom is so dominant that only this type of interaction need be considered, at least as a first approach'' \cite{Hub65}, pp 240. Limit cases include the Hamiltonian with no interactions ($U=0$) and the Hamiltonian of `single site' (no hopping $t_{ij}=0 \, \forall \, i, j$) $H=\mbox{diag}(0, \varepsilon -\mu, \varepsilon -\mu, 2(\varepsilon -\mu)+U)$. In the strong coupling limit the Hubbard model reduces to a system of spins and holes on a two-dimensional square lattice which was already studied by Anderson \cite{And87} and is associated to the so-called t-J model \cite{Zha88}. After excluding the local two-electron states and neglecting the three-site terms (i.e., the cubic forms of $X$-operators), the adjacent sites configuration yields the Hamiltonian
\be
H= -t \sum_{\langle ij \rangle,  \sigma} \left( c^{\dagger}_{i\sigma} c_{j\sigma} + h.c. \right) +  \sum_{\langle ij \rangle} J \left( \vec S_i \cdot \vec S_j - \frac{n_in_j}{4} \right),
\label{tj}
\ee
where $\vec S_i$ is a vector spin-1/2 operator at the site $i$ of a two-dimensional square lattice, and $J= 4t^2/U$ (remember, $t/U <<1$) is the antiferromagnetic coupling between nearest neighbours sites $\langle ij \rangle$ \cite{Ovc97}. Given $i$, this model has only three possible states $\vert i, 0 \rangle$, and $\vert i, \pm \rangle$. At half-filling (each site includes one and only one electron) the Hamiltonian (\ref{tj}) leads to the Heisenberg model \cite{Ovc04} (see also Section~\ref{diagon}).

The t-J model describes electrons on a lattice in correlated motion involving nearest neighbour hopping (t) as well as nearest neighbour spin exchange and charge interactions (J), so the model includes only the energy parameters t and J. In the paper by Zhang and Rice \cite{Zha88} the model is used to describe the cooper-oxide planes in high-Tc superconductors. Such a work attracted increasing interest since the phenomenology of high-Tc superconductivity could be explained in strong correlated electronic systems \cite{Kor94}. Yet, the presence of superconductivity in the Hubbard-like models seems to include some unexpected subtleties \cite{Dag94}. On the other hand, recent works have addressed the problem of developing an $X$-operator Lagrangian approach by assuming that the Lagrange function can be expressed in terms of the Hubbard operators \cite{Fou99a,Fou99b,Fou00}. Using the Faddeev-Jackiw symplectic formalism one can realize that no classical dynamics is consistent with the algebra (\ref{refe1}), (\ref{refe2}), and (\ref{hub5}). Therefore, some constraints must be included to define a consistent classical dynamics by using a path-integral formalism that corresponds to the coherent state representation \cite{Fou99a}. The generalisation embracing the t-J model is appropriate \cite{Fou99b}, even in the perturbative approach \cite{Fou00}. This method was used  to get a large-$N$ expansion of the t-J model that include diagrammatic rules in which the propagators and vertex are written in terms of Hubbard operators \cite{Fou02,Fou04,Bej06}. Some other physical applications of the algebra of the Hubbard operators as this is studied in the present work can be found in the books \cite{Ovc04,Kor93,Kor94}. Next, we are going to develop the algebra for the Kronecker products of the Hubbard operators.


\section{Kronecker algebra in Hubbard representation}
\label{titkro}

We start the analysis of the Kronecker algebra with the definition of the direct product.

\begin{itemize}
\item[]
{\bf Definition~K1.} Let $A=[a_{i,j}]$ and $B =[b_{r,s}]$ be respectively matrices of order ${m\times n}$ and ${k\times \ell}$ over the field $\mathbb K$. The Kronecker product $A\otimes B$ is the matrix of order $mk \times n\ell$ over the field $\mathbb K$ defined as $A\otimes B =[a_{i,j} B]$.
\end{itemize}


\noindent
As a first example consider the basis vectors $\vert e_k^n \rangle$, these are matrices of order $1\times n$ so that $\vert e_k^n \rangle \otimes \vert e_j^n \rangle$ is a matrix of order $1 \cdot 1 \times n\cdot n= 1\times n^2$. Moreover, this $n^2$-tuple has only one unity at $(k-1)n +j$, and is zero in all other entries. In general, the Kronecker product of two basis vectors belonging to different spaces $\vert e_{i_1}^{n_1} \rangle$ and $\vert e_{i_2}^{n_2} \rangle$ gives a tuple of size $n_1 n_2$ that has a single unit among $n_1 n_2-1$ zeros, as this is stated in the following definition.

\begin{itemize}
\item[]
{\bf Definition~K2.} Let $\vert e_{i_1}^{n_1} \rangle$ and $\vert e_{i_2}^{n_2} \rangle$ be basis vectors of $\mathbb K^{n_1}$ and $\mathbb K^{n_2}$ respectively. Then the Kronecker product $\vert e_{i_1}^{n_1} \rangle \otimes \vert e_{i_2}^{n_2} \rangle$ is the $n_1 n_2$-tuple having 1 at $(i_1-1)n_2 +i_2$, and zero in all other entries. 
\end{itemize}

\noindent
It is immediate to verify that the vectors constructed according to Definition~K2 are orthonormal  and that there are only $n_ 1n_2$ of them. Thus, all of them represent an orthonormal basis of the vector space $\mathbb K^{n_1 n_2}$. We have the next proposition without giving a proof.

\begin{itemize}
\item[]
{\bf Proposition~K0.} Let $\mathbb K^{n_1} =\mbox{Sp} \left\{ \vert e_{i_1}^{n_1} \rangle \right\}_{i_1=1}^{n_1}$ and $\mathbb K^{n_2} =\mbox{Sp} \left\{ \vert e_{i_2}^{n_2} \rangle \right\}_{i_2=1}^{n_2}$ be vector spaces. The set of all the Kronecker products $\vert e_{i_1}^{n_1} \rangle \otimes \vert e_{i_2}^{n_2} \rangle$ is orthonormal and spans a vector space of dimension $n_1 n_2$, written $\mathbb K^{n_ 1n_2} = \mbox{Sp} \{ \vert e_{i_1}^{n_1} \rangle \otimes \vert e_{i_2}^{n_2} \rangle,\, i_1=1, \ldots, n_1; i_2=1, \ldots, n_2 \}$, with the following axioms ($\alpha, \beta, \gamma, \eta$ are elements of $\mathbb K$):

(i) $(\alpha \vert e_{i_1}^{n_1} \rangle) \otimes \vert e_{i_2}^{n_2} \rangle = \alpha (\vert e_{i_1}^{n_1} \rangle \otimes \vert e_{i_2}^{n_2} \rangle) = \vert e_{i_1}^{n_1} \rangle \otimes (\alpha \vert e_{i_2}^{n_2} \rangle)$

(ii) $(\alpha \vert e_{i_1}^{n_1} \rangle + \beta \vert e_{i_1}^{n_1} \rangle) \otimes \vert e_{i_2}^{n_2} \rangle = \alpha (\vert e_{i_1}^{n_1} \rangle \otimes \vert e_{i_2}^{n_2} \rangle )+ \beta (\vert e_{i_1}^{n_2} \rangle \otimes \vert e_{i_2}^{n_2} \rangle)$

(iii) $\vert e_{i_1}^{n_1} \rangle \otimes ( \gamma \vert e_{i_2}^{n_2} \rangle + \eta \vert e_{i_2}^{n_2} \rangle)  = \gamma( \vert e_{i_1}^{n_1} \rangle \otimes \vert e_{i_2}^{n_2} \rangle) + \eta(\vert e_{i_1}^{n_1} \rangle \otimes \vert e_{i_2}^{n_2} \rangle)$

\end{itemize}

\noindent
An arbitrary vector $\vert x \rangle \in \mathbb K^{n_1 n_2}$ can be written either as a twice-indexed linear combination
\[
\vert x \rangle = \sum_{i_1=1}^{n_1} \sum_{i_2=1}^{n_2} x_{i_1,i_2} \vert e_{i_1}^{n_1} \rangle \otimes \vert e_{i_2}^{n_2} \rangle, 
\]
or as a single-indexed linear combination
\[
\vert x \rangle = \sum_{k=1}^{n_1 n_2} \widetilde x_k \vert e_k^{n_1 n_2} \rangle, \quad k= (i_1 -1)n_2 +i_2.
\]
Whenever this produces no confusion we shall write $\vert e_k^{n_1 n_2 n_3 \cdots} \rangle$, $k=1, 2, \ldots, n_1 n_2 n_3 \cdots$, to represent the basis vectors $\vert e_{i_1}^{n_1} \rangle \otimes \vert e_{i_2}^{n_2} \rangle  \otimes \vert e_{i_3}^{n_3} \rangle \otimes \cdots$, $i_{\ell} \in \{1,  \ldots, n_{\ell} \}$, that span $\mathbb K^{n_1 n_2 n_3 \cdots}$. In particular, if $n_{\ell} = n$ for all $\ell \in \{1, \ldots, p\}$, then 
\be
\mathbb K^{n^p}= \mbox{Sp} \left\{ \vert e_{i_1}^n \rangle \otimes \cdots \otimes \vert e_{i_p}^n \rangle  \right\}_{i_{\ell =1}}^n
\label{tensor}
\ee
is the space of contra variant tensors of rank $p$ while its dual ${\cal K}^{n^p}$ is the space of covariant tensors of rank $p$ \cite{Bac77}. For instance, in quantum computing it is customary to write $\vert 0 \rangle$ and $\vert 1\rangle$ for the basis vectors of $\mathbb K^2$; using our notation they are $\vert e_1^2 \rangle$ and $\vert e_2^2 \rangle$ respectively. According to Proposition~K0, the four products $\vert e_{i_1}^2 \rangle \otimes \vert e_{i_2}^2 \rangle$, $i_{\ell} \in \{1,2\}$, span $\mathbb K^4$. These can be expressed in binary form as follows
\[
\begin{array}{c}
\vert e_1^4 \rangle = \vert e_1^2 \rangle \otimes \vert e_1^2 \rangle = \vert 0 \rangle \otimes \vert 0 \rangle = \vert 00 \rangle, \quad 
\vert e_2^4 \rangle = \vert e_1^2 \rangle \otimes \vert e_2^2 \rangle = \vert 0 \rangle \otimes \vert 1 \rangle = \vert 01 \rangle,\\[2ex]
\vert e_3^4 \rangle = \vert e_2^2 \rangle \otimes \vert e_1^2 \rangle = \vert 1 \rangle \otimes \vert 0 \rangle = \vert 10 \rangle, \quad 
\vert e_4^4 \rangle = \vert e_2^2 \rangle \otimes \vert e_2^2 \rangle = \vert 1 \rangle \otimes \vert 1 \rangle = \vert 11 \rangle.
\end{array}
\]
The same notation holds for an arbitrary number of two-dimensional factors. For instance,
\[
\vert e_1^{32} \rangle = \underbrace{\vert e_1^2 \rangle \otimes \cdots \otimes \vert e_1^2 \rangle}_{\rm{5 \, times}} =\vert 00000 \rangle.
\]

\subsection{Kronecker algebra of Hubbard operators}
\label{titkro-sub}

The Kronecker product $A\otimes B$ is simple if the factors are Hubbard operators (\ref{hub00}). In this case, the resulting matrix has only one entry different from zero since each of the factors has a unique non zero entry. That is, the Kronecker product is closed in the set of Hubbard operators. 

\begin{itemize}
\item[]
\textbf{Proposition~K1. } Let $X_m^{i,j}$ and $X_n^{k,\ell}$ be two Hubbard operators of order $n$ and $m$ respectively. The Kronecker product $X_m^{i,j}\otimes X_n^{k,\ell}$ is the $mn$-Hubbard operator $X_{mn}^{n(i-1)+k, n(j-1)+\ell}$. That is,
\be
X_m^{i,j}\otimes X_n^{k,\ell}=X_{mn}^{n(i-1)+k, n(j-1)+\ell}.
\label{hub0}
\ee
\end{itemize}

\noindent
{\em Proof. } The proof follows from Definition~K1, explicitly 
\[
\begin{array}{rl}
\lefteqn{
X_m^{i,j}\otimes X_n^{k,\ell} } &\\[2ex]
&={\footnotesize \left(
\begin{array}{ccccccc}
&&& 0 &&&\\
&0_{(i-1) \times (j-1)} & & \vdots & & 0_{(i-1) \times (n-j)} &\\
&& & 0 &&&\\
0 & \cdots & 0 & 1 & 0 & \cdots & 0  \\
&& & 0 &&&\\
&0_{(n-i) \times (j-1)} &  & \vdots & & 0_{(n -i) \times (n-j)} & \\
&&& 0 &&&
\end{array}
\right)}  
\otimes X_n^{k,\ell}\\[12ex]
&=  {\footnotesize
\left(
\begin{array}{ccccccc}
&&& 0 &&&\\
& 0_{[n(i-1)+k-1]\times [n(j-1) +\ell -1]}  & & \vdots & & 0_{[n(i-1) +k-1] \times [mn-n(j-1) -\ell ]} &\\
&& & 0 &&&\\
0 & \cdots & 0 & 1 & 0 & \cdots & 0  \\
&& & 0 &&&\\
& 0_{[mn-n(i-1)-k]\times [n(j-1) +\ell -1]} &  & \vdots & & 0_{[mn-n(i-1)-k]\times [n(j-1) +\ell -1]} & \\
&&& 0 &&&
\end{array}
\right)} \\[12ex]
& = X_{mn}^{n( i-1) +k, n(j-1) +\ell}. \quad\Box
\end{array}                        
\]
Besides the basic property introduced in Proposition~K1, the Kronecker algebra of the Hubbard operators includes the following set of properties.

\begin{itemize}
\item[]
{\bf Proposition~K2.} Let $X_{\alpha}^{\beta, \gamma}$ be Hubbard operators of order $\alpha$ and take $\lambda \in \mathbb K$. Then 
\item[i)] 
$X_m^{i,j} \otimes X_n^{k, \ell}\neq X_n^{k, \ell} \otimes X_m^{i,j}$ in general.

\item[ii)] 
$\left(X_m^{i,j} \otimes X_n^{k, \ell}\right)^T=\left(X_m^{i,j}\right)^T \otimes\left(X_n^{k, \ell}\right)^T$.

\item[iii)] 
$\left(\lambda X_m^{i,j}\right) \otimes X_n^{k, \ell}= \lambda \left(X_m^{i,j} \otimes  X_n^{k, \ell}\right) =X_m^{i,j} \otimes \left( \lambda X_n^{k, \ell} \right)$.

\item[iv)] 
$\left(X_m^{i,j} +X_m^{r,s} \right) \otimes X_n^{k, \ell} =X_m^{i,j} \otimes X_n^{k, \ell}+ X_m^{r,s} \otimes X_n^{k, \ell}$.

\item[v)] 
$X_n^{k, \ell} \otimes \left( X_m^{i,j} + X_m^{r,s} \right)= X_n^{k, \ell} \otimes X_m^{i,j} + X_n^{k, \ell} \otimes X_m^{r,s}$

\item[vi)] 
$\left(X_m^{i,j} \otimes X_n^{k, \ell} \right) \otimes X_p^{r,s}= X_m^{i,j} \otimes \left( X_n^{k, \ell} \otimes X_p^{r, s}\right)$.

\end{itemize}


\noindent
{\em Proof}. Parts iii, iv and v are immediate from Definition~K1.

\vskip1ex
\noindent
i) From Proposition~K1 we see that $X_m^{i,j} \otimes X_n^{k, \ell}= X_n^{k, \ell} \otimes X_m^{i,j}$ requires the roots of the system
\[
n(i-1) + k = m(k-1) +i, \qquad n(j-1) + \ell = m(\ell -1) +j,
\]
which in general has no solution for arbitrary fixed values of $n$ and $m$. A particular solution is obtained if $n=m$, for which one gets $i=k$ and $j=\ell$. The symmetric case $X_m^{i,j} \otimes X_m^{i, j}$ is recovered from this last result.

\vskip1ex
\noindent
ii) From (\ref{hub0}) and (\ref{hub4}),
\[
\begin{array}{ll}
\left( X_m^{i,j} \otimes X_n^{k, \ell} \right)^T  & = \left( X_{mn}^{n(i-1)+k, n(j-1) + \ell}\right)^T = X_{mn}^{n(j-1) + \ell, n(i-1)+ k}\\[2ex]
& =X_m^{j,i} \otimes X_n^{\ell, k}= \left( X_m^{i,j} \right)^T \otimes \left(X_n^{k, \ell} \right)^T.
\end{array}
\]
Remark that this result immediately gives $\left( X_m^{i,j} \otimes X_n^{k, \ell} \right)^{\dagger} = \left( X_m^{i,j} \right)^{\dagger} \otimes \left(X_n^{k, \ell} \right)^{\dagger}$, since the Hubbard operators are real matrices.

\vskip1ex
\noindent
vi) From (\ref{hub0}), 
\[
\begin{array}{ll}
\lefteqn{
\left( X_m^{i,j} \otimes X_n^{k, \ell} \right) \otimes X_p^{r, s} } & \\[2ex]
& =X_{mn}^{n(i-1)+k, \, n(j-1)+ \ell} \otimes X_p^{r, s} \\[2ex]
& = X_{mnp}^{p[n(i-1) + k- 1] + r, \, p[n(j-1)+ \ell -1]+ s} = X_{mnp}^{pn(i-1) + p (k-1) + r, \, pn(j-1) + p( \ell-1 )+ s }\\[2ex]
& =X_m^{i, j} \otimes X_{np}^{p(k-1)+r, \, p(\ell-1)+s} = X_m^{i, j} \otimes  \left(X_n^{k, \ell} \otimes X_p^{r, s}\right). \quad\Box
\end{array} 
\]
Proposition~K2 includes the basic properties of the Kronecker product of Hubbard operators. Applied to the set $\{ X_{\alpha}^{\beta, \gamma} \}$, they mean that the product $\otimes$ is distributive over ordinary matrix addition (iv, v), associative (vi), compatible with ordinary matrix transposition (ii) as well as with matrix multiplication by an scalar (iii) and, in general, non-abelian (i). Next we generalise such properties to the case of arbitrary square matrices while some other algebraic relationships are derived.

\subsection{Kronecker algebra of permutation matrices}
\label{titkro-sub2}

The Kronecker algebra of the Hubbard operators is particularly useful in the operating with, and the construction of permutation matrices. In this section we report some results on permutation matrices that are fundamental in the ensuing applications of the Kronecker product. From Proposition~K1 one has the following result.

\begin{itemize}
\item[]
{\bf Theorem~P1.} The square matrix 
\be
\Pi = \sum_{i, j=1}^n X_n^{i,j} \otimes X_n^{j,i}
\label{sim1}
\ee
is a permutation matrix of order $n^2$, defined by the rule
\be
\pi (p) =n(p+n-1) -(n^2 -1) p', \qquad p=1,2,\ldots, n^2,
\label{per4}
\ee
with
\be
p' =\left\lceil \frac{p}{n} \right\rceil
\label{per5}
\ee
the ceiling function applied on $\frac{p}{n}$ (see Eq. (\ref{ceilfun}) of the appendix).

\end{itemize}

\noindent
{\em Proof}. From (\ref{hub0}) we have
\[
\Pi =  \sum_{i,j=1}^n X_{n^2}^{n(i-1)+j, n(j-1)+i}.
\]
Let us define $p= n(i-1)+j$, then $p=1,2, \ldots, ni$, and $j= p- n(i-1)$. Therefore,
\[
\Pi = \sum_{i=1}^n \, \sum_{p=n(i-1)+1}^{ni}  X_{n^2}^{p, \, n(p+ n -1)-(n^2-1)i},
\]
with
\be
\frac{p}{n} \leq i \leq \frac{p}{n} + 1 - \frac{1}{n} < \frac{p}{n} +1.
\label{cotas}
\ee
Using Lemma~A1(i) of the appendix we get
\[
i = \left\lceil \frac{p}{n} \right\rceil \equiv p',
\]
to write
\[
\Pi = \sum_{p=1}^{n^2} X_{n^2}^{p, \, n(p+ n -1)-(n^2-1) p'}.
\]
Comparing this last result with (\ref{per1}) we arrive at the rule (\ref{per4}). It can be verified that $\pi(p)$ is indeed a bijection on the set $\{1, \ldots,n^2\}$. $\Box$

\vskip2ex
To get some insight on the meaning of the permutation matrix (\ref{sim1}) let us consider an arbitrary contra variant tensor of rank 2:
\be
\vert x_1 \rangle \otimes \vert x_2 \rangle = \sum_{i_1, i_2=1}^n x_{i_1} x_{i_2} \vert e_{i_1}^n \rangle \otimes \vert e_{i_2}^n \rangle = \sum_{i_1, i_2=1}^n x_{i_1} x_{i_2} \vert e_{({i_1}-1)n +i_2}^{n^2} \rangle.
\label{sim2}
\ee
The action of $\Pi$ on this last vector gives
\[
\begin{array}{rl}
\Pi (\vert x_1 \rangle \otimes \vert x_2 \rangle)= &  \left( \displaystyle\sum_{i, j=1}^n  X_n^{i,j} \otimes X_n^{j,i} \right) \left( \displaystyle\sum_{i,_1 i_2=1}^n x_{i_1} x_{i_2} \vert e_{i_1}^n \rangle \otimes \vert e_{i_2}^n \rangle \right)\\[3ex]
 = & \displaystyle\sum_{\substack{i, j,
i_1, i_2=1}}^n x_{i_1} x_{i_2} \delta_{j i_1} \delta_{i i_2} \vert e_{i}^n \rangle \otimes \vert e_{j}^n \rangle = \displaystyle\sum_{i_1, i_2=1}^n x_{i_1} x_{i_2} \vert e_{i_2}^n \rangle \otimes \vert e_{i_1}^n \rangle\\[3ex]
& = \vert x_2 \rangle \otimes \vert x_1 \rangle.
\end{array}
\]
Thus, relative to the indices labelling the contra variant tensor (\ref{sim2}), the operator $\Pi$ corresponds to the bijection $\pi_2: (1,2) \mapsto (2,1)$. Hence $\Pi \equiv P_{\pi_2} \in S_2$. Indeed, there are only $2! = 2$ different permutations on the set $\{1,2\}$, these are the identity $\pi_e \equiv \pi_1$ and $\pi_2$. In Hubbard representation we have 
\be
P_{\pi_1} =\sum_{i=1}^n X_n^{i,i} \otimes X_n^{i,i} \equiv \sum_{i,j=1}^n \delta_{ij} X_n^{i,j} \otimes X_n^{j,i}, \quad P_{\pi_2} = \Pi.
\label{sim3}
\ee
From these results it is easy to verify that 
\be
{\cal S}_{(2)} =\tfrac12( P_{\pi_1} + P_{\pi_2}) = \frac12 \sum_{i,j=1}^n (1+\delta_{ij}) X_n^{i,j} \otimes X_n^{j,i}
\label{sim4}
\ee
is the symmetrization operator for the vectors in $\mathbb K^{n^2}$. Namely,
\[
{\cal S}_{(2)} (\vert x_1 \rangle \otimes \vert x_2 \rangle ) =\frac{\vert x_1 \rangle \otimes \vert x_2 \rangle + \vert x_2 \rangle \otimes \vert x_1 \rangle}{2}
\]
is a symmetric tensor of rank 2. In a similar form, the operator
\be
{\cal A}_{(2)} = \tfrac12  \left[\chi(\pi_1) P_{\pi_1} + \chi(\pi_2) P_{\pi_2} \right] = \frac12 \sum_{i,j=1}^n\left[ \chi(\pi_1)+ \chi(\pi_2) \delta_{ij} \right]  X_n^{i,j} \otimes X_n^{j,i},
\label{sim5}
\ee
with $\chi(\pi)$ the parity of the bijection $\pi$ \cite{Bac77}, produces antisymmetric tensors of rank 2:
\[
{\cal A}_{(2)} (\vert x_1 \rangle \otimes \vert x_2 \rangle ) =  \frac{\vert x_1 \rangle \otimes \vert x_2 \rangle - \vert x_2 \rangle \otimes \vert x_1 \rangle}{2}.
\]
The generalisation of the above results to tensors of arbitrary rank is straightforward. We summarise this in the following proposition without a proof. 

\begin{itemize}
\item[]
{\bf Proposition~P1.} The operators
\be
{\cal S}_{(p)} = \frac{1}{p!} \sum_{\ell =1}^p P_{\pi_{\ell}} \qquad \mbox{and} \qquad {\cal A}_{(p)} = \frac{1}{p!} \sum_{\ell =1}^p \chi(\pi_{\ell}) P_{\pi_{\ell}} 
\label{sim6}
\ee
with $\pi_{\ell} \in S_p$, $\pi_1\equiv \pi_e$, and $P_{\pi_{\ell}}$ a definite linear combination of the Kronecker products 
\[
X_n^{i_1, j_1} \otimes X_n^{i_2, j_2} \otimes \cdots X_n^{i_p, j_p}, \quad i_k, j_k \in \{1, \ldots, n\},
\]
produce respectively the symmetrization and antisymmetrization of the contra variant tensors of rank $p$.

\end{itemize}

\noindent
The operators ${\cal S}_{(p)}$ and ${\cal A}_{(p)}$ are useful in group theory and symmetries \cite{Wey31,Wig59,Fon70,Bac77}. In the literature of combinatorics they appear in connection with the concepts of determinant and permanent of a matrix, these last give rise to entire treatises \cite{Min78} and are associated to the concept of majorization of vectors that is fundamental in the algorithms of quantum computing \cite{Nie00} and in the geometry properties of the quantum states \cite{Ben06} as well. The symmetrization of vectors as this has been indicated above is also useful in the analysis of the Majorana representation of multi-qubit states for the studies of the barycentric measure of quantum entanglement \cite{Gan12}. The major result in Proposition~P1 is that the Kronecker products defined in Theorem~P1 as permutation matrices are noting but the building blocks of the symmetrization and antisymmetrization operators of the contra variant vector space (\ref{tensor}). The result reported in Theorem~P1 has been already included in the works by other authors (see e.g. Eq.~(4) of Ref.~\cite{Bre78}, and Section~2.5 of Ref. \cite{Gra81}) but, as far as we know, such works give no reference to the explicit form of the permutation. Here, equation (\ref{per4}) gives the concrete realisation of such a permutation and the connection of $\Pi$ with the operators (\ref{sim5}) has been also achieved.

On the other hand, the Kronecker product of permutation matrices is also compatible with the composition of permutations described by equation (\ref{per2}). That is, the product $\otimes$ is closed in the set of permutation matrices.

\begin{itemize}
\item[]
{\bf Theorem~P2.} Let $P_{\pi} (n)$ and $P_{\sigma} (m)$ be the $n$ and $m$-permutation matrices defined by the rules $\pi$ and $\sigma$ respectively. The Kronecker product $P_{\pi} (n) \otimes P_{\sigma} (m)$ is the $nm$-permutation matrix $P_{\alpha} (n,m)$ defined by the rule
\be
\alpha(p) = m[\pi (p') -1] + \sigma (p -mp' + m),
\label{per6}
\ee
with $p'= \lceil \frac{p}{m} \rceil$.

\end{itemize}

\noindent
{\em Proof}. From Proposition~K1 and the linearity of $\otimes$ one gets
\[
P_{\alpha} (n,m) = P_{\pi} (n) \otimes P_{\sigma} (m) = \sum_{i=1}^n \sum_{j=1}^m X_{nm}^{m(i-1) + j, \, m[\pi(i) -1] + \sigma(j)}.
\]
Following the proof of Theorem~P1 we realize that the change $p=m(i-1) + j$ gives rise to equation (\ref{cotas}) with $n \leftrightarrow m$, so that $p'= \lceil \frac{p}{m} \rceil$ and the rule (\ref{per6}) follows from the expression
\[
P_{\alpha} (n,m) = \sum_{p=1}^{nm} X_{nm}^{p, \, m[\pi (p') -1] + \sigma (p -mp' + m)}.
\]
Now, using the multiplication rule (\ref{hub2}) and property (\ref{hub8}) we have
\[
P_{\alpha}^T(n,m) P_{\alpha} (n,m) = \sum_{p,q=1}^{nm} X_{nm}^{\alpha(p), p} X_{nm}^{q, \alpha(q)} = \sum_{p=1}^{nm} X_{nm}^{\alpha(p), \alpha(p)} = \mathbb I_{nm},
\]
and a similar procedure shows that $P_{\alpha} (n,m) P_{\alpha}^T(n,m) = \mathbb I_{nm}$. $\Box$

\vskip2ex
\noindent
Now, let us consider Proposition~K2(i). This indicates that the Kronecker product of two Hubbard operators, $X_n^{i,j}$ and $X_m^{k, \ell}$, is non-abelian in general. Such restriction, however, can be relaxed because $X_n^{i,j} \otimes X_m^{k, \ell}$ has only one entry 1, as this is established in Proposition~K1; the same is true for $X_m^{k,\ell} \otimes X_n^{i,j}$. Therefore, it should be possible to arrive at $X_m^{k,\ell} \otimes X_n^{i,j}$ by applying the appropriate permutation of rows and columns in $X_n^{i,j} \otimes X_m^{k, \ell}$. That is, $X_n^{i,j} \otimes X_m^{k, \ell}$ and $X_m^{k,\ell} \otimes X_n^{i,j}$ must be permutation equivalent.


\begin{itemize}
\item[]
{\bf Proposition~P2.} The Kronecker product $X_n^{i,j} \otimes X_m^{k, \ell}$ is permutation equivalent to $X_m^{k,\ell} \otimes X_n^{i,j}$. That is, there exist $P_{\pi}$, a permutation matrix of order $nm$, such that 
\be
P_{\pi}^T  \left( X_n^{i,j} \otimes X_m^{k, \ell} \right) P_{\pi} = X_m^{k,\ell} \otimes X_n^{i,j}.
\label{per7}
\ee
\end{itemize}

\noindent
{\em Proof}. We use the multiplication rule (\ref{hub2}) and the expression (\ref{hub0}), together with the linearity of $\otimes$, to arrive at
\[
P_{\pi}^T  \left( X_n^{i,j} \otimes X_m^{k, \ell} \right) P_{\pi} = X_{nm}^{\pi( mi -m + k), \, \pi(mj -m + \ell)}.
\]
Hence, in order to satisfy (\ref{per7}) we have
\be
\pi( mi -m + k) = nk-n+i.
\label{per9}
\ee
The bijection $\pi$ we are looking for is defined by this last equation (the labels $j$ and $\ell$ satisfy the same equation under the change $i\rightarrow j$ and $k \rightarrow \ell$). $\Box$

\vskip2ex
The permutation equivalence of matrix Kronecker products is of enormous interest in quantum information theory as this is useful in solving the problem of constructing maximally entangled bases of multipartite quantum systems. If dealing with Fourier matrices, it is possible to discriminate whether the permutation equivalence is preserved for the Kronecker products if matrix multiplication by unitary diagonal matrices is also allowed \cite{Tad06}. As discussed in the introduction, this problem deals with equation (\ref{inteq}) for which the permutation matrices $P_1$ and $P_2$ are to be determined. The results we have presented in this section are addressed to show the basic operation rules of the permutation matrices when they are expressed as linear combinations of Hubbard operators. Further insights will be given in the sequel. 

\subsection{Kronecker algebra of square matrices}
\label{titkro2-sub}

We now consider the properties of $n$-square matrices associated with the Kronecker product. One way to phrase the main subject of this section is to say that every $n$-square matrix $A$ is a linear  combination of Hubbard operators, just as this is stated in Eq.~(\ref{hub6}). In such a representation the assertions of the following propositions, theorems and corollaries are readily verified.


\begin{itemize}
\item[]
{\bf Theorem~M1.} The Kronecker product of $A=[a_{i,j}]$ and $B = [b_{k,\ell}]$, respectively $n$ and $m$-square matrices, can be written as
\[
A \otimes B = \sum_{i,j=1}^{nm} c_{i,j} X_{nm}^{i,j} \equiv C.
\]
That is, the Kronecker product $A \otimes B$ is a linear combination of the Hubbard operators of order $nm$.
\end{itemize}


\noindent
{\em Proof}. From Proposition~K1 and the linearity of $\otimes$ we have
\[
A\otimes B= \sum_{i,j=1}^n \, \sum_{k,\ell=1}^m a_{i,j} b_{k,\ell} X_{mn}^{m(i-1)+k, \, m(j-1)+\ell}.
\]
Let us define $p = m (i-1) + k$ and $q = m (j-1)+\ell$, so that $p, q=1, \ldots mn$. Therefore $k = p-m (i-1)$ and $\ell =q- m(j-1)$. Hence,
\[
A\otimes B =\sum_{i, j=1}^n \, \sum_{p = m(i-1)+1}^{ni}\, \sum_{q=m(j-1)+1}^{mj}a_{i,j} b_{p +m -mi,\,  q +m -mj} X_{mn}^{p,q}.
\]
The indexes $i, p$ and $m$ in the third sum are related by Eq.~(\ref{cotas}). The same is true for the indexes $j,q$ and $m$ in the fourth sum, so that according with Lemma~A1(i) of the appendix we get
\be
i= \left\lceil \frac{p}{m} \right\rceil := p', \qquad j= \left\lceil \frac{q}{m} \right\rceil := q'.
\label{ak0}
\ee
Then
\be
A\otimes B =\sum_{p,q=1}^{nm}  c_{p,q} X_{mn}^{p,q}, \qquad c_{p,q}:= a_{p', q'} \, b_{p +m -m p', \,  q +m -m q'}. \quad\Box
\label{ab}
\ee
As a first consequence of this theorem we realise that $A=\mathbb I_n$ and $B= \mathbb I_m$ produce $C= \mathbb I_{nm}$. That is, the Kronecker product of identity matrices is an identity matrix.


\begin{itemize}
\item[]
{\bf Corollary~M1.1.} The identity is preserved in the Kronecker product. That is $\mathbb I_n \otimes \mathbb I_m = \mathbb I_{nm}$

\end{itemize}

\noindent
{\em Proof}. Use Theorem~M1 with $a_{i,j}=\delta_{ij}$ and $b_{k,\ell} = \delta_{k \ell}. \quad\Box$

\vskip1ex
\noindent
Another important consequence of Theorem~M1 is that the nontrivial Kronecker powers of matrix $A$, written $A^{\otimes k+1}$ with $k \geq 1$, also admit a definite expression in Hubbard notation.


\begin{itemize}
\item[]
{\bf Corollary~M1.2.} The Kronecker product of the $n$-square matrix $A=[a_{i,j}]$  with itself $k\geq 1$ times, denoted $A^{\otimes k+1}$, is given by the expression
\be
A^{\otimes k+1} = \sum_{p,q=1}^{n^{k+1}} a_{p,q}^{(k+1)} X_{n^{k+1}}^{p,q}, \quad k \geq 1,
\label{ak}
\ee
with
\be
a_{p,q}^{(k+1)} = a_{p_k,q_k} \prod_{s=0}^{k-1} a_{p_s +n -np_{s+1}, \, q_s +n -n q_{s+1}},
\label{ak1}
\ee
and
\be
p_s= \left\lceil \frac{p}{n^s} \right\rceil, \quad q_s= \left\lceil \frac{q}{n^s} \right\rceil, \quad s =0, 1, \ldots, k.
\label{ak2}
\ee

\end{itemize}

\noindent
{\em Proof}. From Theorem~M1, with $B=A$ in (\ref{ab}), one has
\[
A^{\otimes 2}  =\sum_{p,q=1}^{n^2}  a_{p', q'} \, a_{p +n -n p', \,  q +n -n q'} X_{n^2}^{p,q},
\]
with $p'$ and $q'$ as they have been introduced in (\ref{ak0}). Applying again Theorem~M1 with $A \leftrightarrow A^{\otimes 2}$ and $B \leftrightarrow A$ we get
\be
A^{\otimes 2} \otimes A= \sum_{p,q=1}^{n^3} \left[ a_{(p')', (q')'} \, a_{p' +n -n (p')', \,  q' +n -n (q')'} \right] a_{p +n -n p', \,  q +n -n q'} X_{n^2}^{p,q}
\label{ak3}
\ee
where
\[
(x')' = \left\lceil \frac{x'}{n} \right\rceil =\left\lceil \frac{\left\lceil \frac{x}{n} \right\rceil}{n} \right\rceil = \left\lceil \frac{x}{n^2} \right\rceil, \quad x=p,q.
\]
In the last result we have used Lemma~A1(iv) of the appendix. Here, it is convenient to use the notation introduced in (\ref{ak2}), so that (\ref{ak3}) reads as follows
\[
\begin{array}{rl}
A^{\otimes 3} = & \displaystyle\sum_{p,q=1}^{n^3} a_{p_2, q_2} \, a_{p_1+n-np_2, \, q_1 +n-nq_2} \, a_{p_0+n-np_1, \, q_0 +n-nq_1}X_{n^2}^{p,q}\\[4ex]
 = & \displaystyle\sum_{p,q=1}^{n^3} a_{p_2, q_2} \left( \prod_{s=0}^2 a_{p_s+n-np_{s+1}, \, q_s +n-nq_{s+1}} \right) X_{n^2}^{p,q}\\[4ex]
 = & \displaystyle\sum_{p,q=1}^{n^3} a_{p,q}^{(3)} X_{n^2}^{p,q}.
\end{array}
\]
The proof is completed by induction. $\Box$

\vskip2ex
The action of $A^{\otimes t}$ on the vector space $\text{Sp} \left\{ \vert e_{\ell}^n \rangle^{\otimes t} \right\}_{\ell =1}^n$ represents the parallel action of $t$ operators $A$ on $t$ vector states $\vert \psi \rangle \in \text{Sp} \left\{ \vert e_{\ell}^n \rangle \right\}_{\ell =1}^n$. This manifestation of the quantum parallelism is a fundamental feature of many quantum algorithms \cite{Nie00}. Thus, quantum circuits can be constructed to evaluate a function $f(x)$ for multiple values of $x$ simultaneously. Most of the procedures implemented to calculate functions on an arbitrary number of bits use the Hadamard transform $H^{\otimes n}$. This operation is just $n$ Hadamard operators acting in parallel on $n$ qubits. It is then profitable to get a practical expression of $H^{\otimes n}$ as a particular application of Corollary~M1.2.


\begin{itemize}
\item[]
{\bf Proposition~M1.1.} Let $H$ be the Hadamard matrix (\ref{had1}), then
\be
H^{\otimes k+1} = \frac{1}{\sqrt{2^{k+1}}} \sum_{p,q=1}^{2^{k+1}} (-1)^{\vec p \cdot \vec q } \, X_{2^{k+1}}^{p,q},  \quad k\geq 1,
\label{had3}
\ee
with $p_s = \lceil \frac{p}{2^s} \rceil$, $q_s= \lceil \frac{q}{2^s} \rceil$, and
\be
\vec p \cdot \vec q := \sum_{s=0}^k (p_s-1)(q_s-1).
\label{had3a}
\ee

\end{itemize}

\noindent
{\em Proof}. For $n=2$ and $A=H$, equations (\ref{ak}) and (\ref{ak1}) respectively read as
\be
H^{\otimes k+1} = \sum_{p,q=1}^{2^{k+1}} h_{p,q}^{(k+1)} X_{2^{k+1}}^{p,q}, \quad k\geq 1
\label{had3b}
\ee
and
\[
h_{p,q}^{(k+1)} = h_{p_k, q_k} \prod_{s=0}^{k-1} h_{p_s -2-2p_{s+1}, \, q_s -2 -2q_{s+1}}, \qquad x_s = \left\lceil \frac{x}{2^s} \right\rceil, \quad x=p,q.
\]
From (\ref{had1}) we know that
\[
h_{i,j} = 2^{-1/2} (-1)^{(i-1)(j-1)}, \quad i,j=1,2.
\]
Therefore, 
\[
h_{p_s -2-2p_{s+1}, \, q_s -2 -2q_{s+1}} = \frac{(-1)^{(p_s -2-2p_{s+1} -1) (q_s -2 -2q_{s+1}-1)} }{\sqrt 2} = \frac{(-1)^{(p_s-1)(q_s-1)}}{\sqrt 2}
\]
and
\[
h_{p,q}^{(k+1)} = \frac{1}{\sqrt{2^{k+1}}} (-1)^{\sum_{s=0}^k (p_s -1) (q_s -1)}.
\]
Equation (\ref{had3}) follows from the introduction of this last result in (\ref{had3b}). $\Box$

\vskip2ex
Let $H_n$ denote a Hadamard matrix of order $n$; the matrix $H$ $(\equiv H_2)$ defined in Eq.~(\ref{had1}) and used in (\ref{had3}) is the simplest example. The next case is found for $n=4$ as the orthogonality condition on the rows of $H_n$ forces $n$ to be even \cite{Ban12}. Proposition~M1.1. gives an easy way to construct a Hadamard matrix of any order because $H^{\otimes k+1}$ is of size $2^{k+1}$. For instance, $H_4 =H \otimes H= H^{\otimes 2}$ reads as follows
\[
H_4 =H^{\otimes 2} = \frac12 \sum_{p,q=1}^4 (-1)^{(p-1)(q-1) +(\lceil \frac{p}{2} \rceil -1)
(\lceil \frac{q}{2} \rceil -1)} X_4^{p,q} =\frac12 \left(
\begin{array}{rrrr}
1 & 1 & 1 & 1\\
1 & -1 & 1 & -1\\
1 & 1 & -1 & -1\\
1 & -1 & -1 & 1
\end{array}
\right).
\]
The matrices $H_n$ are known as symmetric multi ports (or Zeilinger matrices) in quantum optics \cite{Jex95} and have applications in combinatorial problems, coding algorithms and quantum engineering, among a diversity of subjects. 

In order to appreciate the significance of Proposition M1.1., let us apply the operator $H^{\otimes k +1}$ on any of the vectors spanning $\mathbb K^{2^{k+1}}$. Using (\ref{had3}) and (\ref{abas})  with $A=H^{\otimes k+1}$  we get
\be
H^{\otimes k+1} \vert e^{2^{k+1}}_j \rangle = \frac{1}{\sqrt{2^{k+1}}} \sum_{p=1}^{2^{k+1}} (-1)^{\sum_{s=0}^k (p_s -1)(j_s -1)} \vert e_p^{2^{k+1}} \rangle,
\label{had3c}
\ee
with $y_s= \lceil \frac{y}{2^s} \rceil$ for $y=p,j$. In particular, if $k=1$ the latter expression gives
\be
H^{\otimes 2} \vert e_j^4 \rangle  = \frac12 \sum_{p=1}^4 (-1)^{(p-1)(j-1) + (p_1 -1)(j_1 -1)} \vert e_p^4 \rangle.
\label{had3d}
\ee
Explicitly,
\[
2H^{\otimes 2} \vert e_j^4 \rangle \rightarrow \left\{
\begin{array}{lr}
\vert e^4_1 \rangle + \vert e^4_2  \rangle +\vert e^4_3 \rangle + \vert e^4_4 \rangle,  & j=1\\[1ex]
\vert e^4_1 \rangle  - \vert e^4_2  \rangle + \vert e^4_3 \rangle - \vert e^4_4 \rangle,  & j=2\\[1ex]
\vert e^4_1 \rangle  + \vert e^4_2  \rangle - \vert e^4_3 \rangle - \vert e^4_4 \rangle, & j=3\\[1ex]
\vert e^4_1 \rangle  - \vert e^4_2  \rangle - \vert e^4_3 \rangle + \vert e^4_4 \rangle, & j=4
\end{array}
\right.
\]
We can see that the action of $H^{\otimes 2}$ on $\vert e^4_1 \rangle$ produces an equal superposition of all basis states. In quantum computing, this corresponds to set an empty quantum register of 2 qubits $\vert 00 \rangle  \equiv \vert e^4_1 \rangle$ into an equally weighted distribution of all the basis states of the register $\vert 00 \rangle$, $\vert 01 \rangle =\vert e_2^4 \rangle$, $\vert 10 \rangle =\vert e_3^4 \rangle$ and $\vert 11 \rangle =\vert e_4^4 \rangle$. At this stage, it would be useful to show the translation  of the results from our notation to the binary one, which is widely used in the quantum computing context. We first give the following definition.

\begin{itemize}
\item[]
{\bf Definition~M1.1.} Consider a positive integer $x\leq 2^{k+1}$ with $k \in \mathbb N$. The expansion of $x$ in powers of 2 is defined by the binary coefficients $x_s \in \{0,1 \}$, $s=0,1,\ldots, k+1$, as follows
\be
x= \sum_{i=0}^{k+1} x_i 2^i.
\label{xbin}
\ee

\end{itemize}

\vskip2ex
Now we pay attention to the coefficients of the linear combination (\ref{had3}), as only these must be rewritten to get equations (\ref{had3}) and  (\ref{had3c}) in binary form. The next proposition is necessary.

\begin{itemize}
\item[]
{\bf Proposition~M1.2.} Let $p$ and $q$ be respectively the $i$-th and $j$-th powers of 2 with $i,j=0,1, \ldots, k+1$, and $k \in \mathbb N$. Then
\be
 (-1)^{\sum_{s=0}^k(\lceil \frac{p}{2^s}\rceil-1)(\lceil \frac{q}{2^s}\rceil-1)}=(-1)^{\sum_{s=0}^k(p-1)_s(q-1)_s},
\label{exppq0}
\ee
where $(p-1)_s$ and $(q-1)_s$ are the $s$-th binary coefficients of $p-1$ and $q-1$ respectively.
\end{itemize}

\noindent
{\em Proof. } We first prove the identity
\be 
(-1)^{\sum_{s=0}^k \lfloor \frac{p}{2^s} \rfloor\lfloor \frac{q}{2^s}\rfloor }=(-1)^{\sum_{s=0}^k p_sq_s},
\label{flobin}
\ee
with $\lfloor x \rfloor$ the floor function of $x$ (see Eq. (\ref{floorfun}) of the appendix). Using the binary expansion of $p$ and $q$ we can write
\be
 \left\lfloor \frac{p}{2^s}\right\rfloor \left\lfloor \frac{q}{2^s}\right\rfloor=\sum_{i,j =0}^{k+1} \lfloor p_i 2^{i-s}\rfloor\lfloor q_j 2^{j-s}\rfloor, \qquad s=0,1,2\ldots, k.
\label{marc1}
\ee
Given $s$ one has $\lfloor x_{\ell} 2^{\ell -s} \rfloor =0$ for $x_{\ell} \in \{0,1\}$ and  $\ell <s$, since $0 \leq x_{\ell} 2^{\ell -s} <1$. Then, all the terms labelled with either $i< s$ or $j<s$ in (\ref{marc1}) are equal to zero. We have four partial sums
\[
 \left\lfloor \frac{p}{2^s}\right\rfloor \left\lfloor \frac{q}{2^s}\right\rfloor = \lfloor p_s \rfloor \lfloor q_s \rfloor +  \lfloor p_s\rfloor \sum_{j =s+1}^{k+1}\lfloor q_j 2^{j-s} \rfloor +   \lfloor q_s\rfloor \sum_{i=s+1}^{k+1}\lfloor p_i 2^{i-s} \rfloor + \sum_{i,j =s+1}^{k+1} \lfloor p_i 2^{i-s}\rfloor\lfloor q_j 2^{j-s}\rfloor.
\]
The second and third terms of this last result are either zero or an even number, so they can be omitted from the exponent of $-1$ in the expression at the left of Eq.~(\ref{flobin}). The same can be said of the fourth term after the change $i-s \rightarrow r$, $j-s \rightarrow t$. Finally, $\lfloor p_s \rfloor \lfloor q_s \rfloor= p_s q_s$ since $p_s, q_s \in \{0,1\}$. Therefore, taking into account only the elements of (\ref{marc1}) that contribute in nontrivial form to the exponent of $-1$ we get (\ref{flobin}). That is,
\[
\left\lfloor \frac{p}{2^s} \right\rfloor \left\lfloor \frac{q}{2^s}\right\rfloor \stackrel{*}{=}  p_s q_s \quad \Rightarrow \quad (-1)^{\sum_{s=0}^k \lfloor \frac{p}{2^s} \rfloor\lfloor \frac{q}{2^s}\rfloor }=(-1)^{\sum_{s=0}^k p_sq_s}.
\]
Using Lemma~A1 (part v) of the appendix and this last result, equation (\ref{exppq0}) follows. $\Box$

\vskip2ex
To verify the compatibility of our results with those obtained in a binary representation let us rewrite the coefficients of the linear combination (\ref{had3c}) according to Proposition~M1.2. For $k=1$ one has
\be
H^{\otimes 2} \vert e_j^4 \rangle  = \frac12 \sum_p^4 (-1)^{(p-1)_0(j-1)_0 + (p
-1)_1(j -1)_1} \vert e_p^4 \rangle,
\label{had3e}
\ee
which explicitly gives the following result
\[
2H^{\otimes 2} \vert e_j^4 \rangle \rightarrow \left\{
\begin{array}{lr}
(-1)^{0\cdot 0+0\cdot 0} \vert e^4_1 \rangle + (-1)^{1\cdot 0+0\cdot 0} \vert e^4_2 
\rangle +(-1)^{0\cdot 0+0\cdot 1} \vert e^4_3 \rangle + (-1)^{1\cdot 0+1\cdot 0}
\vert e^4_4 \rangle,  & j=1\\[2ex]
(-1)^{0\cdot 1+0\cdot 0} \vert e^4_1 \rangle  + (-1)^{1\cdot 1+0\cdot 0} \vert e^4_2
 \rangle + (-1)^{0\cdot 1+1\cdot 0} \vert e^4_3 \rangle + (-1)^{1\cdot 1+1\cdot 0} 
\vert e^4_4 \rangle,  & j=2\\[2ex]
(-1)^{0\cdot 0+0\cdot 1} \vert e^4_1 \rangle  + (-1)^{1\cdot 0+0\cdot 1} \vert e^4_2
 \rangle + (-1)^{0\cdot 0+1\cdot 1} \vert e^4_3 \rangle + (-1)^{0\cdot 1+1\cdot 1}
\vert e^4_4 \rangle, & j=3\\[2ex]
(-1)^{0\cdot 1+0\cdot 1} \vert e^4_1 \rangle  + (-1)^{1\cdot 1+0\cdot 1} \vert e^4_2
 \rangle + (-1)^{0\cdot 1+1\cdot 1} \vert e^4_3 \rangle + (-1)^{1\cdot 1+1\cdot 1}
\vert e^4_4 \rangle, & j=4
\end{array}
\right.
\]
The comparison of (\ref{had3e}) with (\ref{had3c}) shows that the results are consistent in both representations. The final step is to express $\vert e_j^4 \rangle$ in binary form. Using  Proposition~K0 and Definition~M1.1 we make
\[
\vert e_j^{2^{k+1}} \rangle \rightarrow \vert j-1 \rangle_{(k)} := \vert (j-1)_0, (j-1)_1, \ldots, (j-1)_{k} \rangle, 
\]
with $(j-1)_s \in \{0,1\}$ the binary coefficients of $j-1$ up to $2^{k}$. If $k=1$ then $\vert e_j^4 \rangle \rightarrow \vert j-1 \rangle_{(1)} = \vert (j-1)_0 , (j-1)_1\rangle$. Hence $\vert e_1^4 \rangle \rightarrow \vert 0 \rangle_{(1)} =\vert 0,0 \rangle$, so that we can write $\vert e_1^4 \rangle=\vert 00 \rangle$, and so on. Then, we can write equation (\ref{had3c}) in the standard form
\[
H^{\otimes n} \vert x \rangle = \frac{1}{\sqrt{2^n}} \sum_z (-1)^{x\cdot z} \vert z \rangle,
\]
where $n=2^{k+1}$. Here $\vert x \rangle$ and $\vert z \rangle$ are in binary notation.

Coming back to Theorem~M1, we stress  that this can be generalised to an arbitrary number of factors by including the products $A\otimes B$ and $A^{\otimes k+1}$ as particular cases. This is stated in the following proposition.

\begin{itemize}
\item[]
{\bf Proposition M1.3.} Let $A_r = \left[a_{i,j}^{(r)} \right]$ be a square matrix of order $n_r$. The Kronecker product $A_1\otimes A_2\otimes \cdots \otimes A_{k+1}$ is the square matrix of order $n^{(k)}=n_1 n_2\cdots n_{k+1}$, expressed as the following linear combination of Hubbard operators 
\be
A=A_1\otimes A_2\otimes \cdots \otimes A_{k+1}=\sum_{p,q =1}^{n^{(k)}} {\widetilde a}^{~(k)}_{p,q}~X_{n^{(k)}}^{p,q},\quad k\ge 1,
\label{amult}
\ee
where 
\be
\displaystyle {\widetilde a}_{p,q}^{~(k)}=a^{~(1)}_{p_k,q_k}\prod_{s=0}^{k-1} a^{~(k-s+1)}_{p_s+n_{k-s+1}-n_{k-s+1}p_{s+1},~q_s+n_{k-s+1}-n_{k-s+1}q_{s+1}},
\label{atilde}
\ee
and
\be
p_s= \left\lceil \frac{p}{\prod_{\ell=1}^s n_{k-\ell+2}}\right\rceil, \qquad q_s= \left\lceil\frac{q}{\prod_{\ell=1}^s n_{k-\ell+2}}\right\rceil.
\label{pqes}
\ee

\end{itemize}

\noindent
{\em Proof}. This is immediate by following the proof of Theorem~M1 and Corollary M1.2. $\Box$

\vskip2ex
The advantage of having an expression for the matrix Kronecker product as general as the one reported in Proposition~M1.3 relies on the fact that this includes an arbitrary number of factors, the size of which is in turn arbitrary. Immediate results can be obtained as particular cases. For instance, if $k=1$ and $a_{i,j}^{(1)} = a_{i,j}$, $a_{i,j}^{(2)} = b_{i,j}$, the product (\ref{amult}) gives the expression of $A\otimes B$ reported in Theorem~M1. Now, let all factors in (\ref{amult}) be of the same order, namely $n_r=n$ for $r=1,2,\ldots,k+1$, then the coefficient (\ref{atilde}) reads as
\be
{\widetilde a}_{p,q}^{~(k)}=a^{~(1)}_{p_k,q_k}\prod_{s=0}^{k-1} a^{~(k-s+1)}_{p_s+n-n p_{s+1},~q_s+n-n q_{s+1}},
\label{atilde2}
\ee
and the definitions of $p_s$ and $q_s$ in (\ref{pqes}) are reduced to the ones given in (\ref{ak2}). Using these last results and considering the situation in which all the factors are equal, i.e. $A_r = A$ for all $r$, one recovers the expression for $A^{\otimes k+1}$ reported in Corollary~M1.2. In such a case, the super-index could be omitted from all the matrix elements appearing in (\ref{atilde2}). On the other hand, let $A_r$ be Fourier matrices $F_r$, then (\ref{amult}) is the `Fourier Kronecker product' defined in \cite{Tad06b}, expanded in terms of Hubbard operators. In addition, this is `factored' if the size $n_r$ of the matrices $A_r$ are natural powers of prime numbers, and is called `pure factored' if such powers are ordered:
\[
F= F_{a^{\ell_1}} \otimes F_{a^{\ell_2}} \otimes \cdots \otimes F_{a^{\ell_{k+1}}}, \quad \ell_1 >> \ell_2 >> \cdots >>\ell_{k+1}, \quad \ell_j \in \mathbb N,
\]
where $a$ is a prime number. The immediate generalisation considers $n_r =a_r^{\ell_r}$, with $a_r$ standing for a prime number and $r=1,\ldots, k+1$. In Ref.~\cite{Tad06b}, a `multi-index' notation is introduced to operate the above mentioned products. Such notation works well for the necessities indicated by its author in the study of Fourier matrices. However, it could be not easy to generalise this to the Kronecker algebra of square matrices other than the Fourier ones. In contradistinction, the algebraic rules presented here are based on the linear superposition of Hubbard operators, so that they minimise the number of indices that are required to operate with. In this form, the Hubbard representation facilitates the calculation of the Kronecker algebra of the square matrices of any sort.


\begin{itemize}
\item[]
{\bf Theorem~M2.} Let $A =[a_{i,j}], C =[c_{p,q}]$, and $B =[b_{k, \ell}] ,D =[d_{r,s}]$, be pairs of $n$ and $m$-square matrices respectively. The usual matrix product of the $nm$-square matrices $A\otimes B$ and $C\otimes D$ fulfills
\be
(A\otimes B)(C\otimes D)=AC\otimes BD.
\label{m1}
\ee
That is, the Kronecker product of square matrices is compatible with ordinary matrix multiplication.
\end{itemize}


\noindent
{\em Proof}. From Proposition~K1 and the linearity of $\otimes$ one arrives at
\[
\begin{array}{l}
A \otimes B= \displaystyle\sum_{i,j}^n \sum_{k, \ell}^m a_{i,j} b_{k, \ell} \, X_{mn}^{m(i-1)+k, \, m(j-1) + \ell},\\[4ex]
C\otimes D= \displaystyle\sum_{p,q}^n \sum_{r,s}^m c_{p, q} d_{r, s} \, X_{mn}^{m (p-1)+r, \, m(q-1)+s}.
\end{array}
\]
Using (\ref{hub2}) and Lemma~A2 of the appendix we get
\[
\begin{array}{ll}
(A\otimes B)(C\otimes D) & =\displaystyle\sum_{i,j,p,q}^n \, \sum_{k, \ell ,r,s}^m a_{i, j}b_{k, \ell}\, c_{p, q} d_{r, s} \, \delta_{m (p-1) +r, \, m(j-1) + \ell} \, X_{mn}^{m(i-1)+k, \, m(q-1)+ s}\\[3.5ex]
& =\displaystyle\sum_{i,j,q}^n \, \sum_{k,\ell ,s}^m (a_{i, j} c_{j, q})(b_{k, \ell} d_{\ell, s}) X_{mn}^{m (i-1)+ k, \, m(q-1)+ s}\\[3.5ex]
& =\displaystyle\sum_{i,q}^n \sum_{k, s}^m (AC)_{i,q} (BC)_{k, s}\,  X_{mn}^{m (i-1)+ k, \, m(q-1)+ s}\\[3.5ex]
& =\displaystyle\left(\sum_{i,q}^n (AC)_{i,q} X_m^{i, q}\right) \otimes \left(\displaystyle\sum_{k,s}^m (BD)_{k,s} X_n^{k, s}\right) = AC \otimes BD. \quad\Box
\end{array}
\]

\noindent
At this stage we have to stress that Theorem~M2 can be extended to rectangular matrices whenever the involved matrix products make sense (see the discussion of Section~\ref{titrec-sub}). This fact will be taken into account in e.g., Corollary~TM2.2 and Proposition~G.1.


\begin{itemize}
\item[]
{\bf Corollary~TM2.1.} Let $A$ and $D$ be square matrices of order $n$ and $m$ respectively. Then the following relationship holds
\be
A \otimes D = (A\otimes \mathbb I_m)(\mathbb I_n \otimes D).
\label{m2}
\ee
\end{itemize}


\noindent
{\em Proof}. Use $B= \mathbb I_m$ and $C= \mathbb I_n$ in Theorem~M2. $\Box$

\begin{itemize}
\item[]
{\bf Corollary~TM2.2.} Consider the eigenvalue equations $A \vert a_i \rangle = \alpha_i \vert a_i \rangle$ and $B \vert b_j \rangle = \beta_j \vert b_j \rangle$, with $i = 1,2, \ldots, n$ and $j = 1,2, \ldots, m$. Then (i) the $nm$ numbers $\alpha_i \beta_j$ are the eigenvalues of $A\otimes B$ associated to the vectors $\vert a_i \rangle \otimes \vert b_j \rangle$ (ii) The eigenvalues of $A\otimes \mathbb I_m + \mathbb I_n \otimes B$ are the numbers $\alpha_i + \beta_j$.

\end{itemize}


\noindent
{\em Proof}. Using (\ref{ket1}) the vectors $\vert a_i\rangle$ and $\vert b_j\rangle$ can be represented by row tuples of size $n$ and $m$ respectively. Then, Theorem~M2 applies as follows 
\[
(A\otimes B)(\vert a_i\rangle\otimes\vert b_j\rangle)=A\vert a_i\rangle \otimes B\vert b_j\rangle = \alpha_i \beta_j \left( \vert a_i\rangle \otimes \vert b_j \rangle \right),
\]
and the proof of (i) is completed. In similar form, 
\[
(A \otimes \mathbb I_m + \mathbb I_n \otimes B)(\vert a_i\rangle\otimes\vert b_j\rangle) = A\vert a_i\rangle \otimes \mathbb I_m \vert b_j\rangle + \mathbb I_n \vert a_i\rangle \otimes B\vert b_j\rangle =(\alpha_i + \beta_j) (\vert a_i\rangle\otimes\vert b_j\rangle)
\]
completes the proof of (ii). $\Box$


\vskip2ex
\noindent
The following properties are presented here in connection with the Kronecker algebra of Hubbard operators, because any square matrix can be expressed as a linear combination of such operators. In this form, the proof of each item is simple by using equation (\ref{hub6}) and Proposition~K2.


\begin{itemize}
\item[]
{\bf Theorem~M3.} Let $A$, $B$ and $C$ be $n$-square matrices and $\lambda \in \mathbb K$. Then

\item[i)] 
In general, $A\otimes B\neq B\otimes A$.

\item[ii)] 
$( A \otimes B )^{\dagger} =A^{\dagger} \otimes B^\dagger$ and  $( A\otimes B )^T=A^T\otimes B^T$.

\item[iii)] 
$(\lambda A) \otimes B =\lambda ( A \otimes  B )= A \otimes ( \lambda B )$.

\item[iv)] 
$( A+ B ) \otimes C=A\otimes C+B\otimes C$.

\item[v)] 
$A \otimes ( B + C )= A \otimes B + A \otimes C$.

\item[vi)] 
$( A \otimes B ) \otimes C = A\otimes  (B \otimes C )$.

\end{itemize}


\noindent
{\em Proof}. The theorem follows from Proposition~K2 and properties (\ref{hub6}) and (\ref{hub8}). For instance, the proof of both parts in (ii) is quite similar: 
\[
\begin{array}{rl}
( A \otimes B )^{\dagger} & =\displaystyle\left[ \left( \sum_{i,j=1}^n a_{i,j} X_n^{i,j} \right) \otimes \left( \sum_{k,\ell=1}^n b_{k, \ell} X_n^{k, \ell} \right) \right]^{\dagger} = \left( \sum_{i,j,k,\ell =1}^n a_{i,j} b_{k, \ell} X_n^{i,j} \otimes X_n^{k, \ell} \right)^{\dagger}\\[4ex]
& = \displaystyle\sum_{i,j,k,\ell =1}^n \overline{a_{i,j}  b_{k, \ell}} \left(X_n^{i,j} \otimes X_n^{k, \ell} \right)^{\dagger} = \sum_{i,j,k,\ell =1}^n \overline a_{i,j} \overline b_{k, \ell} \left(X_n^{i,j} \right)^{\dagger}  \otimes \left( X_n^{k, \ell} \right)^{\dagger}\\[4ex]
& = A^{\dagger} \otimes B^{\dagger}.
\end{array}
\]
Then $(A \otimes B)^T =(\overline{A \otimes B})^{\dagger} = \overline A^{\dagger} \otimes \overline B^{\dagger} = A^T \otimes B^T. \quad\Box$

\vskip2ex
\noindent
As for Proposition~K2, the above properties mean that the Kronecker product is distributive over ordinary matrix addition (iv, v), associative (vi), compatible with ordinary matrix transposition (ii) and the matrix multiplication by an scalar (iii). Remarkably, although the product $\otimes$ is non-abelian for arbitrary square matrices (i), it is possible to set equivalence classes between Kronecker products that are `abelian' up to a permutation matrix. Our claim is based on the Proposition~P2 as well as the linearity of the Hubbard operators and leads to the following theorem. 


\begin{itemize}
\item[]
{\bf Theorem~M4.} Let $A=[a_{i,j}]$ and $B=[b_{k, \ell}]$ be two square matrices of order $n$ and $m$ respectively. The Kronecker product $A\otimes B$ is permutation equivalent to $B \otimes A$.

\end{itemize}


\noindent
{\em Proof}. From Proposition~P2 we know that there exists a permutation matrix $P$ such that (\ref{per7}) is true. Then, by linearity in the conventional matrix product we have
\[
\begin{array}{rl}
P^T (A\otimes B)P = & \displaystyle\sum_{i,j}^n \sum_{k, \ell}^m a_{i,j} b_{k, \ell} \left[ P^T \left( X_n^{i,j} \otimes X_m^{k, \ell} \right) P \right]\\[4ex]
= & \displaystyle\sum_{i,j}^n \sum_{k, \ell}^m b_{k, \ell} a_{i,j} X_m^{k, \ell} \otimes X_n^{i,j} = B\otimes A. \quad\Box
\end{array} 
\]
The next properties show that the Kronecker product is compatible with the conventional measure properties of square matrices.


\begin{itemize}
\item[]
{\bf Proposition~M1.4.} Let $A=[a_{i,j}]$ and $B=[b_{k, \ell}]$ be two square matrices of order $n$ and $m$ respectively. Then
\[
\text{Tr} (A\otimes B) = \text{Tr}(A) \text{Tr}(B) = \text{Tr}(B \otimes A).
\]
\end{itemize}


\noindent
{\em Proof}. Using Proposition~K1, the linearity of $\otimes$, and property (\ref{hub3a}) we can write
\[
\langle e^{nm}_s \vert A \otimes B \vert e^{nm}_s \rangle = \sum_{i,j}^n \sum_{k, \ell}^m a_{i,j} b_{k, \ell}\,  \delta_{s, m(i-1) +k} \, \delta_{m(j-1) + \ell, s}.
\]
Then, (\ref{hub7c}) gives
\[
\begin{array}{rl}
\text{Tr}(A \otimes B) = & \displaystyle\sum_s^{nm} \langle e^{nm}_s \vert A \otimes B \vert e^{nm}_s \rangle =  \displaystyle\sum_{i,j}^n \sum_{k, \ell}^m a_{i,j} b_{k, \ell}\,  \delta_{m(i-1) +k, \, m(j-1) + \ell}\\[4ex]
= & \displaystyle\sum_{i,j}^n \sum_{k, \ell}^m a_{i,j} b_{k, \ell}\,  \delta_{i,j} \delta_{k, \ell} =\left( \sum_i^n a_{i,i} \right) \left( \sum_k^m b_{k,k} \right)\\[4ex]
= & \text{Tr}(A) \text{Tr}(B) =  \text{Tr}(B) \text{Tr}(A) = \text{Tr}(B \otimes A),
\end{array}
\]
where we have used Lemma~A2 of the appendix. $\Box$


\begin{itemize}
\item[]
{\bf Proposition~M1.5.} Let $A=[a_{i,j}]$ and $B=[b_{k, \ell}]$ be two $n$-square matrices. Then
\be
\text{Det}(A\otimes B) = (\text{Det}\, A)^n (\text{Det} \, B)^n.
\label{aob}
\ee

\end{itemize}


\noindent
{\em Proof}. From Corollary~TM2.1 we have
\[
A\otimes B = (A \otimes \mathbb I_n ) (\mathbb I_n \otimes B).
\]
Then
\[
\text{Det}(A\otimes B) = \text{Det} (A \otimes \mathbb I_n )  \text{Det}(\mathbb I_n \otimes B) = \text{Det} (A \otimes \mathbb I_n ) (\text{Det} \, B)^n.
\]
The last term in the previous expression is because $\mathbb I_n \otimes B$ is a Jordan matrix having $n$ repetitions of $B$ along the diagonal. Now we use Theorem~M4 to get
\[
\text{Det} \left[ P^T (A\otimes \mathbb I_n) P \right] =\text{Det} (A\otimes \mathbb I_n) = \text{Det} (\mathbb I_n \otimes A) = (\text{Det} \, A)^n.
\]
Equation (\ref{aob}) follows from the above results. $\Box$


\vskip2ex
Additional properties (and proofs) of the Kronecker product of matrices can be found in the books \cite{Gra81,Hor91,Ste11}. The most recent summary of the properties of the $\otimes$ operation has been reported in \cite{Van00} (see also \cite{Lan04}). Although these references are addressed to the case of rectangular matrices, most of the useful applications of the product $\otimes$ require square matrices only. Indeed, properties like the permutation equivalence of Kronecker products are fundamental in the simplification of algorithms to process data or in the study of symmetries associated to a given physical system, among other applications of square matrices. 


\subsection{A pair of physical models}
\label{phys}

As  examples of the applications of the above developed formalism we are going to discuss two concrete problems. The first one deals with the diagonalization of the Hamiltonian of a given $n$-level system. This problem corresponds to the spectral decomposition of the Hamiltonian in the appropriate basis and represents a cumbersome calculation in the conventional approaches, the difficulty of which increases as the order of the matrix representation. We shall show that the Hubbard operators are the best mathematical tool that one has at hand to face such a problem in easy manner. We apply the method to diagonalize the Hamiltonian of the Heisenberg model associated to the Hubbard and t-J models discussed in Section~\ref{new}. The second problem consists in solving the Schr\"odinger equation for the well known Jaynes-Cummings model in terms of Hubbard operators. We shall address the cases of a single-atom in a single-mode cavity and two atoms in single-mode cavities.

\subsubsection{Diagonalization in the Heisenberg model}
\label{diagon}

Here, we follow the unitary transformation method introduced in \cite{Val88} (see also \cite{Ovc04}). Consider the Hamiltonian of an $n$-level system in Hubbard notation
\begin{equation}
\label{nlham}
 H = \sum _ {p=1} ^ n \varepsilon _ p ~ X _ n^{ p , p } + \sum _ {\substack{p,q = 1 \\ p \neq q}} V _ {p,q} ~ X _ n^{p,q}, \qquad V_{q,p} = {\overline V}_{p,q}.
\end{equation}
To get (\ref{nlham}) in diagonal form one applies $n-1$ unitary transformations that depend on certain complex parameters. The requirement that all the off-diagonal elements be null in each step leads to a system of equations that determine the parameters of the transformation. To be precise, the unitary operators
\begin{equation}
 U_{k,m}(\alpha) = \exp \left(\alpha X_n^{k,m} - {\overline a} X_n ^{m,k}\right), \quad m > k = 1,2,\ldots, n, \quad \alpha = \vert \alpha \vert e^{i\mu},
\end{equation}
define the similarity transformation
\begin{equation}
H^\prime = U_{k,m}(\alpha) H U_{k,m}^\dagger (\alpha) = \sum _ {p=1}^n \varepsilon _ p^\prime ~ X _ n^{ p , p } + \sum _ {\substack{p,q = 1 \\ p \neq q}} V _ {p,q} ^ \prime  ~ X _ n^{p,q},
\end{equation}
where
\begin{equation}
\label{coefprim}
\begin{array}{ll}
 \varepsilon _ k^\prime = \frac12 \left[ \varepsilon _k  + \varepsilon _ m + (\varepsilon _k - \varepsilon _ m) \cos 2 \vert \alpha \vert + 2  \mbox{Re} ( V_{k,m} e ^ {-i\mu}) \sin 2 \vert \alpha \vert \right],\\[2ex]
 \varepsilon _m ^ \prime = \frac12 \left[ \varepsilon _k + \varepsilon _ m - (\varepsilon _k - \varepsilon _ m) \cos 2 \vert \alpha \vert - 2 \mbox{Re} ( V_{k,m} e ^ {-i\mu}) \sin 2 \vert \alpha \vert \right],\\[2ex]
 V_{k,m} ^ \prime e ^ {i\mu} = \frac12 \left[ \frac12 (\varepsilon _ m - \varepsilon _k) \sin 2 \vert \alpha\vert + V_{ k, m} e^{-i \mu} \cos ^ 2 \vert \alpha \vert - {\overline V}_{ k, m} e ^ {i \mu} \sin ^ 2 \vert \alpha \vert\right], \\[2ex]
 V_{k,p} ^ \prime = V_{k,p} \cos \vert \alpha \vert + V_{m,p} e ^ {i\mu} \sin \vert \alpha \vert, \\[2ex]
 V_{m,p} ^ \prime = V_{p,m} \cos \vert \alpha \vert - V_{p,k} e ^ {i\mu} \sin \vert \alpha \vert, \\[2ex]
 \varepsilon _ p ^ \prime = \varepsilon _ p, \qquad V_{p,q} ^ \prime = V_{p,q} \qquad p,q \neq k,m.
\end{array}
\end{equation}
The condition $ V_{k,m}^\prime = 0$ produces
\begin{equation}
\tan 2 \vert \alpha \vert = \frac{2 (-1) ^ {\kappa+1} \vert V_{k,m} \vert }{\varepsilon _ m - \varepsilon_k}.
\end{equation}
Hence
\begin{equation}
 \varepsilon_k^\prime = \frac12 - \sqrt{\frac14 (\varepsilon_m - \varepsilon_k) ^ 2 + \vert V_{k,m} \vert ^ 2},  \quad \varepsilon_m ^ \prime = \frac12 + \sqrt{\frac14(\varepsilon_m - \varepsilon_k) ^ 2 + \vert V_{k,m} \vert ^ 2}.
 \end{equation}
The Hamiltonian (\ref{nlham}) is diagonalized by iterating the above procedure as many times as necessary.

Consider now a system of $n$ spin-1/2 particles that interact according to the Heisenberg model \cite{Ovc04}. The involved Hamiltonian is
\begin{equation}
\label{heiham}
H = -\frac12 \sum_{ j = 1 }^n \left( J_x \sigma_j ^ x \sigma_{j+1} ^ x + J_y \sigma_j ^ y \sigma_{j+1} ^ y + J_z \sigma_j ^ z \sigma_{j+1} ^ z \right), 
\end{equation}
with
\begin{equation*}
 \sigma _ j ^{k} = \underbrace{\mathbb{I}_2 \otimes \ldots \otimes \mathbb{I}_2}_{j-1} \otimes \sigma^k \otimes \mathbb{I}_2 \otimes \ldots \otimes \mathbb{I}_2.
\end{equation*}
Here $\sigma _ {n+1}^k := \sigma _1^k$, with $\sigma^k$, $k = x, y, z$, standing for the Pauli matrices (see Eq.~(\ref{pauli}) of Section~\ref{titsu2}) . In the XXX model one has $J_x = J_y = J_z = J$, and using $n=2$ the above Hamiltonian is reduced:
\begin{equation}
H = -\frac J2 \left( \sigma ^ x \otimes \sigma ^ x + \sigma ^ y \otimes \sigma ^ y + \sigma ^ z \otimes \sigma ^ z \right). 
\end{equation}
Using the representation of the Pauli matrices in terms of the Hubbard operators (see Eq.~(\ref{pauref}) of Section~\ref{tithub-sub}), and the Proposition~K1 we arrive at the $X$-operator representation of the Hamiltonian which is going to be diagonalized
\begin{equation}
\label{heiham2}
H= - \frac J4 \left( X ^ {1,1} _ 4 - X ^ {2,2} _ 4 - X ^ {3,3} _ 4 + X ^ {4,4} _ 4 \right) - 2J \left( X ^ {2,3} _ 4 + X ^ {3,2} _ 4 \right).
\end{equation}
In this case only the unitary transformation $U_{2,3} (\alpha)$ is required since the off-diagonal element in (\ref{heiham2}) is simple. From (\ref{coefprim}) we get $\cos \left( 2 \vert \alpha \vert \right) = 0$, $\sin \mu = 0$, and the system
\begin{equation}
\begin{array}{ll}
 \varepsilon ^ \prime _ 1 = \varepsilon _ 1 = -\frac J4, \quad  \varepsilon ^ \prime _ 2 = \frac12 (\varepsilon _ 2 + \varepsilon _ 3) - V_{2,3} = \frac 94 J,\\[2ex]
 \varepsilon ^ \prime _ 3 = \frac12 (\varepsilon _ 2 + \varepsilon _ 3) + V_{2,3} = -\frac 74 J, \quad  \varepsilon ^ \prime _ 4 = \varepsilon _ 4 = -\frac J4.
\end{array}
\end{equation}
Therefore we arrive at the diagonal Hamiltonian
\begin{equation}
H^\prime = -\frac J4 \left( X ^ {1,1} _ 4 - 9 X ^ {2,2} _ 4 +7 X ^ {3,3} _ 4 + X ^ {4,4} _ 4 \right).
\end{equation}

\subsubsection{The Jaynes-Cummings model in Hubbard notation}
\label{jay}

The simplest form of describing the interaction between a single atom and a single mode of the quantised electromagnetic field in a cavity is in terms of the Jaynes-Cummings model \cite{Jay63}. This is defined in the Hilbert space associated to the composite system atom+field, so that the vector states of the entire system are spanned by the Kronecker products of the basis vectors belonging to each of the subsystems. The Hamiltonian of the system $H=H_{at}+H_f+H_I$ is integrated by the Hamiltonians of the atom alone $H_{at}$, the field alone $H_f$, and the interaction between the atom and the field $H_I$. The former two operators are diagonal and can be combined as $H_0=H_{at}+H_f$. Using the Kronecker algebra of the Hubbard operators it easy to shown that $H_0$ and $H_I$ are first integrals of the system \cite{Enr13}. The dynamics of the atom is then analysed by using the off-diagonal Hamiltonian 
\begin{equation}
\label{jcham}
H_I = \gamma (\sigma ^ + a + \sigma ^ - a ^ \dagger),
\end{equation}
because $H_0$ adds only a global phase to the time-evolution of the vector states. Here $a$ and $a^\dagger$ are the boson ladder operators and $\sigma^{\pm}$ the spin-1/2 ladder operators. The $X$-operator representation of the Hamiltonian (\ref{jcham}) is written as follows \cite{Enr13}:
\begin{equation}
\label{jcham2}
 H_I = \sum_{ p = 1} ^ 2 N_p~X ^ {p, 3-p}\qquad N_p = \sqrt{ N + 2 - p},
\end{equation}
where the boson number operator $N=a^{\dagger}a$ has been promoted to act on the vector space of the entire atom+field system: $N= \mathbb I_{at} \otimes N$, with $\mathbb I_{at}$ the identity operator in the vector space of the atom. The Hamiltonian (\ref{jcham2}) can be diagonalized by following the method described in the previous section. The Kronecker algebra of the Hubbard operators facilitates the construction of the unitary evolution operator
\begin{equation}\label{timeop}
 U( t ) = e^{-i H_I t} = \sum_{p,q=1} ^ 2 u_{p,q} (N_p) ~ X ^ {p,q},
\end{equation}
with
\begin{equation}
 u_{p,q} (N_p)= e ^ {i \frac \pi 2 \vert p - q \vert} \cos \left( \gamma t N_p - \frac \pi 2 \vert p - q \vert \right).
\end{equation}
The unitary operator (\ref{timeop}) corresponds to the solution of the related Schr\"odinger equation \cite{Enr13}, and is useful in extending the model to an arbitrary number of cavities \cite{Enr13b}. For instance, the dynamics of two non-interacting atoms can be analysed in terms of the unitary operator \begin{equation}
 U(t) = U_1 (t) \otimes U_2 (t)
\end{equation}
where $U_1(t)$ and $U_2(t)$ are expressed in $X$-operator notation \cite{Enr13}:
\begin{equation}
 U_i = \displaystyle \sum_{ p , q = 1 } ^ 2 u_{ p , q } ( N_{i;p} ) ~ X ^ { p , q } _ i, \quad i=1,2.
 \end{equation}
The operators $N_{i;p}$ are equivalent to the ones defined in (\ref{jcham2}) for each mode of the field. By virtue of (\ref{hub0}), the time evolution operator of the whole system reads
\begin{equation}
  U (t) = \sum_{ p , q = 1} ^ 4 u _ {p ^ \prime, q ^ \prime} \left ( N _ { p ^ \prime } \right ) ~ u _ {p + 2 - 2 p^ \prime, q + 2 - 2 q^ \prime } \left ( M _ { q + 2 - 2 q^ \prime } \right ) ~ X ^ { p , q}.
\end{equation}
This last operator gives the solutions to the Schr\"odinger equation associated to the two non-interacting atoms in separated cavities. 


\section{Basics of the Clebsch-Gordan decomposition}
\label{titbas}

In this section we consider the Kronecker product of two irreducible representations of a given group. Our interest here is to analyse the generalities of the reduction of such a product as a sum of irreducible representations. To start with, we require some basic definitions of the direct sum of vector spaces and the linear representation of groups. 


\subsection{Direct sum of vector spaces}
\label{titdir-sub}


\begin{itemize}
\item[]
{\bf Definition~S1.} Let ${\cal H} = {\cal H}' \oplus {\cal H}''$ be the direct sum vector space of ${\cal H}'$ and ${\cal H}''$ over the field $\mathbb K$. Then ${\cal H}' \cap {\cal H}'' = \mbox{Sp} \{ \vert \emptyset \rangle \}$, with $\vert \emptyset \rangle$ the null vector of ${\cal H}$. Assuming $\mbox{Dim} ({\cal H}) < \infty$, we have $\mbox{Dim} ({\cal H}) = \mbox{Dim} ({\cal H}') + \mbox{Dim} ({\cal H}'')$. In general, any vector $\vert x \rangle \in {\cal H}$ can be written in the form of a two-vector tuple $\vert x \rangle = (\vert x' \rangle, \vert x'' \rangle)^T$, with $\vert x' \rangle \in {\cal H}'$ and $\vert x'' \rangle \in {\cal H}''$. The following matrix notation will often be used
\[
\vert x \rangle = \left(
\begin{array}{c}
x_1\\
\vdots\\
x_{n'}\\
x_{n'+1}\\
\vdots\\
x_{n'+n''}
\end{array}
\right) \equiv \vert x'\rangle \oplus \vert x'' \rangle = \left(
\begin{array}{c}
\vert x'\rangle\\[1ex]
\vert x'' \rangle
\end{array}
\right)  \equiv \left(
\begin{array}{c}
x_1'\\
\vdots\\
x_{n'}'\\
x_1''\\
\vdots\\
x_{n''}''
\end{array} \right).
\]
\end{itemize}

\noindent
Now, let $A': {\cal H}' \rightarrow {\cal H}'$ and $A'': {\cal H}'' \rightarrow {\cal H}''$ be respectively automorphisms of ${\cal H}'$ and ${\cal H}''$. Assume their action is as follows
\[
A'\vert x' \rangle = \vert y' \rangle, \quad A''\vert x'' \rangle = \vert y'' \rangle.
\]
They together can be expressed in a single matrix operator acting on the entire space ${\cal H}' \oplus {\cal H}''$ in block-diagonal form
\[
\left(
\begin{array}{c|c}
A' & 0_{n'\times n''}\\
\hline
0_{n'' \times n'} & A''
\end{array}
\right) \left(
\begin{array}{c}
\vert x'\rangle\\[1ex]
\vert x'' \rangle
\end{array}
\right) = \left(
\begin{array}{c}
\vert y'\rangle\\[1ex]
\vert y'' \rangle
\end{array}
\right).
\]
We shall write in this case $A=A' \oplus A''$, and we shall say that the automorphism $A: {\cal H} \rightarrow {\cal H}$ decomposes into the direct sum of $A'$ and $A''$. 


\begin{itemize}
\item[]
{\bf Definition~S2.} Let ${\cal H} = {\cal H}'\oplus {\cal H}''$ be the direct sum of vector spaces ${\cal H}'$ and ${\cal H}''$, with $\text{Dim}({\cal H}') =n'$ and $\text{Dim}({\cal H}'') =n''$. The direct sum operator $A=A' \oplus A''$, written in diagonal matrix form as
\[
A= A' \oplus A'' = \left(
\begin{array}{c|c}
A' & 0_{n'\times n''}\\
\hline
0_{n'' \times n'} & A''
\end{array}
\right),
\]
is defined to act on $\vert x \rangle = \vert x'\rangle \oplus \vert x'' \rangle \in {\cal H}$ as $A \vert x \rangle = (A' \oplus A'') \vert x \rangle = A'\vert x' \rangle \oplus A'' \vert x'' \rangle$, with $A' \in \text{Aut} ({\cal H}')$ and $A'' \in \text{Aut} ({\cal H}'')$ respectively.

\end{itemize}

\noindent
Hereafter $\mbox{Aut}(X)$ denotes the set of all automorphisms of $X$. The next proposition allows to distinguish between operators that can be decomposed as a direct sum when they act on any vector space $\bigoplus_k {\cal H}_k$, and those that act on $\bigoplus_k {\cal H}_k$ in a more complicated manner.


\begin{itemize}
\item[]
{\bf Proposition~S1.} The most general operator acting on the direct sum space ${\cal H} = {\cal H}' \oplus {\cal H}''$ is of the form
\[
A = \left(
\begin{array}{c|c}
A' & B\\
\hline
C & A''
\end{array}
\right),
\]
with $B$ and $C$ matrices of order $n'\times n''$ and $n''\times n'$ which respectively maps ${\cal H}''$ into ${\cal H}'$ and ${\cal H}'$ into ${\cal H}''$. $A'$ and $A''$ are $n'$- and $n''$-square matrices respectively.

\end{itemize}

\noindent
{\em Proof.} The proof is immediate in matrix notation
\[
A \vert x \rangle = \left(
\begin{array}{rr}
A' & B\\
C & A''
\end{array}
\right) \left(
\begin{array}{l}
\vert x'\rangle\\
\vert x''\rangle
\end{array}
\right)= \left(
\begin{array}{l}
A'\vert x'\rangle + B \vert x'' \rangle\\
C \vert x'\rangle + A'' \vert x'' \rangle
\end{array}
\right),
\]
where the products $B \vert x'' \rangle$ and $C \vert x'\rangle$ are $n'\times 1$ and $n'' \times 1$ matrices respectively. $\Box$


\subsection{Linear representation of groups}
\label{titlin-sub}

\begin{itemize}
\item[]
{\bf Definition~G1.} Let $G$ and ${\cal H}$ be a group and a vector space respectively. The homomorphism
\[
\begin{array}{rl}
T: & G \rightarrow \text{Aut} ({\cal H})\\
& g \mapsto T(g)
\end{array}
\]
is a linear representation of $G$ on ${\cal H}$. The vector space ${\cal H}$ is called the representation space induced by $T$. The operator $T(g)$ fulfils
\[
T(g_2 g_1) = T(g_2) T(g_1), \quad \forall g_1, g_2 \in G.
\]

\end{itemize}

\noindent
Henceforth all the operators will be expressed in matrix form, so that the representations considered will be linear.

\begin{itemize}
\item[]
{\bf Definition~G2.} Given a representation $T$ of $G$ on ${\cal H}$, a subspace ${\cal H}_s$ of ${\cal H}$ is said to be $G$-invariant if, for all $\vert x \rangle \in {\cal H}_s$ and all $g \in G$, we have $T(g) \vert x \rangle \in {\cal H}_s$, i.e., $T(g) \in \text{Aut}({\cal H}_s)$ for all $g \in G$. In any representation there exist two trivial invariant subspaces; ${\cal H}$ itself and the null vector space $\mbox{Sp}\{ \vert \emptyset \rangle\}$. A representation $T$ of $G$ on ${\cal H}$ is irreducible, written $\Delta(G)$, if the only $G$-invariant subspaces of ${\cal H}$ are $\mbox{Sp}\{ \vert \emptyset \rangle\}$ and  ${\cal H}$ itself. 
\end{itemize}

\begin{itemize}
\item[]
{\bf Proposition~G1.} Let $\Delta'(G)$ and $\Delta''(G)$ be two irreducible representations of a given group $G$. Let $\text{Dim}({\cal H}') =n'$ and $\text{Dim}({\cal H}'') = n''$ be the dimensions of the corresponding representation spaces. Given $g \in G$, the Kronecker product of matrices $\Delta'(g)$ and $\Delta''(g)$ is a square matrix of order $n' n''$ acting on ${\cal H}' \otimes {\cal H}''$. That is, the set of matrices
\[
(\Delta' \otimes \Delta'') (g) = \Delta'(g) \otimes \Delta''(g)  
\]
defines a linear representation of the product $G\times G$, with ${\cal H}' \otimes {\cal H}''$ as the representation space.
\end{itemize}


\noindent
{\em Proof}. Given $\vert x' \rangle \in {\cal H}'$ and $\vert x'' \rangle \in {\cal H}''$, one has $\Delta'(g) \vert x' \rangle = \vert y' \rangle \in {\cal H}'$ and $\Delta''(g) \vert x'' \rangle = \vert y'' \rangle \in {\cal H}''$. Using Theorem~M2 we obtain
\[
\left( \Delta'(g) \otimes \Delta''(g) \right) (\vert x' \rangle \otimes \vert x'' \rangle) = \Delta'(g) \vert x' \rangle \otimes \Delta''(g) \vert x'' \rangle = \vert y' \rangle \otimes \vert y'' \rangle = \vert y \rangle.
\]
Therefore, given $g \in G$, there exists a matrix $T(g) = \Delta'(g) \otimes \Delta''(g)  \in \text{Aut}({\cal H}' \otimes {\cal H}'')$ such that $T(g) \vert x \rangle = \vert y \rangle$, with $\vert x \rangle = \vert x' \rangle \otimes \vert x'' \rangle$ and $\vert y \rangle = \vert y' \rangle \otimes \vert y'' \rangle. \quad \Box$


\subsection{Clebsch-Gordan decomposition}
\label{titcle-sub}

Let us stress that in general the Kronecker product $\Delta' \otimes \Delta''$ is not irreducible, even though $\Delta'$ and $\Delta''$ are irreducible. This unpleasant situation defines the problem of reducing $\Delta' \otimes \Delta''$ as much as possible to a representation where the vector space ${\cal H}' \otimes {\cal H}''$ is $G$-invariant. The best we can do is to split the representation space ${\cal H}' \otimes {\cal H}''$ into a set of subspaces ${\cal H}_k \cap {\cal H}_j =\mbox{Sp} \{\vert \emptyset \rangle \}$, $k\neq j$, each of them being $G$-invariant. Then, $\Delta' \otimes \Delta''$ is decomposed into a set of irreducible representations $\Delta_k$, one per each subspace ${\cal H}_k$.

\begin{itemize}
\item[]
{\bf Definition~G3.} Let $\Delta'(G)$ and $\Delta''(G)$ be two irreducible representations of a given group $G$. The Clebsch-Gordan decomposition of $(\Delta' \otimes \Delta'')(g)$ is the reduction of the Kronecker product $\Delta' (g) \otimes \Delta'' (g)$ into a direct sum of $\rho$ irreducible representations $\Delta_k(g)$ defined as
\[
(\Delta' \otimes \Delta'')(g) = \bigoplus_{k=1}^{\rho} \Delta_k (g) = \Delta_1(g) \oplus \Delta_2(g) \oplus \cdots \oplus \Delta_{\rho}(g), \quad \forall g\in G.
\]
\end{itemize}

\vskip2ex
\noindent
Note that the representation space of the Clebsch-Gordan decomposition $\bigoplus_{k=1}^{\rho} \Delta_k$ is the direct sum $\bigoplus_{k=1}^{\rho} {\cal H}_k$. Therefore, Definition~G3 implies that the vector space ${\cal H}' \otimes {\cal H}''$ is decomposed into the direct sum ${\cal H}' \otimes {\cal H}'' = \bigoplus_{k=1}^{\rho} {\cal H}_k$. If it is possible to solve the Clebsch-Gordan decomposition problem for a given product $\Delta' \otimes \Delta''$ we say that such representation is {\em completely reducible}. Remark that only finite dimensional representations are always completely reducible, for infinite dimensional representations this is not generally true \cite{Wig59}.

\section{The group $SU(2)$ in Hubbard representation}
\label{titsu2}

Let us consider the group of unimodular (i.e., with determinant $+1$) unitary $2\times 2$ matrices $SU(2)$. The general form of one of these matrices is 
\be
A = \left(
\begin{array}{rr}
a & b\\
-\overline b & \overline a
\end{array}
\right), \quad \vert a \vert^2 + \vert b \vert^2 =1.
\label{su2-1}
\ee
In terms of the identity $\mathbb I_2$ and the Pauli matrices $\vec \sigma =(\sigma_x, \sigma_y, \sigma_z)$, with
\be
\sigma_x =\left(
\begin{array}{rr}
0& 1\\
1&0
\end{array}
\right), \quad \sigma_y =\left(
\begin{array}{rr}
0 & -i\\
i & 0
\end{array}
\right), \quad \sigma_z =\left(
\begin{array}{rr}
1 & 0\\
0& -1
\end{array}
\right),
\label{pauli}
\ee
equation (\ref{su2-1}) reads
\[
A= a_0 \mathbb I_2 + i \vec a \cdot \vec \sigma, \quad \vec a = (a_1, a_2, a_3), \quad a=a_0+ia_3, \quad b= a_2+ia_1.
\]
The well known correspondence between the elements of $SU(2)$ and the points of a sphere of radius $\pi$ in the three-dimensional euclidean space makes clear that this Lie group is not only compact and connected, but it is also simply connected \cite{Fon70}. The Lie algebra of $SU(2)$ is usually defined in terms of the Hermitian operators $T_k = \frac12 \sigma_k$, $k=1,2,3$. That is, the basis of the representation $T(SU(2))$ satisfies the angular momentum algebra
\be
[T_k, T_{\ell} ] = i \epsilon_{k \ell m} T_m,
\label{irr1}
\ee
with $\epsilon_{k\ell m}$ the Levi-Civita symbol. The $SU(2)$ group is of rank 1 (there is not a pair of independent elements of the algebra (\ref{irr1}) that commute among themselves), and the analysis of the structure constants $c_{k \ell}^m = i\epsilon_{k\ell m}$ shows that this is actually simple. Therefore, one can confine the study of $SU(2)$ to the construction of the corresponding finite-dimensional irreducible representations. Remark that the set $\{ T_1, T_2, T_3\}$ is indeed an irreducible two-dimensional representation, as this is defined on the vector space $\text{Sp} \{ \vert e^2_1 \rangle, \vert e^2_2 \rangle \}$. We are interested in the more general situation where the algebra (\ref{irr1}) induces complex representation spaces ${\cal H}$ with $\text{Dim} ({\cal H}) =n \geq 2$. 

Let $\Delta$ be an $n$-dimensional irreducible complex representation of $T(SU(2))$ in ${\cal H}$, then the operators $\Delta(T_k)$ are complex $n^2$-square matrices. We define
\be
J_3 = \Delta(T_3), \quad J_{\pm} = [\Delta(T_1) \pm i \Delta(T_2)],
\label{irr2}
\ee
so that
\be
[J_+, J_-] = 2 J_3, \quad [J_3, J_{\pm}] = \pm J_{\pm},
\label{irr3}
\ee
and
\be
J_3^{\dagger} = J_3, \quad J_{\pm}^{\dagger}= J_{\mp}.
\label{irr4}
\ee

\subsection{Irreducible representation of $SU(2)$}
\label{titirr2-sub}

From (\ref{irr3}) one realises that $J_{\pm}$ are raising and lowering operators for the eigenvectors of $J_3$. Indeed, let $\vert \varphi \rangle$ be such that $J_3 \vert \varphi \rangle = \alpha \vert \varphi \rangle$, then
\be
J_3 J_{\pm} \vert \varphi \rangle = (\alpha \pm 1) J_{\pm} \vert \varphi \rangle \quad \Rightarrow \quad J_3 J_{\pm}^r \vert \varphi \rangle = (\alpha \pm r) J_{\pm}^r \vert \varphi \rangle, \quad r \in \mathbb N.
\label{irr5}
\ee
Remark that the non-vanishing vectors $J_{\pm}^r \vert \varphi \rangle$ are linearly independent as they belong to different eigenvalues of $J_3$. Let $\vert \varphi_{ext} \rangle$ be such that $J_3 \vert \varphi_{ext} \rangle = j \vert \varphi_{ext} \rangle$ and $J_+ \vert \varphi_{ext} \rangle =0$. The eigenvalue $j$ corresponds to the highest weight of the representation since the extremal state $\vert \varphi_{ext} \rangle$ is annihilated by $J_+$ (otherwise the representation would be not finite dimensional). In the same context we should have $J_-^m \vert \varphi_{ext} \rangle =0$ for some positive integer $m$. Let $s+1$ be the smallest value of $m$ for which this is true, then one can verify that necessarily $s=2j$. In this form, according to $s$, the highest weight $j$ is either a half integer or an integer. The representation $\Delta$ is therefore $2j+1$-dimensional and this is characterised by the highest weight $j$. For the related representation space we have
\be
{\cal H} =\text{Sp} \{ \vert \varphi_{ext} \rangle, J_- \vert \varphi_{ext} \rangle, \ldots, J_-^{2j-1} \vert \varphi_{ext} \rangle, J_-^{2j} \vert \varphi_{ext} \rangle \}, \quad \text{Dim}({\cal H}) = n= 2j+1.
\label{repspace1}
\ee
Now let $\vert \varphi_r \rangle := J_-^r \vert \varphi_{ext} \rangle$, $r=0,1, \ldots, 2j$. Using (\ref{irr3}) and (\ref{irr4}) one gets 
\[
J_+ \vert \varphi_r \rangle = (2J_3 + J_-J_+) J_-^{r-1} \vert \varphi_{ext} \rangle = 2(j+1-r) \vert \varphi_{r-1} \rangle + J_-J_+ \vert \varphi_{r-1} \rangle.
\]
After $r$ iterations we obtain
\[
J_+ \vert \varphi_r \rangle = 2\left[ r(j+1-r) + \sum_{\ell =1}^{r-1} \ell \right] \vert \varphi_{r-1} \rangle,
\]
so that
\be
J_+ \vert \varphi_r \rangle = r (2j +1 -r) \vert \varphi_{r-1} \rangle.
\label{irr6}
\ee
Then, applying (\ref{irr4}) and after $r-1$ iterations, from (\ref{irr6}) we obtain the normalisation constant
\be
\langle \varphi_r \vert \varphi_r \rangle = \langle \varphi_{r-1} \vert J_+ \vert \varphi_r \rangle = r (2j +1 -r) \langle \varphi_{r-1} \vert \varphi_{r-1} \rangle = \frac{r! (2j)!}{(2j-r)!} \langle \varphi_{ext} \vert \varphi_{ext} \rangle = C_r^2.
\label{irrck}
\ee
Hereafter we shall assume $\langle \varphi_{ext} \vert \varphi_{ext} \rangle =1$. 

The normalised states $C_{r=j-m}^{-1} \vert \varphi_{r=j-m} \rangle$, $m=-j, \ldots, j$, integrate an orthonormal set of eigenvectors belonging to $J_3$. For these, it is customary to use the following notation
\be
\vert j, m \rangle := \frac{1}{C_{j-m}} \vert \varphi_{j-m} \rangle = \sqrt{\frac{(j+m)!}{(2j)! (j-m)!}} \, \vert \varphi_{j-m} \rangle, \qquad m=-j, \ldots, j,
\label{irrbas}
\ee
where the first entrance of $\vert j, m\rangle$ refers to the highest weight $j$ and the second one to the eigenvalue $m$ of $J_3$ that labels the specific member of the basis we are dealing with. The representation space (\ref{repspace1}) is then rewritten as
\be
{\cal H} = \text{Sp} \{ \vert j,j, \rangle, \vert j, j-1 \rangle, \ldots, \vert j, -j+1 \rangle, \vert j,-j \rangle \} = \text{Sp} \{ \vert j, m \rangle \}_{m=j}^{-j}.
\label{repspace2}
\ee
The basis operators (\ref{irr2}) act on this last vector space as follows
\be
\begin{array}{l}
J_3 \vert j, m \rangle = m \vert j, m \rangle, \\[1.5ex]
J_+ \vert j, m \rangle =  \sqrt{(j-m)(j+m+1)} \vert j, m +1 \rangle, \\[1.5ex]
J_- \vert j, m \rangle =   \sqrt{(j+m)(j-m+1)} \vert j, m-1 \rangle.
\end{array}
\label{irraction}
\ee

\subsection{Hubbard operators}
\label{tithub-sub}

In order to express the set $\{J_3, J_{\pm}\}$ in Hubbard notation let us introduce the change 
\be
m \leftrightarrow m_k =j+1-k, \quad \text{with} \quad  k=1, \ldots, 2j+1.
\label{irrm}
\ee
Then $\vert j, m \rangle \leftrightarrow \vert j, m_k \rangle \equiv \vert j, j+1-k \rangle$, and we can define
\be
 X_n^{p,q}:= \vert j, m_p \rangle \langle j, m_q \vert \equiv \vert j, j+1-p \rangle \langle j, j+1 -q \vert, \quad n=2j+1.
\label{irrhub}
\ee
In this notation the diagonal matrix $J_3$ reads in simple form
\be
J_3 =  \sum_{k=1}^n m_k \, X_n^{k,k}, \qquad n=2j+1.
\label{jz}
\ee
Using the linearity of the Hubbard operators and  property (\ref{hub3}), it is straightforward to verify the first of equations (\ref{irraction}):
\[
J_3 \vert j, m \rangle = \sum_{k=1}^n m_k \left( X_n^{k,k} \vert j, m_s \rangle \right) = \sum_{k=1}^n m_k \delta_{k,s} \vert j, m_k\rangle = m_s \vert j, m_s \rangle = m \vert j, m \rangle,
\]
where we have used (\ref{irrm}). In a similar form we have the irreducible representation of the raising and lowering operators
\be
J_+ = \sum_{k=1}^{n-1} \sqrt{k(2j+1-k)} \, X_n^{k, k+1}, \quad J_- = \sum_{k=1}^{n-1} \sqrt{k(2j+1-k)} \, X_n^{k+1, k}.
\label{jpm}
\ee
For completeness, let us express equations (\ref{irraction}) in the ``$k$''-representation defined in (\ref{irrm})-(\ref{irrhub}). We have
\be
\begin{array}{l}
J_3 \vert j, m_k \rangle = m_k \vert j, m_k \rangle, \\[1.5ex]
J_+ \vert j, m_k \rangle =  \sqrt{(k-1)(2j+2-k)} \vert j, m_{k-1} \rangle, \\[1.5ex]
J_- \vert j, m_k \rangle =   \sqrt{k(2j+1-k)} \vert j, m_{k+1} \rangle.
\end{array}
\label{irraction2}
\ee
Since the Hubbard operators $X_n^{i,j}$ are real matrices, using (\ref{hub4}) in (\ref{jz}) and (\ref{jpm}), it is now easy to verify the relationships (\ref{irr4}). For simplicity, it is also convenient to introduce the notation
\be
c_k^2 = k(2j+1-k).
\label{ck}
\ee
Remark that
\[
\prod_{i=1}^k c_i^2= C_k^2,
\] 
with $C_k^2$ defined in (\ref{irrck}). Hence, the expressions (\ref{jpm}) can be rewritten as
\be
J_+ = \sum_{k=1}^{n-1} c_k \, X_n^{k, k+1} \quad \text{and} \quad J_- = \sum_{k=1}^{n-1} c_k \, X_n^{k+1, k}.
\label{jpm2}
\ee
The first three lowest dimensional irreducible representations of $SU(2)$ are reported below. In all cases there is agreement with the matrix representation obtained in conventional form reported in e.g. \cite{Pei03}.

\subsubsection{Highest weight $j=1/2$.}

The lowest dimension of the representation space ${\cal H}$ is obtained for the weight $j=1/2$. Thus ${\cal H}_{1/2} = \text{Sp} \{ \vert \frac12, \frac12 \rangle, \vert \frac12, -\frac12 \rangle \}$, with $\text{Dim}({\cal H}_{1/2}) =2$ and
\be
\begin{array}{c}
J_z^{(1/2)} = \frac12 X_2^{1,1} - \frac12 X_2^{2,2}= \left(
\begin{array}{cc}
\frac12  & 0\\
0 & -\tfrac12 
\end{array}
\right),\\[4ex]
J_+^{(1/2)} =X_2^{1,2} =  \left(
\begin{array}{cc}
0 & 1\\
0 & 0
\end{array}
\right), \qquad J_-^{(1/2)} = X_2^{2,1} = \left(
\begin{array}{cc}
0 & 0\\
1 & 0
\end{array}
\right).
\end{array}
\label{pauref}
\ee

\subsubsection{Highest weight $j=1$.}

For $j=1$ we have ${\cal H}_1 = \text{Sp} \{ \vert 1,1 \rangle, \vert 1,0 \rangle, \vert 1,-1 \rangle\}$ and $\text{Dim}({\cal H}_1) = 3$, with
\be
\begin{array}{c}
J_z^{(1)} = X_3^{1,1} -X_3^{3,3} =\left(
\begin{array}{rrr}
1 & 0 & 0\\
0 & 0 & 0\\
0 & 0 & -1
\end{array}
\right),\\[5ex]
J_+^{(1)} = \sqrt 2  \, X_3^{1,2} +  \sqrt 2 \, X_3^{2,3} = \left(
\begin{array}{ccc}
0 & \sqrt 2 & 0\\
0 & 0 & \sqrt 2\\
0 & 0 & 0
\end{array}
\right) = \left( J_-^{(1)}\right)^{\dagger}.
\end{array}
\ee

\subsubsection{Highest weight $j=3/2$.}

For $j=3/2$ we have ${\cal H}_{3/2} = \text{Sp} \{ \vert \frac{3}{2}, \frac{3}{2} \rangle, \vert \frac32, \frac12 \rangle, \vert \frac32, -\frac12 \rangle, \vert \frac32, -\frac32 \rangle\}$, with $\text{Dim}({\cal H}_{3/2}) = 4$ and
\be
\begin{array}{c}
J_z^{(3/2)} =\frac32 X_4^{1,1} + \frac12 X_4^{2,2} - \frac12 X_4^{3,3} - \frac32 X_4^{4,4} = \left( 
\begin{array}{rrrr}
\tfrac32 & 0 & 0 & 0\\
0 & \tfrac12 & 0 & 0\\
0 & 0 & -\tfrac12 & 0\\
0 & 0 & 0 & -\tfrac32
\end{array}
\right),\\[6ex]
J_+^{(3/2)} = \sqrt 3  X_4^{1,2} + 2 X_4^{2,3} + \sqrt 3 X_4^{3,4} = \left(
\begin{array}{cccc}
0 & \sqrt 3 & 0 & 0\\
0 & 0 & 2 & 0\\
0 & 0 & 0 & \sqrt 3\\
0 & 0 & 0 & 0
\end{array}
\right) = \left( J_-^{(3/2)} \right)^{\dagger}.
\end{array}
\ee

\section{$SU(2) \times SU(2)$ in Hubbard notation}
\label{titsu2-2}

In this section we solve the Clebsch-Gordan problem associated to the Kronecker product of two irreducible representations of $SU(2)$. That is, we are going to find the invariant subspaces that integrate the complete representation space of $SU(2) \times SU(2)$ as a direct sum (see Section~\ref{titbas}).

\subsection{The product of irreducible representations}
\label{titpro-sub}

Let $\Delta_1$ and $\Delta_2$ be two irreducible representations of $T(SU(2))$ with $j_1$ and $j_2$ the corresponding highest weights. Denote
\[
{\cal H}_{j_s} = \text{Sp} \{ \vert j_s, m^{(j_s)} \rangle \, \vert \,  m^{(j_s)} = -j_s, \ldots j_s\}, \quad s=1,2,
\]
the related representation spaces. Therefore 
\be
{\cal H}_{(j_1, j_2)} = \text{Sp} \{\vert j_1, m^{(j_1)}  \rangle \otimes \vert j_2, m^{(j_2)}  \rangle \, \vert \, m^{(j_s)} =-j_s, \ldots, j_s, \, s=1,2 \}
\label{base2su2}
\ee
is the $n_1 n_2$-dimensional representation space of $\Delta^{(j_1, j_2)} = \Delta_1 \otimes \mathbb I_{n_2} + \mathbb I_{n_1} \otimes \Delta_2$, with $n_1= 2j_1+1$ and $n_2 =2 j_2+1$. Remark that  $\Delta^{(j_1, j_2)}$ is not necessarily an irreducible representation. As the basis of operators we shall use in each space
\be
J_3^{(j_s)} = \Delta_s (T_3), \quad J_{\pm}^{(j_s)} = [\Delta_s (T_1) \pm i \Delta_s (T_2)], \quad s=1,2.
\ee
Following (\ref{irrm}-\ref{jpm2}), in Hubbard notation we write
\be
J_3^{(j_s)} = \displaystyle\sum_{k=1}^{n_s} m_k^{(j_s)} X_{n_s}^{k,k}, \qquad  J_+^{(j_s)} = \displaystyle\sum_{k=1}^{n_s-1} c_k^{(j_s)} X_{n_s}^{k, k+1}, \qquad J_-^{(j_s)} = \displaystyle\sum_{k=1}^{n_s-1} c_k^{(j_s)} X_{n_s}^{k+1, k},
\label{j2su}
\ee
with 
\be
m_k^{(j_s)} = (j_s +1-k), \quad c_k^{(j_s)} = \sqrt{ k(2j_s + 1 -k)}, \quad n_s=2j_s+1, \quad s=1,2.
\label{j2sub}
\ee
We now promote the latter operators to act on the entire vector space (\ref{base2su2}). Thus, the operators
\be
J_3^{(j_1, j_2)} = J_3^{(j_1)} \otimes \mathbb I_{n_2} + \mathbb I_{n_1} \otimes J_3^{(j_2)}, \quad J_{\pm}^{(j_1, j_2)} = J_{\pm}^{(j_1)} \otimes \mathbb I_{n_2} + \mathbb I_{n_1} \otimes J_{\pm}^{(j_2)}
\label{op2su}
\ee
correspond to the basis of $SU(2) \times SU(2)$ in the representation defined by the vector space ${\cal H}_{(j_1, j_2)}$. It is straightforward to verify the commutation rules
\be
[J_+^{(j_1, j_2)}, J_-^{(j_1, j_2)}] = 2 J_3^{(j_1, j_2)}, \quad [J_3^{(j_1, j_2)}, J_{\pm}^{(j_1, j_2)}] = \pm J_{\pm}^{(j_1, j_2)},
\ee
and the relationships
\be
\left( J_3^{(j_1, j_2)} \right)^{\dagger} = J_3^{(j_1, j_2)}, \quad \left( J_{\pm}^{(j_1, j_2)} \right)^{\dagger}= J_{\mp}^{(j_1, j_2)}.
\ee
In Hubbard notation the matrix representation of the operators (\ref{op2su}) read as
\be
J_3^{(j_1, j_2)}= \sum_{k=1}^{n_1} \sum_{\ell =1}^{n_2} \left( m^{(j_1)}_k + m^{(j_2)}_{\ell} \right)  X_{n_1 n_2}^{n_2(k-1) + \ell, \, n_2(k-1) + \ell}, 
\label{op2sub}
\ee
and
\be
J_+^{(j_1, j_2)}=  \sum_{k=1}^{n_1 -1} \,  \sum_{\ell =1}^{n_2 } c_k^{(j_1)} X_{n_1 n_2}^{n_2 (k-1) + \ell, \, n_2 k+ \ell} + \sum_{r=1}^{n_1} \,  \sum_{t =1}^{n_2 -1}  c_t^{(j_2)} X_{n_1 n_2}^{n_2 (r-1) +t, \, n_2 (r -1) + t +1}.
\label{op2suc}
\ee
Using Theorem~M1 one can simplify the above expressions to get
\be
J_3^{(j_1, j_2)} = \sum_{p=1}^{n_1 n_2} \left[ m_{p'}^{(j_1)} + m_{p+n_2 -n_2p'}^{(j2)}\right] X_{n_1 n_2}^{p,p}
\label{simple1}
\ee
and
\be
J_+^{(j_1, j_2)} = \sum_{p=1}^{n_2(n_1 -1)} c_{p'}^{(j_1)} X_{n_1n_2}^{p, \,p+n_2} + \sum_{p=1}^{n_1 n_2 -1} c_{p+ n_2 -n_2 p'}^{(j_2)} X_{n_1 n_2}^{p, \, p+1},
\label{simple2}
\ee
where $p'= \lceil \frac{p}{n_2} \rceil$. 

As an example let us show the explicit form of some of the matrices (\ref{simple2}). In all the following examples we use $c_1^{(\tfrac12)} =1$ and $c_1^{(1)} = c_2^{(1)} = \sqrt 2$. Our results can be compared with those obtained in conventional form reported in e.g. \cite{Pei03}.

\subsubsection{Weights $j_1=\tfrac12$ and $j_2 = \tfrac12$}
\label{iguales}

\be
J_+^{(\tfrac12, \tfrac12)} = c_1^{(\tfrac12)} \left(X_4^{1,3} + X_4^{2,4} + X_4^{1,2} + X_4^{3,4}
\right)=
\left(
\begin{array}{cc|cc}
0 &1 & 1 & 0\\
0 & 0 & 0 & 1\\
\hline
0 & 0 & 0 & 1\\
0 & 0 & 0 & 0
\end{array}
\right).
\ee

\subsubsection{Weights $j_1= \tfrac12$ and $j_2=1$}

\be
\begin{array}{rl}
J_+^{(\tfrac12 ,1)} = & c_1^{(\tfrac12)} \left( X_6^{1,4} +X_6^{2,5} + X_6^{3,6} \right) + c_1^{(1)} \left( X_6^{1,2} + X_6^{4,5} \right) + c_2^{(1)} \left( X_6^{2,3} + X_6^{5,6} \right)\\[2ex]
= &  \left(
\begin{array}{ccc|ccc}
0 & \sqrt 2 & 0 & 1 & 0 & 0\\
0 & 0 & \sqrt 2 & 0 & 1 & 0\\
0 & 0& 0 & 0 & 0 & 1\\
\hline
0 & 0 & 0 & 0 & \sqrt 2 & 0\\
0 & 0 & 0 & 0 & 0 & \sqrt 2\\
0 & 0 & 0 & 0 & 0 & 0
\end{array}
\right).
\end{array}
\ee

\subsubsection{Weights $j_1=1$, $j_2=\tfrac12$}
\label{diferentes}

\be
\begin{array}{rl}
J_+^{(1,\tfrac12)} = &  c_1^{(1)} \left( X_6^{1,3} + X_6^{2,4} \right) + c_2^{(1)} \left( X_6^{3,5} + X_6^{4,6} \right) + c_1^{(\tfrac12)} \left( X_6^{1,2} + X_6^{3,4} \right)
\\[2ex]
= &  \left(
\begin{array}{ccc|ccc}
0 & 1& \sqrt 2 & 0 & 0 & 0\\
0 & 0 &  0 & \sqrt 2 & 0 & 0\\
0 & 0 & 0 & 1 & \sqrt 2 & 0\\
\hline
0 & 0 & 0 & 0 & 0& \sqrt 2\\
0 & 0 & 0 & 0 & 0 & 1\\
0 & 0 & 0 & 0 & 0 & 0
\end{array}
\right).
\end{array}
\ee

\subsection{Irreducible representation}
\label{titirr-sub}

We want to decompose the $n_1 n_2$-dimensional space ${\cal H}_{(j_1, j_2)}$ into a direct sum of irreducible subspaces. Using the additivity of eigenvalues in the Kronecker product (see Corollary~TM2.2), we realise that $j_1 + j_2$ is the highest weight of $J_3^{(j_1, j_2)}$. Given $j_1 + j_2 -k+1$, there are $k$ different pairs of weights $m^{(j_1)}$ and $m^{(j_2)}$ such that $m^{(j_1)} +m^{(j_2)} = j_1 + j_2 -k+1$. For $k=1$ there is only one way to get $m^{(j_1)} + m^{(j_2)} =j_1+j_2$, therefore we have a single state $\vert j_1, j_1 \rangle \otimes \vert j_2, j_2 \rangle$ in ${\cal H}_{(j_1, j_2)}$ belonging to the highest weight $j=j_1+j_2$. According to the discussion of Section~\ref{titirr2-sub}, $\vert j_1, j_1 \rangle \otimes \vert j_2, j_2 \rangle$ is the extremal state of a vector space $ {\cal H}_{j_1 + j_2} \subset {\cal H}_{(j_1, j_2)}$ in which the matrix basis $\{ T_1, T_2, T_3 \}$ is irreducible. Denote $\Delta_{j_1+j_2}$ such an irreducible representation and write $\{ J_3^{(j_1+j_2)}, J_{\pm}^{(j_1+j_2)} \}$ for the basis operators. Then
\[
{\cal H}_{j_1 + j_2} = \text{Sp} \{ \vert j,j \rangle, J_-^{(j_1+j_2)} \vert j, j \rangle,  J_-^{(j_1+j_2)+1} \vert j, j \rangle, \ldots, J_-^{2(j_1+j_2)} \vert j,j \rangle \}
\]
with $\vert j,j \rangle \equiv \vert j_1, j_1 \rangle \otimes \vert j_2, j_2 \rangle$. In this form, $\text{Dim}({\cal H}_{j_1+ j_2}) = 2(j_1 + j_2) + 1$ and the vector space (\ref{base2su2}) is rewritten as the direct sum ${\cal H}_{(j_1, j_2)} = {\cal H}_{j_1+j_2} \oplus {\cal H}_R$, with ${\cal H}_R \subset {\cal H}_{(j_1, j_2)}$ such that ${\cal H}_{j_1 + j_2} \cap {\cal H}_R =\mbox{Sp} \{\vert \emptyset \rangle \}$.

For $k=2$ there are two different forms of solving $m^{(j_1)} + m^{(j_2)} = j_1 + j_2 -1$. Thus, the pair of states $\vert j_1, j_1 \rangle \otimes \vert j_2, j_2 -1 \rangle$ and  $\vert j_1, j_1 -1 \rangle \otimes \vert j_2, j_2 \rangle$ belong to the same weight $j_1 + j_2 -1$ in ${\cal H}_{(j_1, j_2)}$. One of them has been already included in ${\cal H}_{j_1+j_2}$, so that this does not belong to ${\cal H}_R$. The remaining vector is the extremal state of an irreducible representation $\Delta_{j_1 + j_2 -1}$ with $\vert j, j-1 \rangle = \vert j_1, j_1 \rangle \otimes \vert j_2, j_2-1 \rangle$ for $j_1 \geq j_2$, or $\vert j, j-1 \rangle = \vert j_1, j_1-1 \rangle \otimes \vert j_2, j_2 \rangle$ for $j_2 \geq j_1$, and
\[
{\cal H}_{j_1 + j_2 -1} = \text{Sp} \{ \vert j, j-1 \rangle, J_-^{(j_1 + j_2 -1)}  \vert j, j-1 \rangle, \ldots, J_-^{2(j_1 + j_2 -1)}  \vert j, j-1 \rangle \}.
\]
Hence, the vector space (\ref{base2su2}) is rewritten as ${\cal H}_{(j_1, j_2)} = {\cal H}_{j_1+j_2} \oplus {\cal H}_{j_1+j_2 -1} \oplus {\cal H}_{R'}$, with ${\cal H}_{R'} \cap {\cal H}_{j_1 + j_2} \oplus {\cal H}_{j_1 + j_2 -1} = \mbox{Sp} \{ \vert \emptyset \rangle \}$. The procedure can be repeated at will by noticing that each irreducible representation $\Delta_j$ is $2j+1$-dimensional with $j = j_1+ j_2, j_1 + j_2 -1, \ldots, \vert j_1 - j_2 \vert -1, \vert j_1 -j_2 \vert$. One then arrives at the expression
\[
{\cal H}_{(j_1, j_2)} =  {\cal H}_{j_1+ j_2} \oplus {\cal H}_{j_1+ j_2-1}  \oplus \cdots \oplus {\cal H}_{\vert j_1 - j_2 \vert -1} \oplus  {\cal H}_{\vert j_1 -j_2 \vert} \equiv \widetilde{{\cal H}}_{(j_1, j_2)}
\]
The $k$-th term in the above direct sum has the dimension 
\be
d_k =2(j_1+j_2 +1-k) +1= 2(j_1 +j_2)+ 3-2k= n_1 +n_2 +1-2k.
\label{dimk}
\ee 
The following property shows that the dimension $d_k$ of the representation space ${\cal H}_k$ is reduced in two units as the value of $j=j_1 + j_2 +1-k$ increases in one unit. This will be useful in the sequel 
\be
d_k = d_{k-1} -2, \quad k\geq 2.
\label{dimk2}
\ee
Then, as expected, for the dimension of the entire space we have
\[
\text{Dim}(\widetilde{{\cal H}}_{(j_1, j_2)})=\left\{
\begin{array}{ll}
\displaystyle\sum_{k=1}^{2j_2+1} d_k  = (2j_2+1)(2j_1+1)=n_2n_1, & j_1 \geq j_2\\[4ex]
\displaystyle\sum_{k=1}^{2j_1+1} d_k  = (2j_1+1)(2j_2+1)=n_1n_1, & j_2 \geq j_1
\end{array}
\right.
\]
Let $j_0 =\text{min} \{ j_1, j_2\}$, then $n_0 = 2j_0 +1 = \text{min} \{n_1, n_2\}$, and
 \be
\widetilde{{\cal H}}_{(j_1, j_2)} =   \underbrace{{\cal H}_{j_1+ j_2} \oplus {\cal H}_{j_1+ j_2-1}  \oplus \cdots \oplus {\cal H}_{\vert j_1 - j_2 \vert -1} \oplus  {\cal H}_{\vert j_1 -j_2 \vert}}_{n_0 \, \text{terms}}.
\label{spaces}
\ee
We now look for the operators $\Delta_j$ leaving invariant the subspaces ${\cal H}_j$ in (\ref{spaces}), with $j_1+j_2 \geq j \geq \vert j_1 - j_2 \vert$. That is, we want to construct the solution to the Clebsch-Gordan problem $\Delta^{(j_1, j_2)} \rightarrow \widetilde{\Delta}^{(j_1, j_2)}$, with $\widetilde{\Delta}^{(j_1, j_2)}$ the direct sum operator
\be
\widetilde{\Delta}^{(j_1, j_2)} = \Delta_{j_1+j_2} \oplus \Delta_{j_1 + j_2 -1} \oplus \cdots \oplus \underbrace{\Delta_{j_1 + j_2 +1 -k}}_{\text{$k$-th term}} \oplus \cdots \oplus \Delta_{\vert j_1 -j_2 \vert -1} \oplus \Delta_{\vert j_1 - j_2\vert}.
\label{cg1}
\ee
The first $k$ terms in the direct sum (\ref{cg1}) integrate a square matrix of order
\be
z_k = \sum_{\ell =1}^{k}d_{\ell} = k(d_k + k-1), \quad k= 1,2, \ldots, 2j_2+1.
\label{z}
\ee
Therefore, $\Delta_{j_1+j_2+1-k}$ in (\ref{cg1}) is a square matrix of order $d_k$, the first element of which is at the $(z_{k-1}+1, z_{k-1}+1)$ entrance of $\widetilde \Delta^{(j_1, j_2)}$ with 
\be
z_{k-1}= (k-1)(d_k + k), \quad z_0:=0.
\label{z2}
\ee
In this form, the basis elements $\widetilde J_{\alpha}^{(j_1, j_2)}, \alpha =3, \pm$, should read
\be
\widetilde J_{\alpha}^{(j_1, j_2)} = J_{\alpha}^{(j_1 + j_2)} \oplus \cdots \oplus \underbrace{J_{\alpha}^{(j_1 + j_2 + 1-k)}}_{\text{$k$-th term}} \oplus \cdots \oplus J_{\alpha}^{(\vert j_1 - j_2 \vert)}, \quad J_{\alpha}^{(0)} :=0.
\label{j3d}
\ee
For $\alpha = 3$, the Hubbard representation of the $k$-th term in (\ref{j3d}) is given by
\[
J_3^{(j_1 +j_2 +1 -k)} = \sum_{p=1}^{d_k} \left( m_k^{(j_1)} + m_p^{(j_2)} \right) X_{n_1 n_2}^{z_{k-1} +p, \, z_{k-1} +p}, \quad z_0 =0,
\]
where we have used (\ref{j2su}). Hence, the entire operator reads
\be
\widetilde J_3^{(j_1, j_2)}= \sum_{k=1}^{n_0} \, \sum_{p=1}^{d_k} \left( m_k^{(j_1)} + m_p^{(j_2)} \right) X_{n_1 n_2}^{z_{k-1} +p, \, z_{k-1} +p}, \quad n_0 = \text{min} \{n_1, n_2\}.
\label{j3comp}
\ee
In a similar form we get
\be
\widetilde J_+^{(j_1, j_2)} = \sum_{k=1}^{n_0} \, \sum_{p=1}^{d_k -1} c_{k,p}^{(j_1, j_2)} \, X_{n_1 n_2}^{z_{k-1} + p, \, z_{k-1} + p +1}, \quad c_{k,p}^{(j_1, j_2)} =\sqrt{p [2(j_1 + j_2 -k) + 3 -p]}.
\label{j+comp}
\ee
Let us give a pair of examples. The following results can be compared with those we have reported in Sections~\ref{iguales} and \ref{diferentes}.

\subsubsection{Weights $j_1=j_2=1/2$}

\[
\widetilde J_3^{(\tfrac12, \tfrac12)} = X_4^{1,1} - X_4^{3,3} =\left( 
\begin{array}{ccc|c}
1 & 0 & 0 & 0\\
0 & 0 & 0 & 0\\
0 & 0 & -1 & 0\\
\hline
0 & 0 & 0 & 0
\end{array}
\right) = \left(
\begin{array}{c|c}
J_3^{(1)} & 0_{3\times 1}\\
\hline
0_{1\times 3} & 0
\end{array}
\right) = J_3^{(\tfrac12 + \tfrac12)} \oplus J_3^{(\left\vert \tfrac12 - \tfrac12 \right\vert)},
\]

\[
\widetilde J_+^{(\tfrac12, \tfrac12)} = \displaystyle\sum_{p=1}^2 c_{1,p}^{(\tfrac12, \tfrac12)}\, X_4^{p, p+1} =\left(
\begin{array}{ccc|c}
0 & \sqrt 2 \!\! & 0 & 0\\
0 & 0 & \sqrt 2 \!\! & 0\\
0 & 0 & 0 & 0\\
\hline
0 & 0 & 0 & 0
\end{array}
\right) \equiv \left(
\begin{array}{c|c}
J_+^{(1)} & 0_{3\times 1}\\
\hline
0_{1\times 3} & 0
\end{array}
\right) = J_+^{(\tfrac12+\tfrac12)} \oplus J_+^{(\left\vert \tfrac12 - \tfrac12 \right\vert)}.
\]

\subsubsection{Weights $j_1 =1$ and $j_2=\tfrac12$}

\[
\begin{array}{rl}
\widetilde J_3^{(1,\tfrac12)} & = \displaystyle\sum_{p=1}^4 \left( \tfrac52 -p \right) X_6^{p,p} + \sum_{p=1}^2 \left(\tfrac32 -p \right) X_6^{4+p, 4+p} =\left(
\begin{array}{cccc|cc}
\tfrac32 & 0 & 0 & 0 & 0 & 0\\
0 & \tfrac12 &0 & 0 & 0 & 0\\
0 & 0 & -\tfrac12 & 0 & 0 & 0\\
0 & 0 & 0 & -\tfrac32 & 0 & 0\\
\hline
0 & 0 & 0 & 0 & \tfrac12 & 0\\
0 & 0 & 0 & 0 & 0 & -\tfrac12
\end{array}
\right)\\[3ex]
&  = \left(
\begin{array}{c|c}
J_3^{(\tfrac32)} & 0_{4\times 2}\\
\hline
0_{2\times 4} & J_3^{(\tfrac12)}
\end{array}
\right) = J_3^{(1 +\tfrac12)} \oplus J_3^{(\left\vert 1-\tfrac12 \right\vert )} = J_3^{(\tfrac12 + 1)} \oplus J_3^{(\left\vert \tfrac12 -1\right\vert)} =\widetilde J_3^{(\tfrac12,1)},
\end{array}
\]

\[
\begin{array}{rl}
\widetilde J_+^{(1, \tfrac12)} & = \displaystyle\sum_{p=1}^3 c_{1,p}^{(1,\tfrac12)} X_6^{p, p+1} + c_{2,1}^{(1,\tfrac12)} X_6^{5,6} = \left(
\begin{array}{cccc|cc}
0 & \sqrt 3 & 0 & 0 & 0 & 0\\
0 & 0 & 2 & 0 & 0 & 0\\
0 & 0 & 0 & \sqrt 3 & 0 & 0\\
0 & 0 & 0 & 0 & 0 & 0\\
\hline
0 & 0 & 0 & 0 & 0 & 1\\
0 & 0 & 0 & 0 & 0 & 0
\end{array}
\right)\\[3ex]
& = \left( 
\begin{array}{c|c}
J_+^{(\tfrac32)} & 0_{3\times 2}\\
\hline
0_{2\times 3} & J_+^{(\tfrac12)}
\end{array}
\right) = J_+^{(1+ \tfrac12)} \oplus J_+^{(\left\vert 1-\tfrac12 \right\vert)}  = J_+^{(\tfrac12 +1)} \oplus J_+^{(\left\vert \tfrac12 -1 \right\vert)} = \widetilde J_+^{(\tfrac12,1)}.
\end{array}
\]

\section{Clebsch-Gordan decomposition of $SU(2) \times SU(2)$}
\label{titcle2}

Let us consider the transformation matrix
\be
S= \sum_{k,q=1}^{n_1 n_2} S_{k,q} X_{n_1 n_2}^{k,q}
\label{s1}
\ee
From (\ref{hub7}) and (\ref{simple1}) we obtain
\be
J_3^{(j_1, j_2)} S = \sum_{p,q=1}^{n_1 n_2} \left( m_{p'}^{(j_1)} + m_{p+n_2-n_2 p'}^{(j_2)} S_{p,q}\right)  X_{n_1 n_2}^{p,q}, \quad p'= \lceil \tfrac{p}{n_2} \rceil.
\label{s2}
\ee
In similar form, equation~(\ref{j3comp}) leads to
\[
S \widetilde J_3^{(j_1, j_2)} = \sum_{t=1}^{n_1 n_2} \sum_{k=1}^{n_0} \sum_{r=1}^{d_k} S_{t, z_{k-1} +r} \left( m_k^{(j_1)} + m_r^{(j_2)} \right) X_{n_1 n_2}^{t, \, z_{k-1}+r}.
\]
The $(p,q)$-entrance of this last matrix is then given by
\be
\left[ S \widetilde J_3^{(j_1, j_2)} \right]_{p,q} = \sum_{k=1}^{n_0} \sum_{r=1}^{d_k} S_{p, z_{k-1} +r} \left( m_k^{(j_1)} + m_r^{(j_2)} \right) \delta_{z_{k-1}+r, q},
\label{s3}
\ee
where we have used (\ref{hub7c}). Given $q$, the double sum in (\ref{s3}) is reduced to a single term for  the labels $k_0 \in \{1,2,\ldots, n_0\}$ and $r_0  \in \{1,2, \ldots, d_{k_0} \}$ such that 
\be
z_{k_0-1}+r_0= q.
\label{s3a}
\ee
Then, the equality  $J_3^{(j_1, j_2)} S = S \widetilde J_3^{(j_1, j_2)}$ holds whenever that
\be
\left( m_{p'}^{(j_1)} + m_{p+n_2-n_2 p'}^{(j_2)} - m_{k_0}^{(j_1)} - m_{r_0}^{(j_2)} \right) S_{p,q}=0.
\label{s4}
\ee
On the other hand, the matrix elements of $J_+^{(j_1, j_2)} S$ are obtained from (\ref{simple2}) to read
\be
\left[ J_+^{(j_1, j_2)} S\right]_{p,\ell}= c^{(j_1)}_{p'}S_{p+n_2, \ell} + c^{(j_2)}_{p+ n_2 -n_2p'} S_{p+1, \ell}, \quad p'= \lceil \tfrac{p}{n_2} \rceil.
\label{s5a}
\ee
Now, departing from (\ref{j+comp}) one arrives at the matrix elements
\[
\left[ S \widetilde J_+^{(j_1, j_2)} \right]_{p, \ell} = \sum_{k=1}^{n_0} \, \sum_{r=1}^{d_k -1} S_{p, z_{k-1}+r} \, c_{k,r}^{(j_1, j_2)}\, \delta_{z_{k-1}+r+1, \ell}.
\]
Given $\ell$, the double sum in the latter equation reduces to a single term for $k_* \in \{1,2,\ldots, n_0\}$ and $r_*  \in \{1,2, \ldots, d_{k_*}-1 \}$ such that $z_{k_*-1}+r_*+1 =\ell$. Let us take $\ell = q +1$, then this last condition is reduced to (\ref{s3a}) with $k_* = k_0$ and $r_* =r_0$. Therefore
\be
\left[ S \widetilde J_+^{(j_1, j_2)} \right]_{p, q+1} = S_{p, q} \, c_{k_0, \, r_0}^{(j_1, j_2)}, \quad r_0 \in \{1,2, \ldots, d_{k_0}-1 \}.
\label{s5b}
\ee
Using this last result, and (\ref{s5a}) with $\ell =q+1$, one realises that $J_+^{(j_1, j_2)} S= S \widetilde J_+^{(j_1, j_2)}$ is fulfilled if 
\be
c^{(j_1)}_{p'} S_{p+n_2, q+1} + c^{(j_2)}_{p+ n_2 -n_2p'} S_{p+1,q+1} = S_{p, q} \, c_{k_0, \, r_0}^{(j_1, j_2)}, \quad r_0 \in \{1, 2, \ldots, d_{k_0}-1 \}.
\label{s5}
\ee
The similar procedure shows that the elements of $J_-^{(j_1, j_2)} S= S \widetilde J_-^{(j_1, j_2)}$ are conditioned to
\be
c^{(j_1)}_{(p-n_2)'} S_{p-n_2, q} + c^{(j_2)}_{p-1+ n_2 -n_2(p-1)'} S_{p-1,q} = S_{p,q+1} \, c_{k_0, r_0}^{(j_1, j_2)}, \quad r_0 \in \{1, 2, \ldots, d_{k_0}-1 \}.
\label{s6}
\ee
Now, we use Lemma~A3 of the appendix and make the change 
\be
p= \alpha n_0 + \beta, \quad \alpha= 0, 1, \ldots, n_1 +n_2 -n_0 -1, \quad \beta=1,2, \ldots, n_0,
\label{p}
\ee
to rewrite equation (\ref{s4}) as
\[
\left( m_{\alpha + 1}^{(j_1)} + m_{\beta}^{(j_2)}- m_{k_0}^{(j_1)} - m_{r_0}^{(j_2)} \right) S_{\alpha n_2 + \beta, \,q}= (k_0 + r_0 -\alpha -\beta -1)S_{\alpha n_2 + \beta, \, z_{k_0 -1}+r_0}= 0,
\]
where we have used (\ref{j2sub}) and (\ref{s3a}). It is convenient to introduce the shortcut notation 
\be
S_{p,q} = S_{\alpha n_0 + \beta, \, z_{k_0 -1} + r_0} = S(\alpha, \beta; k_0, r_0) =S_{\alpha, \beta}^{k_0, r_0},
\label{s}
\ee
so that the latter expression reads in a simpler form
\be
(k_0 + r_0 -\alpha -\beta -1) S_{\alpha, \beta}^{k_0, r_0} =0.
\label{corto1a}
\ee
Remark that given $p$ and $q$ in (\ref{s}), equation (\ref{corto1a}) holds for all the values of $k_0$ and $r_0$ fulfilling (\ref{s3a}), as well as all the values of $\alpha$ and $\beta$ solving (\ref{p}). Thus, for another set of labels $\widetilde k_0, \widetilde r_0$, and $\widetilde \alpha, \widetilde \beta$, fulfilling respectively (\ref{s3a}) and (\ref{p}), we get
\[
(\widetilde k_0 + \widetilde r_0 -\widetilde\alpha - \widetilde\beta -1) S_{\widetilde\alpha, \widetilde\beta}^{\widetilde k_0, \widetilde r_0} =0.
\]
This last property leads to a further simplification in the notation since $k_0$ and $r_0$ can be written without the subindex ``0'', whenever they satisfy equation (\ref{s3a}). In this context, the nontrivial matrix elements $S_{\alpha, \beta}^{k, r}$ are now identified by using (\ref{s}). That is, the roots of the equation
\be
k + r -\alpha -\beta -1=0,
\label{corto1b}
\ee
with $k$, $r$, and $\alpha$, $\beta$, fulfilling (\ref{s3a}) and (\ref{p}) respectively, are the labels of the matrix elements $S_{\alpha, \beta}^{k,r} = S_{p,q}$ that can be different from zero. Using the same notation, eqs. (\ref{s5}) and (\ref{s6}) are rewritten as
\be
c_{\alpha +1}^{(j_1)}\, S_{\alpha +1, \beta}^{k, r +1} + c_{\beta}^{(j_2)}\, S_{\alpha, \beta+1}^{k, r+1} = c_{k,r}^{(j_1, j_2)} \, S_{\alpha, \beta}^{k, r}, \quad 
c_{\alpha}^{(j_1)} \, S_{\alpha -1, \beta}^{k, r } +  c_{\beta-1}^{(j_2)} \, S_{\alpha, \beta-1}^{k, r} = c_{k,r}^{(j_1, j_2)} \, S_{\alpha, \beta}^{k, r+1}
\label{corto2}
\ee
with $r \in \{1, \ldots, d_{k} -1 \}$. To recover the conventional expression for the recurrence relations (\ref{corto2}) we use (\ref{j2sub}) and (\ref{j+comp}), so that the equations to solve are given by
\be
{\footnotesize
\begin{array}{c}
\sqrt{(\alpha +1)(2j_1 - \alpha)} \, S_{\alpha +1, \beta}^{k, r +1} + \sqrt{\beta(2 j_2 - \beta +1)} \, S_{\alpha, \beta+1}^{k, r+1} = \sqrt{r(2j -2k-r+3)} \, S_{\alpha, \beta}^{k, r},
\end{array}
}
\label{corto3a}
\ee
and
\be
{\footnotesize
\begin{array}{c}
\sqrt{\alpha ( 2j_1 -\alpha +1)} \, S_{\alpha -1, \beta}^{k, r } + \sqrt{(\beta -1) (2j_2 - \beta +2)} \, S_{\alpha, \beta-1}^{k, r} = \sqrt{r(2j -2k-r+3)} \, S_{\alpha, \beta}^{k, r+1},
\end{array}
}
\label{corto3b}
\ee
where $j=j_1 + j_2$. 

\subsection{Transformation into the highest dimension invariant subspace.}

\vskip1ex
\noindent
Let us rewrite (\ref{corto3a}) as
\[
\begin{array}{c}
\sqrt{\frac{(\alpha +1)! (2j_1 - \alpha)!}{\alpha! (2j_1 - \alpha -1)!}} \, S_{\alpha +1, \beta}^{k, r +1} + \sqrt{\frac{\beta! (2 j_2 - \beta +1)!}{(\beta -1)! (2j_2-\beta)!}} \, S_{\alpha, \beta+1}^{k, r+1} = \sqrt{\frac{r! (2j -2k-r+3)!}{(r-1)!(2j-2k-r+2)!}} \, S_{\alpha, \beta}^{k, r}
\end{array}
\]
It is convenient to take $\widetilde \alpha = \alpha+1$, $\widetilde \beta = \beta+1$, and $\widetilde r = r+1$ to write
\[
\begin{array}{c}
\sqrt{\frac{\widetilde \alpha! (2j_1 - \widetilde \alpha +1)!}{\alpha! (2j_1 - \alpha -1)!}} \, S_{\widetilde \alpha, \beta}^{k, \widetilde r} + \sqrt{\frac{(\widetilde \beta -1)! (2 j_2 - \widetilde \beta +2)!}{(\beta -1)! (2j_2-\beta)!}} \, S_{\alpha, \widetilde \beta}^{k, \widetilde r} = \sqrt{\frac{(\widetilde r -1)! (2j -2k -\widetilde r+4)!}{(r-1)!(2j-2k-r+2)!}} \, S_{\alpha, \beta}^{k, r}
\end{array}
\]
Now, we multiply this last equation by
\[
\begin{array}{c}
\sqrt{\frac{\alpha! (2j_1 - \alpha -1)! (\beta -1)! (2j_2 - \beta)!}{(\widetilde r -1)! (2j-2k-\widetilde r + 4)!}}
\end{array}
\]
to get
\[
\begin{array}{l}
\sqrt{\frac{2j_1 -\widetilde \alpha +1}{(2j_2 -\beta +1)(2j -2k -\widetilde r +4)}} \,
\sqrt{\frac{\widetilde \alpha! (2j_1 -\widetilde \alpha)! (\beta -1)! (2j_2 -\beta +1)!}{(\widetilde r -1)! (2j -2k -\widetilde r +3)!}}  \, S_{\widetilde \alpha, \beta}^{k, \widetilde r}\\[2ex]
\mbox{\hskip2cm} + \sqrt{\frac{2j_2 -\widetilde \beta +2}{(2j_1 -\alpha)(2j -2k -\widetilde r +4)}} \, \sqrt{\frac{\alpha! (2j_1 - \alpha)! (\widetilde \beta -1)! (2j_2 -\widetilde \beta +1)!}{(\widetilde r -1)! (2j -2k -\widetilde r +3)!}}  \, S_{\alpha, \widetilde \beta}^{k, \widetilde r} \\[2ex]
\mbox{\hskip3cm} = \sqrt{\frac{2j-2k-r+3}{(2j_1 -\alpha)(2j_2 -\beta +1)}} \,
\sqrt{\frac{\alpha! (2j_1 - \alpha)! (\beta -1)! (2j_2 -\beta +1)!}{(r -1)! (2j -2k - r +3)!}}  \, S_{\alpha, \beta}^{k, r}
\end{array}
\]
Thereby, we can take
\be
S_{\alpha, \beta}^{k, r}=\mbox{const.} \times \sqrt{\frac{(r -1)! (2j -2k - r +3)!}{\alpha! (2j_1 - \alpha)! (\beta -1)! (2j_2 -\beta +1)!}}, \qquad k+r=\alpha + \beta +1,
\label{sk1a}
\ee
to arrive at the equation
\[
\begin{array}{c}
\sqrt{\frac{2j_1 -\widetilde \alpha +1}{(2j_2 -\beta +1)(2j -2k -\widetilde r +4)}} +
\sqrt{\frac{2j_2 -\widetilde \beta +2}{(2j_1 -\alpha)(2j -2k -\widetilde r +4)}} =
\sqrt{\frac{2j-2k-r+3}{(2j_1 -\alpha)(2j_2 -\beta +1)}}.
\end{array}
\]
After multiplying by the square-root of $(2j-2k-r+3)(2j_1 -\alpha)(2j_2 -\beta +1)$ one gets
\[
k+r = \alpha + \beta +1 +(k-1).
\]
We realise that condition (\ref{corto1b}) is fulfilled whenever $k=1$. Then the equation (\ref{sk1a}) reduces to
\be
S_{\alpha, \beta}^{1, r}=\mbox{const.} \times \sqrt{\frac{(r -1)! (2j - r +1)!}{\alpha! (2j_1 - \alpha)! (\beta -1)! (2j_2 -\beta +1)!}}, \qquad r=\alpha + \beta.
\label{sk1b}
\ee
To fix the constant in (\ref{sk1b}) let us complete the binomial coefficients in the radicand. Thus, we take $\mbox{const.} = \sqrt{(2j_1)! (2j_2)!/(2j)!}$ so that
\be
S_{\alpha, \beta}^{1, r} = \left[\frac{\left(
\begin{array}{c}
2j_1\\
\alpha
\end{array}
\right) \left(
\begin{array}{c}
2j_2\\
\beta -1
\end{array} \right)}{\left(
\begin{array}{c}
2j\\
r-1
\end{array}
\right)}\right]^{1/2}, \quad r=\alpha + \beta,
\label{sk1c}
\ee
with
\[
\left(
\begin{array}{c}
n\\m
\end{array}
\right) = \frac{n!}{(n-m)! \, m!}
\]
the binomial coefficient. Finally, it is a matter of substitution to verify that (\ref{sk1c}) is also a root of (\ref{corto3b}) for $k=1$. Then, the matrix elements $S^{1,r}_{\alpha, \beta}$ transform the representation of $J_{\alpha}^{(j_1, j_2)}$, $\alpha = \pm, z$, from the appropriate sectors of ${\cal H}_{(j_1, j_2)}$ into the first invariant subspace ${\cal H}_{j_1+j_2}$ of the Clebsch-Gordan decomposition (\ref{spaces}-\ref{cg1}). Moreover, the dimension $d_1 =2j+1$ of ${\cal H}_{j_1+j_2}$ is the highest of the dimensions of all the invariant subspaces ${\cal H}_j$, $j_1 + j_2 \geq j \geq \vert j_1-j_2 \vert$, since $d_k > d_{k+1}$ implies $d_1 >d_2 > \cdots > d_{n_0}$ (see eqs.~\ref{dimk} and \ref{dimk2}). Next, we are going to derive the expressions for the matrix elements $S^{k,r}_{\alpha, \beta}$ involving the invariant subspaces of dimension lower than $d_1$.

\subsection{Transformation into invariant subspaces of lower dimension.} 

\vskip1ex
\noindent
In this section we derive the matrix elements associated to $k\geq 2$. Let us change first $\alpha \rightarrow \alpha +1$, and then $\beta \rightarrow \beta+1$ in (\ref{corto3b}) to get the pair of equations
\[
\begin{array}{c}
\sqrt{(\alpha +1) ( 2j_1 -\alpha)} S_{\alpha, \beta}^{k, r } + \sqrt{(\beta -1) (2j_2 - \beta +2)} S_{\alpha +1, \beta-1}^{k, r} =  \sqrt{r(2j -2k-r+3)}  S_{\alpha +1, \beta}^{k, r+1},\\[3ex]
\sqrt{\alpha ( 2j_1 -\alpha +1)} \, S_{\alpha -1, \beta+1}^{k, r } + \sqrt{\beta (2j_2 - \beta +1)} \, S_{\alpha, \beta}^{k, r} = \sqrt{r(2j -2k-r+3)} \, S_{\alpha, \beta+1}^{k, r+1}.
\end{array}
\]
The substitution of this system in (\ref{corto3a}) leads to the recurrence relation,
\be
\begin{array}{l}
\sqrt{(\alpha +1)(\beta -1)(2j_1 - \alpha) (2j_2 -\beta +2)} \, S_{\alpha +1, \beta}^{k,r} \\[2ex]
\mbox{\hskip1cm} = \sqrt{\alpha \beta (2j_1 -\alpha +1)(2j_2 -\beta +1)} \, S_{\alpha -1, \beta+1}^{k,r}\\[2ex]
\mbox{\hskip2cm} + [\beta(2j_2 -\beta +1) +(\alpha +1)(2j_1 -\alpha) -r(2j-2k-r+3)] \, S_{\alpha, \beta}^{k,r},
\end{array}
\label{inv1}
\ee
where $k+r=\alpha +\beta +1$. To avoid square-roots in the coefficients of (\ref{inv1}) we make also the change 
\be
S_{\alpha, \beta}^{k,r}= \left[\alpha ! (\beta)_{\overline{\alpha}}(2j_1)_{\underline \alpha}(2j_2-\beta +1)_{\underline \alpha}\,\right]^{-1/2} S_{0,\, k+r-1}^{k,r} E_{\alpha, \beta}^{k,r}.
\label{inv2}
\ee
Here, $S_{0,\, k+r-1}^{k,r}$ and $E_{\alpha, \beta}^{k,r}$ are to be determined while the Pochhammer symbols
\be
(x)_{\underline n} =x(x-1) \cdots (x-n+1) \quad \text{and} \quad (x)_{\overline{n}} = x(x+1) \cdots (x+n-1)
\label{inv2a}
\ee
respectively represent the falling and rising factorials with 
\be
(x)_{\overline 0} =(x)_{\underline 0} =1, \quad (x)_{\overline n} = (x+n-1)_{\underline n}, \quad (x)_{\underline n} =(x-n+1)_{\overline n}.
\label{inv2b}
\ee
Notice that $\alpha =0$ in (\ref{inv2}) leads to $\beta =k+r-1$, and necessarily $E_{0, k+r-1}^{k,r} =1$. After introducing (\ref{inv2}) in (\ref{inv1}) one  arrives at the recurrence relation obeyed by $E_{\alpha, \beta}^{k,r}$ in terms of the $\alpha$-parameter:
\be
{\footnotesize
\begin{array}{l}
E_{\alpha +1,\, k+r-\alpha -2}^{k,r} =  -\alpha (k+r-\alpha-1)(2j_2 -k -r +\alpha +2)(2j_1 -\alpha +1) E_{\alpha-1,\, k+r-\alpha}^{k,r}\\[2ex]
\mbox{\hskip1cm} + [(k+r-\alpha-1)(2j_2-k-r+\alpha+2) + (\alpha+1)(2j_1 -\alpha) -r(2j-2k-r+3)] E_{\alpha, \beta}^{k,r},
\end{array}
}
\label{inv3}
\ee
with $\beta = k+r-\alpha-1$. In the solving of (\ref{inv3}) we take $r=1$ and proceed by induction on $\alpha$. The lowest value $\alpha =0$ gives rise to the expression
\be
E_{1, k-1}^{k,1} = -(k-1)(2j_2 -k+2) =-(k-1)_{\overline 1} (2j_2 -(k-1) +1)_{\underline 1}.
\label{inv4a}
\ee
For $\alpha =1$ one gets
\be
E_{2,k-2}^{k,1} = (k-2)(k-1) (2j_2 -k+3)(2j_2 -k +2) = (-1)^2 (k-2)_{\overline 2} (2j_2 -(k-2) +1)_{\underline 2}.
\label{inv4b}
\ee
In general, for $\alpha = \ell$,
\be
E_{\ell, \, k-\ell}^{k,1} = (-1)^{\ell} (k-\ell)_{\overline \ell} (2j_2 -(k-\ell) +1)_{\underline \ell}.
\label{inv4c}
\ee
Remark that $\ell =k$ produces $E_{k,0}^{k,1} =0$, henceforth $\ell \leq k-1$. On the other hand, allowing $\ell =0$ in (\ref{inv4c}) we get $E_{0,k}^{k,1} =1$, which is consistent with the constraint indicated after Eq.~(\ref{inv2}). Thus, making $k-\ell = \beta$  and $\ell =\alpha$ with $0 \leq \ell \leq k-1$, equation (\ref{inv4c}) reads as
\be
E_{\alpha, \, \beta}^{k,1} = (-1)^{\alpha} (\beta)_{\overline \alpha} (2j_2 -\beta+1)_{\underline \alpha}, \quad 0 \leq \alpha \leq k-1, \quad 1 \leq \beta \leq k.
\label{inv4d}
\ee
The substitution of this last result in (\ref{inv2}) gives the expression of $S_{\alpha, \beta}^{k,1}$ in terms of $S_{0,k}^{k,1}$,
\be
\begin{array}{rl}
S_{\alpha, \beta}^{k,1} & = (-1)^{\alpha} \displaystyle\left[ \frac{(\beta)_{\overline \alpha}}{\alpha!} \, \frac{(2j_2 -\beta +1)_{\underline \alpha}}{ (2j_1)_{\underline \alpha}} \right]^{1/2} S_{0,k}^{k,1}\\[3ex]
& = (-1)^{\alpha} \displaystyle\left[ \left(
\begin{array}{c}
k-1\\ 
\alpha
\end{array}
\right) \frac{(2j_2 -k+2)_{\overline \alpha}}{(2j_1 -\alpha +1)_{\overline \alpha}}
\right]^{1/2} S_{0,k}^{k,1},
\end{array}
\label{inv5}
\ee
where we have used (\ref{inv2b}). Since $S$ is unitary, we can fix the value of $S_{0,k}^{k,1}$ as follows
\[
\begin{array}{rl}
1={\displaystyle\sum_{\alpha =0}^{k-1}} \left( S_{\alpha, \beta}^{k,1} \right)^2 & = \left( S_{0,k}^{k,1} \right)^2 \left[ 1 + 
\left(
\begin{array}{c}
k-1\\1
\end{array} \right) \frac{(2j_2 -k+2)_{\overline 1}}{(2j_1)_{\overline 1}}+ \cdots  + \left(
\begin{array}{c}
k-1\\k-1
\end{array} \right) \frac{(2j_2 -k+2)_{\overline{k-1}}}{(2j_1 -k+2)_{\overline{k-1}}}
\right]\\[3ex]
& = \displaystyle\frac{\left( S_{0,k}^{k,1} \right)^2}{(2j_1 -k+2)_{\overline{k-1}}} \sum_{\alpha=0}^{k-1} \left(
\begin{array}{c}
k-1\\ 
\alpha
\end{array}
\right) (2j_1 -k+2)_{\overline{\alpha}} (2j_1 -k+2)_{\overline{k-1-\alpha}}.
\end{array}
\]
Using the addition formula Lemma~A4
\be
(a+b)_{\overline n} = \sum_{s=0}^n \left(
\begin{array}{c}
n\\k
\end{array}
\right) (a)_{\overline s} (b)_{\overline{n-s}}
\label{add1}
\ee
we finally get the roots
\be
S_{0,k}^{k,1} = \pm \sqrt{\frac{(2j_1 -k+2)_{\overline{k-1}}}{(2j -2k+4)_{\overline{k-1}}}} = \pm \sqrt{\frac{(2j_1)_{\underline{k-1}}}{(2j-k+2)_{\underline{k-1}}}}.
\label{inv6}
\ee
As a convention, hereafter we take the positive expression in (\ref{inv6}) as the definition of the matrix elements $S_{0,k}^{k,1}$. Therefore, (\ref{inv5}) becomes
\be
S_{\alpha,\beta}^{k,1} =(-1)^{\alpha} \displaystyle\left[ \frac{(\beta)_{\overline \alpha}}{\alpha!} \, \frac{(2j_2 -\beta +1)_{\underline \alpha} (2j_1 -\alpha)_{\underline{k-1}}
 }{ (2j -k+2)_{\underline{k-1}}} \right]^{1/2}.
\label{inv7}
\ee
To construct the remanent matrix elements we now use the recurrence relation (\ref{corto3b}),
\[
S_{\alpha, \beta}^{k, r+1} = \sqrt{\frac{\alpha (2j_1 -\alpha +1)}{r(2j-2k+3)}} S_{\alpha -1, \beta}^{k,r} \, + \sqrt{\frac{(\beta -1)(2j_2-\beta +2)}{r(2j-2k-r+3)}} S_{\alpha, \beta-1}^{k,1}.
\]
For $r=1$, the straightforward calculation produces
\be
S_{\alpha, \beta}^{k,2} = (-1)^{\alpha} F_{\alpha, \beta}^{k,2}\, \Theta_{\alpha -1, \beta}^{k,1}, 
\label{s-1}
\ee
with
\be
\Theta_{\alpha -1, \beta}^{k,1} = \left[ \frac{(\beta)_{\overline{\alpha -1}} (2j_2 -\beta +1)_{\underline{\alpha-1}} (2j_1+\beta-k)_{\underline{k-1}}}{1! \, (k-\beta)! \, (2j-k+2)_{\underline{k-1}} (2j-2k+2)_{\underline 1}} \,
\right]^{1/2},
\label{s-1b}
\ee
and
\be
\begin{array}{rl}
F_{\alpha, \beta}^{k,2}& = (\beta-1)(2j_2 -\beta +2) -\alpha (2j_1 -\alpha +1)\\[2ex]
& = [(\beta -1)_{\overline 1} (2j_2 -\beta +2)_{\overline 1}\,] \, {}_3F_2 \left[
\begin{array}{l}
-1, \, -\alpha, \, 2j_1 -\alpha +1\\[1ex]
\beta -1, \, -2j_2 + \beta -2
\end{array}
\right].
\end{array}
\label{s-1a}
\ee
To derive (\ref{s-1a}) we have used Lemma~A5 of the appendix with ${}_3F_2$ is the generalised hypergeometric function
\be
{}_3F_2(a,b,c;d,e;z) \equiv {}_3F_2 \left[
\begin{array}{l}
a, b, c; \, z\\[1ex]
d,e  
\end{array}
\right] = \sum_{s=0}^{+\infty} \frac{(a)_{\overline s} (b)_{\overline s} (c)_{\overline s}}{(d)_{\overline s} (e)_{\overline s}} \frac{z^s}{s!}.
\label{hyper1}
\ee
Here we are adopting the convention ${}_3F_2(a,b,c;d,e) \equiv  {}_3F_2(a,b,c;d,e;1)$. Now we take $r=2$ to arrive at
\be
S_{\alpha, \beta}^{k,3} = (-1)^{\alpha}F_{\alpha, \beta}^{k,3}\, \Theta_{\alpha -2, \beta}^{k,2}.
\label{s-2}
\ee
The expression for $\Theta_{\alpha -2, \beta}^{k,2}$ can be obtained from $\Theta_{\alpha -1, \beta}^{k,1}$ in (\ref{s-1b}) after the change $1 \rightarrow 2$. For the first factor in (\ref{s-2}) we have
\be
\begin{array}{rl}
F_{\alpha, \beta}^{k,3} & =(\beta-1)(2j_2 -\beta +2) F_{\alpha, \beta-1}^{k,2} -\alpha (2j_1 -\alpha +1) F_{\alpha-1, \beta}^{k,2}\\[3ex]
& =\displaystyle\sum_{s=0}^2 (-1)^s \left(
\begin{array}{c}
2\\s
\end{array}
\right) (\alpha)_{\underline s} (\beta-1)_{\underline{2-s}} (2j_1 -\alpha+1)_{\overline s}(2j_2 -\beta +2)_{\overline{2-s}}\\[3ex]
& = [(\beta -2)_{\overline 2} (2j_2 -\beta +2)_{\overline 2}\,] \, {}_3F_2 \left[
\begin{array}{l}
-2, \, -\alpha, \, 2j_1 -\alpha +1\\[1ex]
\beta -2,\,  -2j_2 + \beta -3
\end{array}
\right].
\end{array}
\label{s-2a}
\ee
In general, one can apply induction on $r$ to verify that the roots of the system (\ref{corto3a}-\ref{corto3b}) for $k\geq 1$ are given by
\be
S_{\alpha, \beta}^{k,r+1} = (-1)^{\alpha}F_{\alpha, \beta}^{k,r+1}\, \Theta_{\alpha -r, \beta}^{k,r},
\label{s-3a}
\ee
with $\Theta_{\alpha -r, \beta}^{k,r}$ the immediate generalisation of (\ref{s-1b}),
\be
\Theta_{\alpha -r, \beta}^{k,r} = \left[ \frac{(\beta)_{\overline{\alpha -r}} (2j_2 -\beta +1)_{\underline{\alpha-r}} (2j_1+\beta-k)_{\underline{k-r}}}{r! \, (k-\beta)! \, (2j-k+2)_{\underline{k-r}} (2j-2k+2)_{\underline r}} \,
\right]^{1/2}
\label{s-3b}
\ee
and $F_{\alpha, \beta}^{k,r+1}$ the generalisation of (\ref{s-2a}), see Lemma~A5 of the appendix,
\be
F_{\alpha, \beta}^{k,r+1}= [(\beta -r)_{\overline r} (2j_2 -\beta +2)_{\overline r}\,] \, {}_3F_2 \left[
\begin{array}{l}
-r, \, -\alpha, \, 2j_1 -\alpha +1\\[1ex]
\beta -r, \, -2j_2 + \beta -r-1
\end{array}
\right].
\label{s-3c}
\ee
To close this section we give the explicit form of the matrix $S$ for two of the cases discussed in the previous sections. If $j_1=j_2=\tfrac12$ one has
\[
S=\left(\begin{array}{ccc|c}
S^{1,1}_{0,1} & 0 & 0 & 0\\
0 & S^{1,2}_{0,2} & 0 & S^{2,1}_{0,2} \\[1ex] \hdashline
0 & S^{1,2}_{1,1} & 0 & S^{2,1}_{1,1}\\
0 & 0 & S^{1,3}_{1,2} & 0
\end{array}\right)
=\left(\begin{array}{ccc|c}
1 & 0 & 0 & 0\\
0 & \frac{1}{\sqrt{2}} & 0 & \frac{1}{\sqrt{2}}\\[1ex] \hdashline
0 & \frac{1}{\sqrt{2}} & 0 & -\frac{1}{\sqrt{2}}\\
0 & 0 & 1 & 0
\end{array}\right),
\]
while $j_1=1$ and $j_2=\tfrac12$ leads to
\[
S=\left(\begin{array}{cccc|cc}
S^{1,1}_{0,1} & 0 & 0 & 0 & 0 & 0\\
0 & S^{1,2}_{0,2} & 0 & 0 & S^{2,1}_{0,2} & 0 \\[0.6ex] \hdashline
0 & S^{1,2}_{1,1} & 0 & 0 & S^{2,1}_{1,1} & 0\\
0 & 0 & S^{1,3}_{0,2} & 0 & 0 & S^{2,2}_{1,2} \\[0.6ex]\hdashline
0 & 0 & S^{1,3}_{2,1} & 0 & 0 & S^{2,2}_{2,1}\\
0 & 0 &0 & S^{1,4}_{2,2} & 0 & 0
\end{array}\right)
=\left(\begin{array}{cccc|cc}
1 & 0 & 0 & 0 & 0 & 0\\
0 & \sqrt{ \frac{1}{3}} & 0 & 0 & \sqrt{ \frac{2}{3}} & 0 \\[0.6ex] \hdashline
0 & \sqrt{ \frac{2}{3}} & 0 & 0 & -\sqrt{ \frac{1}{3}} & 0\\
0 & 0 & \sqrt{ \frac{2}{3}} & 0 & 0 & \sqrt{ \frac{1}{3}} \\[0.6ex]\hdashline
0 & 0 & \sqrt{ \frac{1}{3}} & 0 & 0 & -\sqrt{ \frac{2}{3}}\\
0 & 0 &0 & 1 & 0 & 0
\end{array}\right).
\]

\subsection{Addition of angular momenta}

Consider a bipartite system integrated by independent subsystems ${\cal S} ={\cal S}_1 + {\cal S}_2$. Any of the observables belonging to ${\cal S}$ should be constructed in terms of the Kronecker products $A\otimes B$, with $A$  and $B$ observables (identities included) of ${\cal S}_1$ and ${\cal S}_2$ respectively. The product $SU(2) \times SU(2)$ defines the symmetry for the coupled system ${\cal S}={\cal S}_1+{\cal S}_2$, where the Lie group $SU(2)$ represents the symmetry of each of the subsystems ${\cal S}_1$ and ${\cal S}_2$. That is, the component ${\cal S}_k$ is characterised by the requirement that the $n_k$-dimensional vector space $\mbox{Sp} \{ \vert e_{i_k}^{n_k} \rangle \}_{i_k=1}^{n_k}$ defines an irreducible representation of $SU(2)$ for $k=1,2$ (This is the situation for the ${}^1H$ and ${}^{13}C$ nuclei which are both $SU(2)$ nuclear spin systems, and ${}^1H^{13}C$ which is a $SU(2) \times SU(2)$ system \cite{Kri99}). Then, the angular momenta $\vec J_1$, $\vec J_2$, as well as their sum $\vec J= \vec J_1 + \vec J_2$, are conserved. These quantities are respectively the generators of the rotations acting on the first system alone, the second one alone, and both systems simultaneously \cite{Bac77}. The diagonalization of $\vec J \cdot \vec J$ and $\vec J_z = \vec J_{1z} + \vec J_{2z}$ gives rise to the orthonormal basis vectors $\vert J, M \rangle$ belonging to the eigenvalues $J(J+1)$ and $M$ respectively. The two subsystems in turn are described by a state vector which belongs to $\mbox{Sp} \{\vert j_1,m^{(j_1)} \rangle \otimes \vert j_2, m^{(j_2)} \rangle \}$, with $J=j_1+j_2$ and $M= m^{(j_1)} + m^{(j_2)}$. This is therefore useful to know the coefficients of the change of basis
\be
\vert J,M \rangle = \sum_{m^{(j_1)}, m^{(j_2)}} \langle j_1,m^{(j_1)}; j_2, m^{(j_2)}\vert J,M \rangle \vert j_1,m^{(j_1)}; j_2, m^{(j_2)}\rangle.
\label{cgcoef}
\ee

\subsubsection{Clebsch-Gordan coefficients}

Let $\{\vert J,M\rangle\}_{M=-J}^{J}$ be the eigenvectors of the operator (87) and
$S$ the unitary matrix defined in (\ref{s1}). Since $S$ is unitary we may write 
\[
\begin{array}{ll}
 M= \langle J,M\vert \tilde{J}_3^{(j_1,j_2)}\vert J,M\rangle & =\langle J,M \vert
S^\dagger J_3^{(j_1,j_2)}S\vert J,M\rangle\\[2ex]
& =\langle j_1,m^{(j_1)}; j_2, m^{(j_2)}\vert
J_3^{(j_1,j_2)}\vert j_1,m^{(j_1)}; j_2, m^{(j_2)}\rangle =m^{(j_1)}+m^{(j_2)},
\end{array}
\]
where $\vert j_1,m^{(j_1)}; j_2, m^{(j_2)} \rangle = \vert j_1,m^{(j_1)} \rangle \otimes \vert j_2, m^{(j_2)}\rangle$ and the Corollary~TM2.2 was used. Therefore, up to a global phase (fixed as 1), we must have
\[
 \vert j_1,m^{(j_1)}; j_2, m^{(j_2)}\rangle=S\vert J,M\rangle.
\]
Hence
\[
\langle J', M' \vert j_1,m^{(j_1)}; j_2, m^{(j_2)}\rangle=\langle J',M'\vert S\vert
J,M\rangle
\]
defines the Clebsch-Gordan coefficient required in (\ref{cgcoef}). Thus, the entries of the (unitary) transformation matrix $S$ are associated to the coupling coefficients of $SU(2)\otimes SU(2)$.

\section{Concluding remarks}

We have applied the Hubbard operators in the study of the Kronecker product of square matrices. The former represent a shorthand notation for the direct product that transforms complicated calculations involving large matrices or a large number of factors into simple relations of subscripts. Thus, the Hubbard representation is compact enough to facilitate the study of the algebra and group properties of the observables defining a multipartite quantum system, no matter the order or the number of the corresponding matrices. In particular, we have shown that the construction of permutation matrices, the identification of the corresponding permutation classes of equivalence and the construction of symmetrization operators is straightforward. All the basic properties of the Kronecker product of square matrices have been revisited in the Hubbard representation. In this framework the proofs of the corresponding theorems, lemmas and corollaries are achieved in easy form. As an immediate application we have constructed irreducible representations of $SU(2)$ by giving concrete expressions for the involved matrices in Hubbard notation. The same has been done for the product group $SU(2) \times SU(2)$. The solution of the Clebsch-Gordan decomposition of $SU(2) \times SU(2)$ by using the Hubbard representation lead to definite expressions for the Clebsch-Gordan coefficients of the addition of angular momenta in terms of the hypergeometric function ${}_3F_2$. Some connections can be found with the results already reported in \cite{Var88}. In this context we like to stress that our results are in contraposition to the common affirmation that ``it is not possible to give a general expression for the Clebsch-Gordan coefficients'' (see e.g. \cite{Coh78}, pp 1023). Therefore, the expressions derived in Section~\ref{titcle2} of the present work give an alternative to obtain the Clebsch-Gordan coefficients that is different from the calculating by iteration or the checking in numerical tables. We hope our results have shed some light on the matter.

\subsection*{Acknowledgments} 

The support of CONACyT project 152574 and IPN project SIP-SNIC-2011/04 is acknowledged. M.E. gratefully acknowledges the funding received towards his PhD through the CONACyT Scholarship. 

\appendix
\section{Appendix}
\numberwithin{equation}{section}
\setcounter{equation}{1}

Some properties of the ceiling function $\lceil\cdot\rceil$ and the floor function $\lfloor \cdot \rfloor$ used through this paper are proven. For a given real number
$x$ the ceiling function
\be
\lceil x\rceil=\min\{z\in\mathbb{Z}:x\leq z\}
\label{ceilfun}
\ee
yields the smallest integer greater than or equal to $x$. On the another hand, the floor function
\be
\lfloor x\rfloor=\max\{z\in\mathbb{Z}:z\leq x\}
\label{floorfun}
\ee
gives the largest previous integer to $x$.

\begin{itemize}
\item[]
{\bf Lemma~A1.} For $x\in\mathbb{R}$ and $m, n\in\mathbb{N}$ it is fulfilled

\item[]
(i) $\lceil x\rceil = n$ if and only if $x \leq n < x+1$
 
\item[]
(ii) $\lceil x + n \rceil = \lceil x \rceil + n$.
 
\item[]
(iii) $\lfloor x + n \rfloor = \lfloor x \rfloor + n$.

\item[]
(iv) $\displaystyle \left\lceil \frac{x}{m n} \right\rceil = \left\lceil
\frac{\left\lceil x/n\right\rceil}{m} \right\rceil.$

\item[]
(v) $\displaystyle \left\lceil \frac{n+1}{m} \right\rceil = \left\lfloor
\frac{n}{m} \right\rfloor + 1$.

\end{itemize}


\noindent
{\em Proof.} 

\vskip1ex
\noindent
(i) From the definition of ceiling function it is clear that
\be
\label{desceil}
\lceil x \rceil = n \mbox{ if and only if } n-1 < x \leq n.
\ee
Rearranging these last inequalities we get $x\leq n<x+1$.

\vskip1ex
\noindent
(ii) By virtue of (\ref{desceil}) we may write
\[
\lceil x\rceil-1 < x \leq\lceil x\rceil.
\]
Then
\[
\lceil x\rceil+n-1 < x + n\leq\lceil x\rceil + n,
\]
and (\ref{desceil}) implies $\lceil x + n \rceil = \lceil x \rceil + n$.

\vskip1ex
\noindent
(iii) Departing from 
\be
\lfloor x \rfloor\leq x < \lfloor x \rfloor + 1,
\label{desfloor}
\ee
one gets
\[
\lfloor x\rfloor+n\leq x+n<\lfloor x\rfloor+n+1,
\]
and $\lfloor x + n\rfloor =\lfloor x\rfloor + n$.

\vskip1ex
\noindent
(iv) Consider a continuous, monotonically increasing function $f$. We know that if $f(x)$ is an integer then $x$ is an integer \cite{Gra89}. Therefore 
\be
\lceil f(x)\rceil = \lceil f(\lceil x\rceil)\rceil.
\label{ceilfn}
\ee
Let us take the continuous and monotonically increasing function $f(x) = \frac xm$. According to the statement above, if $f(x) = k$ with $k\in\mathbb{Z}$, then $x = k m$ is an integer. Hence, from equation (\ref{ceilfn}) it follows
\[
\left\lceil\frac xm \right\rceil = \left\lceil\frac{\lceil x\rceil}{m}\right\rceil,
\]
Make $x\rightarrow x/n$ in the last equation the proof is completed.

\vskip1ex
\noindent
(v) According to (\ref{desceil}) we may write
\be
\left\lceil \frac{n+1}{m}\right\rceil - 1 < \frac{n+1}{m} \leq \left \lceil
\frac{n+1}{m} \right\rceil.
\label{ceiflo1}
\ee
From the right inequality we get
\[ 
\frac nm\leq \left\lceil \frac{n+1}{m} \right\rceil - \frac1m < \left\lceil
\frac{n+1}{m}\right\rceil.
\]
Adding 1 in both sides one has
\be
\frac nm + 1 < \left\lceil \frac{n+1}{m} \right\rceil + 1.
\label{ceiflo0}
\ee
The left inequality in (\ref{ceiflo1}) implies $\left\lceil
\frac{n+1}{m}\right\rceil < \frac nm + 1 + \frac1m$, this is true if and only if
\be
\left\lceil \frac{n+1}{m}\right\rceil \leq \frac nm + 1.
\label{ceiflo2}
\ee
Combining (\ref{ceiflo0}) and (\ref{ceiflo2}) we arrive at
\[
\left\lceil \frac{n+1}{m}\right\rceil\leq \frac{n}{m} + 1 < \left\lceil
\frac{n+1}{m} \right\rceil + 1.
\]
From equation (\ref{desfloor}) we see that
\be
\left\lfloor \frac nm + 1\right\rfloor = \left\lceil\frac{n+1}{m}\right\rceil.
\ee
Using (iii) in the left side of this last equation one gets the result we are looking for. $\Box$


\begin{itemize}
\item[]
{\bf Lemma~A2.} Let $F(i,j,k,l)$ be a function of the indices $i,j,k,l$. Then
\be
\sum_{i,j}^m\sum_{k,l}^n
F(i,j,k,l)\delta_{n(j-1)+l}^{n(i-1)+k}=\sum_{i,j}^m\sum_{k,l}^nF(i,j,k,l)\delta_i^j\delta_k^l.
\ee

\end{itemize}


\noindent
{\em Proof}. Let us define $p=n(i-1)+k$ and $q=n(j-1)+l$. Note that
$p,q=1,2,\ldots,mn$ Thus
\[
\begin{array}{ll}
\displaystyle\sum_{i,j}^m\sum_{k,l}^n F(i,j,k,l)\delta_{n(j-1)+l}^{n(i-1)+k}\\[3.5ex]
\hspace*{1cm}=\displaystyle\sum_{i,j}^m\sum_{ p=n(i-1)+1}^{ni}\sum_{\quad
q=n(j-1)+1}^{nj} F(i,k,p-n(i-1),q-n(i-1))\delta_p^q\\[3.5ex]
\hspace*{1cm}=\displaystyle\sum_{p,q}F\left(p',q',p'',q''\right)\delta_p^q=\displaystyle\sum_{p}F\left(p',p',p'',p''\right)\\[3.5ex]
\hspace*{1cm}=\displaystyle\sum_{i=1}^m\sum_{p=n(i-1)+1}^{ni}
F(i,i,p-n(i-1),p-n(i-1))\\[3.5ex]
\hspace*{1cm}=\displaystyle\sum_{i=1}^m\sum_{k=1}^n
F(i,i,k,k)=\sum_{i,j}^m\sum_{k,l}^n F(i,j,k,l)\delta_i^j\delta_k^l.
\end{array}
\]
Where we have used Lemma~A1(i). $\Box$

\begin{itemize}
\item[]
{\bf Lemma~A3.} Let $\alpha,\beta, n_1,n_2\in\mathbb{N}$ be such that $\alpha=1,\ldots
n_1-1; $ $\beta=1,\ldots n_2$, with $n_1\ge n_2$. The ceiling function satisfies the following properties

(i) $\left\lceil\frac{\beta}{n_2}\right\rceil=1$.

(ii) $p'=\alpha+1$, where $p'=\lceil\frac{p}{n_2}\rceil$

(iii)  $(p-n_2)'=\alpha$, for $\alpha>0$.

\end{itemize}

\noindent
{\em Proof. }

\vskip1ex
\noindent
(i) For any $x\in\mathbb{R}$ we have $\left\lceil x\right\rceil=1$ if $x\le 1$. The proof follows by noticing that $\frac{\beta}{n_2}\le 1$.

\vskip1ex
\noindent
(ii) Using (i) and Lemma~A1(iii) one gets
\[
p'=\left\lceil \frac{\alpha n_2+\beta}{n_2}\right\rceil =\left\lceil \alpha+\frac{\beta}{n_2}\right\rceil =\alpha +\left\lceil\frac{\beta}{n_2} \right\rceil =\alpha+1,
\]
where $\beta= p-n_2p'+n_2$.

\vskip1ex
\noindent
(iii) The proof is similar to the one of (ii). Yet,
\[
(p-n_2)'= \left\lceil \frac{\alpha n_2 +\beta-n_2}{n_2} \right\rceil = \left\lceil \alpha-1 + \frac{\beta}{n_2} \right\rceil =\alpha-1+1= \alpha.\quad\Box
\]

\begin{itemize}
\item[]
{\bf Lemma A4.} Let $n,a, b \in \mathbb{N}$. We have
\begin{equation}
\sum_{i=0}^{n}\frac{(a)_{\overline{n-i}}~(b)_{\overline
n}}{(n-i)!~i!}=\frac{(a+b)_{\overline n}}{n!},
\end{equation}

where $(x)_{\overline{n}}=x(x+1)\cdots(x+n-1)$.
\end{itemize}

\noindent
{\em Proof.} From the binomial expansion we have
\[
(1-x)^{-a}=\sum_{j=0}^\infty\frac{(a)_{\overline j}}{j!}x^j.
\]
Note
\[
(1-x)^{-a}(1-x)^{-b}=\left(\sum_{l=0}^\infty\frac{(a)_{\overline
l}}{l!}\right)\left(\sum_{m=0}^\infty\frac{(b)_{\overline
m}}{m!}\right)=\sum_{l,m}^\infty\frac{(a)_{\overline l}~(b)_{\overline
m}}{l!m!}x^{l+m}\\[3ex]
\]
We make the change $i=l+m$ to get
\[
(1-x)^{-a}(1-x)^{-b}=\sum_{m=0}^\infty\sum_{i=m}^\infty\frac{(a)_{\overline{i-m}}~(b)_{\overline
m}}{(i-m)!~m!}x^i=\sum_{i=0}^\infty\sum_{m=0}^i\frac{(a)_{\overline{i-m}}~(b)_{\overline
m}}{(i-m)!~m!}x^i.
\]
On the other hand,
\[
(1-x)^{-(a+b)}=\sum_{n=0}^\infty\frac{(a+b)_{\overline n}}{n!}x^n.
\]
Comparing term by term the two last equations the proof is completed. $\Box$

\begin{itemize}
\item[]
{\bf Lemma A5.} The generalised hypergeometric function $_3F_2$ satisfies
\be
(d)_{\overline r} (e)_{\underline r}~{}_3F_2(-r,-b,c;d,-e)=\displaystyle
\sum_{s=0}^r(-1)^s\binom{r}{s}~(b)_{\underline s}~(c)_{\overline
s}~(d+1)_{\underline{r-s}}~(e)_{\underline r},
\ee
with $r,b,c,d,e\in\mathbb{Z}$. The right and left Pochhammer symbols are defined as
$(x)_{\overline{n}}=x(x+1)\cdots(x+n-1)$ and
$(x)_{\underline{n}}=x(x-1)\cdots(x-n+1)$, respectively.
\end{itemize}

\noindent
{\em Proof.} Let us take $r=1$, then
\[
{}_3F_2(-1,-b,c;d,-e)=\frac{1}{de}\left(de-bc\right)=\displaystyle\frac{1}{(d)_{\overline
1}(e)_{\underline 1}}\sum_{s=0}^1(-1)^s\binom{1}{s}(b)_{\underline s}~(c)_{\overline
s}~(d+1)_{\underline{1-s}}~(e)_{\underline 1}.
\]
If now we set $r=2$,
\[
\begin{array}{ll}
{}_3F_2(-2,-b,c;d,-e)=\displaystyle{1-\frac{2bc}{de}+\frac{b(b-1)c(c+1)}{d(d+1)e(e-1)}}\\[3.5ex]
\hspace*{3.7cm}=\displaystyle\frac{1}{(d)_{\overline 2}(e)_{\underline
2}}\sum_{s=0}^2(-1)^s\binom{2}{s}(b)_{\underline s}~(c)_{\overline
s}~(d+1)_{\underline{2-s}}~(e)_{\underline 2}.
\end{array}
\]
The proof is completed by induction on $r$. $\Box$


\end{document}